\documentclass[a4paper,11pt]{article}
\pdfoutput=1 % if your are submitting a pdflatex (i.e. if you have
\usepackage{jheppub} % for details on the use of the package, please
\usepackage{graphicx,array}
\usepackage{float}
\usepackage{hyperref}
\usepackage{xcolor}
\usepackage{amsmath,amssymb,latexsym}
\usepackage{subcaption}
\usepackage{slashed}
\usepackage{bm}

\providecommand{\ket}[1]{\left|#1\right\rangle}
\providecommand{\bra}[1]{\left\langle#1\right|}
\makeatletter
\def\braket#1{\@ifnextchar\bgroup{\@braket{#1}}{\left\langle#1\right\rangle}}
\def\@braket#1#2{\left\langle#1\,\middle|\,#2\right\rangle}
\makeatother

\usepackage{array}
\newcolumntype{P}[1]{>{\centering\arraybackslash}p{#1}}
\newcolumntype{M}[1]{>{\centering\arraybackslash}m{#1}}
\definecolor{darkgreen}{rgb}{0.0, 0.8, 0.0} 
\definecolor{darkblue}{cmyk}{1,0.4,0,0.3}
\definecolor{violet}{cmyk}{0,1,0,0.2}
\hypersetup{colorlinks, bookmarksnumbered, citecolor=darkblue, linkcolor=darkblue, pdfstartview=FitH, urlcolor=darkblue, linktocpage}

\newcommand{\GeV}{\mathrm{GeV}}
\newcommand{\TeV}{\mathrm{TeV}}

\def\beq{\begin{equation}}
\def\eeq{\end{equation}}
\def\beqa{\begin{eqnarray}}
\def\eeqa{\end{eqnarray}}

\newcommand{\eg}{\textit{e.g.}}
\newcommand{\ie}{\textit{i.e.}}

\title{Next-to-Leading-Order Electroweak Corrections to Quantum Observables in Lepton-Lepton Collisions}

\author[1]{Spencer Chang,}
\author[1]{Thomas Driscoll,}
\author[2,3]{Sokratis Trifinopoulos}

\affiliation[1]{Department of Physics and Institute for Fundamental Science,\\
University of Oregon, Eugene, Oregon 97403, USA}
\affiliation[2]{Theoretical Physics Department, CERN, 1211 Geneva 23, Switzerland}
\affiliation[3]{Department of Physics and Astronomy, Northwestern University, Evanston, IL 60208, USA}

\abstract{
We study how virtual next-to-leading-order electroweak corrections affect the quantum observables of entanglement and magic in lepton--lepton collisions. Treating the outgoing particles as multi-qudit systems, \ie, qubits for spin-$\tfrac{1}{2}$ fermions and qutrits for massive vector bosons, we compute entanglement entropy and the stabilizer second R\'enyi entropy for the processes
$\ell^+\ell^- \to \tau^-\tau^+,\, ZZ$ 
at one-loop accuracy. We show that radiative corrections coherently modify the quantum structure of the final state, shifting regions of maximal entanglement and altering the generated magic relative to leading order.
Our results demonstrate that these quantum observables provide novel, precision-sensitive probes of electroweak dynamics at future lepton colliders and open a path toward their use in searches for new physics.
}

\begin{document}

\maketitle

\flushbottom

\section{Introduction}
\label{sec:intro}
The scattering of elementary particles is, at its core, a quantum process.  
Every collider event originates from the coherent superposition of helicity amplitudes that interfere according to the symmetries of the Standard Model (SM).  
Yet, when cross sections are computed, most of this quantum structure is integrated out, leaving only classical probabilities.  
A growing effort aims to recover and quantify the genuinely quantum content of high-energy reactions by treating the amplitudes themselves as carriers of quantum information (see Refs.~\cite{Larkoski:2022lmv,Afik:2025ejh} for related perspectives and a recent review).  
In this view, each spin is treated as a qudit, \ie, a qubit for spin-$\tfrac{1}{2}$ particles and a qutrit for massive vector bosons, and the outgoing state of a scattering process can therefore be regarded as a multi-qudit quantum state. Collider observables can then be recast by applying standard tools from quantum information (QI) theory directly to the building blocks of quantum field theory.

In practice, one proceeds by constructing observables that witness quantum resources---most notably \emph{entanglement}---from measurable spin--polarization structures that isolate genuinely quantum correlations generated by interference among helicity amplitudes. When feasible, the relevant spin/polarization density matrix can be reconstructed via \emph{quantum tomography} from differential angular data, enabling direct tests of entanglement at the level of the reconstructed state. This strategy has been developed and tested in concrete settings, most notably in $t\bar t$ production, where bipartite qubit correlations are now theoretically controlled and experimentally accessible at the LHC~\cite{Afik:2020onf,Severi:2021cnj,Afik:2022kwm,Aguilar-Saavedra:2022uye,Severi:2022qjy,Dong:2023xiw,Aoude:2022imd,Fabbrichesi:2021npl,Fabbrichesi:2022ovb,Afik:2022dgh,Aguilar-Saavedra:2023hss,Han:2023fci,Simpson:2024hbr,Cheng:2024btk,Han:2024ugl,Aguilar-Saavedra:2024hwd,Fabbrichesi:2025psr,Maltoni:2024tul,ATLAS:2023fsd,ATLAS:2023jzs,CMS:2024pts,CMS:2025cim,Barr:2024djo}. The same general strategy has also been extended to other channels such as the ones featuring diboson and $\tau^-\tau^+$ final states, while automated tools for computing the corresponding spin-density matrices have also begun to appear~\cite{Barr:2021zcp,Barr:2022wyq,Fabbrichesi:2023cev,Aguilar-Saavedra:2022wam,Aguilar-Saavedra:2022mpg,Aoude:2023hxv,Grabarczyk:2024wnk,Fabbri:2023ncz,Morales:2023gow,Morales:2024jhj,Altomonte:2024upf,Fabbrichesi:2023jep,Bernal:2023ruk,Bernal:2024xhm,Goncalves:2025mvl,Bi:2023uop,Du:2024sly,Aguilar-Saavedra:2024vpd,Aguilar-Saavedra:2024whi,Cheng:2024rxi,Subba:2024aut,Subba:2024mnl,Cheng:2025cuv,Ehataht:2023zzt,Han:2025ewp,LoChiatto:2024dmx,DelGratta:2025qyp,Grossi:2024jae,Durupt:2025wuk}. Alongside entanglement, we will also track \emph{magic} (non-stabilizerness) as a complementary resource; one can reconstruct or constrain the relevant spin/polarization density matrix from differential data and then evaluate certified monotones or lower bounds. Collider applications of magic and related collider diagnostics of quantum resources have so far focused on top production, QED scattering, new-physics sensitivity studies, and methodological questions surrounding collider Bell analyses~\cite{White:2024nuc,Liu:2025qfl,Aoude:2025jzc,Fabbrichesi:2025ywl,Liu:2025frx,Abel:2025skj,Bechtle:2025ugc,Low:2025aqq}.

The next step in this program is to move beyond leading-order predictions and investigate how radiative corrections modify the quantum resources generated in SM processes.  
While most collider observables depend only on squared matrix elements, entanglement and magic are sensitive to relative phases between helicity amplitudes.  
Direct higher-order investigations of collider observables sensitive to quantum resources have only recently begun: sizable next-to-leading-order (NLO) electroweak effects were identified in Higgs decays to vector bosons, in $h\to WW^\ast$, and in the extraction of $Z$-boson spin-density matrices through quantum tomography~\cite{DelGratta:2025qyp,Goncalves:2025xer,DelGratta:2025xjp}. More broadly, Ref.~\cite{Aguilar-Saavedra:2025byk} emphasized that decay-based quantum tomography is intrinsically a leading-order (LO) construct and that higher-order contributions must be modeled carefully rather than absorbed into a naive tree-level interpretation, while Ref.~\cite{Grossi:2024jae} showed from a phenomenological diboson analysis that higher-order, off-shell and fiducial effects can substantially reshuffle which angular-coefficient combinations provide the most reliable diagnostics of quantum resources.
The related loss of spin entanglement induced by radiation has also been analyzed from an open-quantum-system perspective~\cite{Aoude:2025ovu}.

In this work, we perform the first systematic study of electroweak NLO corrections to quantum observables in \emph{lepton-lepton} collisions---specifically entanglement and magic, \ie, the quantum resources of the reconstructible final-state helicity density matrix\footnote{We use the term ``Observables'' for functions of the measurable helicity density matrix, and not in the strict operator sense.}---~\cite{Passarino:1978jh,Bohm:1987ck,Denner:1988tv,Denner:2000bj}. 
Future machines such as \emph{Future Circular Collider (FCC-ee)}~\cite{FCC:2018evy} and the multi-TeV \emph{Muon Collider}~\cite{InternationalMuonCollider:2024jyv,AlAli:2021let} will provide high luminosities and, in several designs, significant control over the initial spin configuration, enabling a detailed exploration of the electroweak sector in a clean, fully perturbative environment (for related studies of quantum resources in future lepton-collider settings, see Refs.~\cite{Altakach:2022ywa,Fabbrichesi:2024wcd,Breuning:2024wxg,Maltoni:2024csn,Ruzi:2024cbt,Wu:2024ovc,Ding:2025mzj,Altakach:2026fpl,Guo:2026yhz,Zhang:2026nwm,Fang:2026ddi,Gabrielli:2026tnl}).  
Leptonic initial states are ideal for this purpose: they are free of QCD uncertainties, allow for controlled spin preparation, and produce final states whose quantum structure can be directly related to the underlying helicity amplitudes.  
We consider representative processes,
\begin{equation} \label{eq:processes}
\ell^-\ell^+ \;\to\; \tau^-\tau^+,~ ZZ,
\end{equation}
which respectively cover fermion-pair and vector-boson-pair final states, and thus probe qubit and qutrit quantum correlations. While calculations are performed for electron-positron initial states throughout this work, the results are equally valid for $\mu^- \mu^+$ scattering up to $\mathcal{O}(m_\mu/\sqrt s)$ corrections, where $\sqrt s$ denotes the center-of-mass (CM) energy\footnote{While scattering channels with same-helicity initial states are suppressed by overall factors of $m_f/\sqrt{s}$, this is removed by the normalization of the density matrix, thus the statement holds for all channels we consider.}. Our analysis focuses on the production-side quantum state before decay reconstruction and spin-analyzing powers are folded in, and is thus complementary to the kinematic-versus-decay tomography framework of Ref.~\cite{Cheng:2024rxi}.

The remainder of this paper is organized as follows. In Sec.~\ref{sec:QI_and_scattering} we set up the scattering framework, define the quantum resources and the observables used to quantify them, and separate the virtual fixed-Born density matrix from the power-suppressed contribution of unresolved real emission. Sections~\ref{sec:results_tautau} and~\ref{sec:results_zz} then present our results for $\tau^-\tau^+$ and $ZZ$ production, highlighting when the virtual corrections mainly shift the positions of extrema in the quantum observables and when they instead reshape the high-energy helicity structure. We conclude in Sec.~\ref{sec:conclusions}. Appendix~\ref{app:tau_slices} collects complementary plots omitted from the main discussion. Appendix~\ref{app:pert_expansion} gives the perturbative expansion of the normalized quantum observables, while App.~\ref{app:loop_Comp} collects the technical details of the one-loop computation and software workflow.

\section{Quantum Information and Scattering Theory}
\label{sec:QI_and_scattering} 

\subsection{$2 \to 2$ Scattering }
\label{sec:scattering} 

Scattering processes are defined in the Fock space of asymptotic 
multi-particle states,
\begin{equation}
\mathcal F = \bigoplus_{n\ge 0} \mathcal H^{(n)},
\end{equation}
where $\mathcal H^{(n)}$ denotes the $n$-particle Hilbert space.
The dynamics are encoded in the unitary $S$-matrix,
\begin{equation} \label{eq:S_operator}
S : \mathcal F_{\rm in} \to \mathcal F_{\rm out} ,
\qquad
S^\dagger S = \mathbf{1},
\qquad
S = \mathbf{1} + iT.
\end{equation}

\paragraph{Preparation of the initial state.}

We consider $2\to2$ scattering processes and therefore work in the 
two-particle sector of the incoming Fock space, $\mathcal H^{(2)}_{\mathrm{in}} \subset \mathcal F_{\mathrm{in}}$.
Within this sector the Hilbert space factorizes into two 
single-particle subspaces associated with the incoming particles,
\begin{equation}
\mathcal H^{(2)}_{\mathrm{in}}
\simeq
\mathcal H_{1} \otimes \mathcal H_{2}.
\end{equation}
Each single-particle Hilbert space decomposes into helicity and 
momentum degrees of freedom,
\begin{equation}
\mathcal H_{i}
\simeq
\mathcal H_{\mathrm{hel},i}
\otimes
\mathcal H_{\mathrm{kin},i},
\qquad
i=1,2,
\end{equation}
where $\mathcal H_{\mathrm{kin},i}\simeq L^2(\mathbb{R}^3)$ 
describes the continuous momentum degrees of freedom.

The helicity space $\mathcal H_{\mathrm{hel},i}$ is spanned by 
eigenstates of the helicity operator\\ $\hat h = (\hat{\mathbf S}_{\rm spin}\!\cdot\!\mathbf p)/|\mathbf p|$,
with eigenvalue equation and eigenstate normalizations
\begin{equation}
\hat h \ket{\lambda} = \lambda \ket{\lambda},
\qquad
\braket{\lambda|\lambda'}=\delta_{\lambda\lambda'}.
\end{equation}

For spin-$\tfrac12$ fermions, the helicity eigenvalues are 
$\lambda=\pm \tfrac12$. 
We denote these states by
\begin{equation}
\ket{R} \equiv \ket{+\tfrac12},
\qquad
\ket{L} \equiv \ket{-\tfrac12},
\end{equation}
where $R$ and $L$ refer to right- and left-helicity states 
(\ie, spin aligned or anti-aligned with the momentum direction), 
not to chirality for massive fermions. The two-fermion helicity space is therefore
$
\mathcal H_{\mathrm{hel}}^{\rm f}
\simeq
\mathbb C^2 \otimes \mathbb C^2
$,
with basis
$\{ \ket{RR},\, \ket{RL},\, \ket{LR},\, \ket{LL} \}$.

For massive vector bosons (spin-1), the helicity eigenvalues are 
$\lambda=-1,0,+1$, corresponding to the two transverse 
($\pm1$) and one longitudinal ($0$) polarisation states. 
The single-particle helicity basis is
\begin{equation}
\{ \ket{+1},\, \ket{0},\, \ket{-1} \},
\end{equation}
and the two-particle helicity space is
$
\mathcal H_{\mathrm{hel}}^{\rm V}
\simeq
\mathbb C^3 \otimes \mathbb C^3$
with basis
$\{ \ket{\lambda_A \lambda_B} \,|\, \lambda_{A,B}\in\{-1,0,+1\} \}$.

We prepare a separable incoming state of definite helicities and momenta,
\begin{equation}
\ket{\mathrm{in}}
=
\frac{1}{\sqrt{V}}
\ket{\lambda_1\lambda_2;\mathbf{p}_1\mathbf{p}_2},
\end{equation}
where
\begin{equation}
V=
4E_1E_2(2\pi)^6
\delta^{(3)}(0)\delta^{(3)}(0)
\end{equation}
ensures $\braket{\mathrm{in}|\mathrm{in}}=1$, since momentum eigenstates are normalized covariantly,
\begin{equation}
\braket{\mathbf{p}_1\mathbf{p}_2|\mathbf{p}_3\mathbf{p}_4}
=
4E_1E_2(2\pi)^6
\delta^{(3)}(\mathbf{p}_3-\mathbf{p}_1)
\delta^{(3)}(\mathbf{p}_4-\mathbf{p}_2).
\end{equation}

\paragraph{Projection onto fixed two-particle kinematics.}

The outgoing state is then
\begin{equation}
\ket{\mathrm{out}} = S \ket{\mathrm{in}} .
\end{equation}
The full outgoing state generally contains components in different 
Fock sectors, $\ket{\mathrm{out}} \in \mathcal F_{\mathrm{out}}$. 
To isolate the exclusive two-particle final state with fixed 
external kinematics, we perform a projective measurement via the operator
\begin{equation}
\Pi_{AB}
=
\sum_{\lambda_3,\lambda_4}
\ket{\lambda_3\lambda_4;\mathbf{p}_3\mathbf{p}_4}
\bra{\lambda_3\lambda_4;\mathbf{p}_3\mathbf{p}_4}.
\end{equation}
This projector acts on the helicity degrees of freedom $\lambda_3$ and $\lambda_4$ of the 
final state particles with momenta $\mathbf{p}_3$ and $\mathbf{p}_4$, respectively.

The projected state
\begin{equation} \label{eq:psi_proj}
\ket{\psi_{AB}}
=
\Pi_{AB} S \ket{\mathrm{in}}
\end{equation}
belongs to the outgoing two-particle Hilbert space
\begin{equation}
\mathcal H^{(2)}_{\mathrm{out}}
\simeq
\mathcal H_A \otimes \mathcal H_B .
\end{equation}
where $A$ and $B$ denote the 
subsystem labels associated with the tensor-product structure 
of the two-particle Hilbert space. 

Using the definition of the helicity amplitudes,
\begin{equation}
(2\pi)^4 \delta^{(4)}(p_1+p_2-p_3-p_4)
\, i\mathcal{M}_{\lambda_3\lambda_4,\lambda_1\lambda_2}
=
\braket{\lambda_3\lambda_4;\mathbf{p}_3\mathbf{p}_4
|\, iT \,
|\lambda_1\lambda_2;\mathbf{p}_1\mathbf{p}_2},
\end{equation}
the projected state may be written as
\begin{align}
\ket{\psi_{AB}}
&=
\frac{(2\pi)^4 \delta^{(4)}(p_1+p_2-p_3-p_4)}{\sqrt{V}}
\sum_{\lambda_3,\lambda_4}
i\mathcal{M}_{\lambda_3\lambda_4,\lambda_1\lambda_2}
\ket{\lambda_3\lambda_4;\mathbf{p}_3\mathbf{p}_4}.
\end{align}
Notice that the identity piece of the $S$ operator in Eq. \eqref{eq:S_operator} vanishes after the projection to fixed outgoing momenta (as well as different particle species).

After normalization, the post-selected pure helicity state is
\begin{equation}
\ket{\psi_{AB}}\to\frac{1}{\sqrt{\braket{\psi_{AB}|\psi_{AB}}}}\ket{\psi_{AB}}
=
\frac{
\sum_{\lambda_3,\lambda_4}
\mathcal{M}_{\lambda_3\lambda_4,\lambda_1\lambda_2}
\ket{\lambda_3\lambda_4}
}{
\sqrt{
\sum_{\lambda_3,\lambda_4}
|\mathcal{M}_{\lambda_3\lambda_4,\lambda_1\lambda_2}|^2
}
},
\end{equation}
where the fixed external kinematics have been suppressed in the notation.

\subsection{Infrared Structure and the Normalized Density Matrix}
\label{sec:IR_structure}

\paragraph{IR factorization.}
All helicity amplitudes are ultraviolet (UV)-renormalized in the on-shell scheme detailed in App.~\ref{app:loop_Comp}. The hard $2\to2$ external momenta and helicity labels are held at their Born values; no real photon is included as an extra measured particle, is recombined with a final-state charged lepton, or is absorbed into initial-state radiation to redefine the partonic kinematics. To regulate the QED infrared (IR) singularity we introduce a fictitious photon mass $\lambda$, and we retain the physical masses of the charged external leptons as collinear cutoffs. We denote the full tree-level density matrix by $\rho_0$, so the virtual interference satisfies $\delta\rho_V=\mathcal O(\alpha\,\rho_0)$ and terms beyond NLO start at $\mathcal O(\alpha^2\rho_0)$. The IR-singular part of the one-loop amplitude factorizes~\cite{Bohm:1986rj,Denner:1988tv,Yennie:1961ad,Weinberg:1965nx,Trifinopoulos:2026irc} as a c-number multiple of the LO amplitude,
\begin{equation}\label{eq:M1loop_factor}
\mathcal{M}^{\text{1-loop}}_{kl}
\;=\;
\mathcal{Z}^{(1)}_{\rm IR}(\lambda,\mu)\,\mathcal{M}^{\text{LO}}_{kl}
\;+\;\mathcal{M}^{\text{1-loop,fin}}_{kl}(\mu)\;+\;\mathcal{O}(\alpha^2\mathcal M^{\rm LO}_{kl}),
\end{equation}
where $\mathcal Z^{(1)}_{\rm IR}$ depends only on the external momenta and the charge assignment of the process, not on the final-state helicity labels $(k,l)$. The split is scheme-dependent at the level of $\mathcal M^{\rm 1-loop}$ itself. In the conventional (undressed) amplitude, the coefficient $\mathcal Z^{(1)}_{\rm IR}$ receives contributions from all charged external legs present in the Born process: for $e^-e^+\to\tau^-\tau^+$ these include the incoming $e^\pm$ and outgoing $\tau^\pm$ eikonal dipoles, while for $e^-e^+\to ZZ$ only the incoming charged leptons contribute to the virtual soft factor. This does not remove initial-state real radiation in the $ZZ$ channel; it only specifies the charged-leg content of the universal soft factor. In both cases the only property of $\mathcal Z^{(1)}_{\rm IR}$ we will use is its helicity-independent multiplicative form in Eq.~\eqref{eq:M1loop_factor}.

\paragraph{Trace projection of the virtual correction.}
At finite $\lambda$ the virtual two-body density matrix contains the scalar IR factor in Eq.~\eqref{eq:M1loop_factor}; by itself, the corresponding unnormalized rate is finite at fixed $\lambda$ but not regulator-independent as $\lambda\to0$. We use this regulated virtual matrix only inside the normalized helicity density matrix,
\begin{equation}\label{eq:rhohat_def}
\hat\rho \;\equiv\; \frac{\rho}{\mathrm{Tr}\,\rho}\,,\qquad
\rho \;=\; \rho_0+\delta\rho_V+\mathcal O(\alpha^2\rho_0)\,,
\end{equation}
where hats denote trace normalization throughout, $\rho_0\propto \mathcal M^{\rm LO}\mathcal M^{\rm LO\,\dagger}$ is the tree-level density matrix, and $\delta\rho_V$ is the virtual interference correction.\footnote{The subscript ``$V$'' is for ``virtual''. Strictly $\delta\rho_V$ is the $\mathcal O(\alpha\rho_0)$ correction linear in the one-loop amplitude, not a pure $|\mathcal M^{\rm 1-loop}|^2$ piece, which is part of the $\mathcal O(\alpha^2\rho_0)$ contribution. Later on we will use the untruncated calculation including this  $|\mathcal M^{\rm 1-loop}|^2$  term to estimate our sensitivity to NNLO effects.} Substituting~\eqref{eq:M1loop_factor} splits $\delta\rho_V$ into an IR-singular and a finite piece,
\begin{equation}\label{eq:drhoV_split}
\delta\rho_V \;=\; 2\,\mathrm{Re}\,\mathcal{Z}^{(1)}_{\rm IR}\,\rho_0 \;+\;\delta\rho_V^{\rm fin}\,,
\qquad
\delta\rho_V^{\rm fin}\propto \mathcal M^{\rm LO}\mathcal M^{\rm 1-loop,fin\,\dagger}+\mathrm{h.c.},
\end{equation}
and the IR-singular piece is by construction proportional to $\rho_0$ \emph{entry-by-entry} on the helicity Hilbert space. It therefore divides out identically when the ratio in Eq.~\eqref{eq:rhohat_def} is taken. Explicitly, writing $\hat\rho_0\equiv\rho_0/\mathrm{Tr}\,\rho_0$ for the unit-trace tree-level density matrix,
\begin{equation}\label{eq:rhohat_at_NLO}
\hat\rho\;=\;\hat\rho_0\;+\;\frac{1}{\mathrm{Tr}\,\rho_0}\Bigl[\delta\rho_V^{\rm fin}-(\mathrm{Tr}\,\delta\rho_V^{\rm fin})\,\hat\rho_0\Bigr]\;+\;\mathcal O(\alpha^2)\,,
\end{equation}
which is manifestly $\lambda$-independent: the scalar virtual term $2\,\mathrm{Re}\,\mathcal Z^{(1)}_{\rm IR}\,[\rho_0-(\mathrm{Tr}\,\rho_0)\hat\rho_0]\equiv 0$ vanishes identically by the definition of $\hat\rho_0$. This is the algebraic trace-projection statement for the normalized fixed-Born helicity matrix. It is separate from the Bloch--Nordsieck/YFS cancellation~\cite{Bloch:1937pw,Yennie:1961ad} of IR poles in unnormalized inclusive rates, where the leading unresolved-real contribution cancels the virtual pole before normalization.

\paragraph{Unresolved real emission.}
Real radiation enters only when one specifies a collider measurement. For unresolved soft photons with energy $\omega<\Delta E$ (the detector energy resolution defining which real photons are unresolved and integrated over), the measurement would add a correction $\delta\rho_R(\Delta E)$ to the two-body density matrix. We organize $\delta\rho_R$ as an expansion in the soft-photon energy $\omega$---the Low--Burnett--Kroll soft expansion~\cite{Low:1958sn,Burnett:1967km,Weinberg:1965nx}---whose successive terms we label leading power (LP; the eikonal contribution scaling as $1/\omega$), next-to-leading power (NLP), and next-to-next-to-leading power (NNLP), corresponding to contributions of order $\ln\Delta E$, $\Delta E$, and $\Delta E^2$, respectively, after integration over the unresolved phase space. Before combination with the virtual pole, the LP part of $\delta\rho_R$ carries the same photon-mass dependence with the opposite sign.

The LP real-emission contribution cancels the virtual IR pole in the unnormalized rate and leaves a finite term proportional to $\rho_0$~\cite{Bloch:1937pw,Yennie:1961ad}, which divides out of $\hat\rho$ by the same trace projection as above. At NLP the contribution is purity preserving, \ie, it moves the state along the Born trajectory rather than mixing it, so decoherence is delayed to the next order in the soft expansion~\cite{Trifinopoulos:2026irc}. For massless emitters the NLP kernel does not modify $\hat\rho$ at all: chirality conservation confines the Born state to a single helicity block, within which the NLP soft coefficients are equal, so the kernel acts as an overall rescaling that trace normalization removes.

Two distinct effects survive at NLP, each evading the rescaling argument above by a different route. The first is kinematic and does not require the degeneracy of the NLP soft coefficients across the helicity blocks to be lifted. %that argument holds $\hat\rho_0$ fixed, which an $s$-channel topology does not permit.%
 Since $\hat\rho_0$ is an explicit function of the collision energy, the unresolved initial-state radiation that removes an energy $\Delta E$ displaces the state along that trajectory. Expanding the QED structure function~\cite{Kuraev:1985hb,Altarelli:1977zs} to first order in the energy loss induces a purity-preserving \emph{coherent} shift,
\begin{equation}\label{eq:QI_diff_coh}
\delta\hat\rho_R^{\rm NLP}
\;\simeq\;
-\,\beta\,\frac{\Delta E}{\sqrt s}\;\frac{\partial\hat\rho_0}{\partial\ln\sqrt s}\,,
\qquad
\beta \;=\; \frac{2\alpha}{\pi}\left(\ln\frac{s}{m_e^2}-1\right),
\end{equation}
with $\beta$ the coefficient of the Kuraev--Fadin exponent~\cite{Kuraev:1985hb}. Where $\hat\rho_0$ varies slowly the derivative is $\mathcal O(1)$; near the $Z$ pole, or wherever a quantum observable approaches a stationary point or a zero, it is enhanced and Eq.~\eqref{eq:QI_diff_coh} is the dominant unresolved effect.

The second lifts the coefficient degeneracy directly. Retaining the final-state fermion mass $m_f$, the same-helicity final-state amplitudes switch on, the NLP soft coefficients cease to be degenerate, and a coherent shift $\mathcal{O}\!\left(\alpha\,\Delta E\,m_f/s\right)$ is induced. Relative to Eq.~\eqref{eq:QI_diff_coh} this is suppressed, \eg, by approximately $\sim 10^{3}$ at $\sqrt s\simeq M_Z$, and thus we do not track it further.

The first \emph{decohering} shift appears only at NNLP in the soft expansion, giving
\begin{equation}\label{eq:QI_diff_dec}
\left\|\delta\hat\rho_R^{\rm NNLP}\right\|
\;=\;
\mathcal{O}\!\left(\alpha\,\frac{\Delta E^2}{s}\,\ln\frac{s}{m_l^2}\right),
\end{equation}
where $m_l=\min_i m_i$ ranges over the charged external legs, and the norm may be taken to be any fixed finite-dimensional matrix norm. The logarithm is the mass-regulated remnant of the collinear enhancement in the NNLP soft kernel; this scaling follows from the soft-kernel power expansion. 

Equations~\eqref{eq:QI_diff_coh} and~\eqref{eq:QI_diff_dec} bound the shift in $\hat\rho$, while what a measurement records is the induced shift in a quantum observable. Any smooth observable inherits a correction of the same order through its first variation with respect to $\hat\rho$, though the constant of proportionality is not generic.  Where the first variation vanishes, as at a stationary point of the observable, the leading correction starts one order higher and the estimates are conservative. Conversely, near a zero of the observable, or where $\hat\rho_0$ varies rapidly with the collision energy, the induced correction is enhanced relative to the norms quoted. 

A separate condition governs whether the estimates apply at all. Both are soft-expansion results normalized to the Born rate, thus both presuppose that the unresolved contribution is small relative to this leading order result, for the given the initial-state helicity configuration. This holds for the opposite-helicity configurations (LR/RL), to which the numerical estimates below refer. It fails for the same-helicity configurations (LL/RR), where the Born amplitude is chirality-suppressed while the radiative contribution is not, so that the ratio the expansion is built on is not small; those configurations are treated separately at the end of this subsection.

Numerically, at $\sqrt s\simeq M_Z$ one has $\beta\simeq0.11$. For a realistic FCC-ee photon-reconstruction threshold, $\Delta E_{\rm FCC}\simeq 1\;\GeV$~\cite{Aleksa:2021ztd,Bacchetta:2019fmz}, Eq.~\eqref{eq:QI_diff_coh} gives $\left\|\delta\hat\rho_R^{\rm NLP}\right\|\simeq1.2\times10^{-3}\,\left\|\partial\hat\rho_0/\partial\ln\sqrt s\right\|$, while Eq.~\eqref{eq:QI_diff_dec} gives $\left\|\delta\hat\rho_R^{\rm NNLP}\right\|\sim2\times10^{-5}$. The coherent term therefore dominates the decohering one for any threshold below several tens of GeV and for an $\mathcal O(1)$ derivative it is roughly a tenth of the $\mathcal O(\alpha)\sim10^{-2}$ virtual corrections we compute---not a negligible fraction of them. Where $\hat\rho_0$ varies rapidly with energy the ratio is correspondingly larger. Taking instead $\Delta E=2m_e$, the QED-internal scale below which a soft photon cannot pair-produce an on-shell $e^-e^+$ rather than a physical experimental resolution, the two shifts fall to $\sim10^{-6}$ and $\sim10^{-11}$: at that threshold both are far below the virtual corrections, and the fixed-Born state is an excellent approximation. The results presented below are therefore exclusive quantities, and the comparison against the virtual corrections must be made at the threshold of the measurement being considered.

For $e^-e^+\to ZZ$ the charged-leg content of the soft factor is restricted to the initial-state leptons. The cross-leg interference that generates channel-dependent structure is therefore initial--initial here, rather than the initial--final configuration of Ref.~\cite{Trifinopoulos:2026irc}. At leading power the dipole is a c-number on the $ZZ$ helicity space, by the same soft universality as the virtual factor in Eq.~\eqref{eq:M1loop_factor}~\cite{Weinberg:1965nx,Denner:1988tv}, and cancels in $\hat\rho$ identically. Beyond leading power the channel dependence is present and the diagonal soft coefficients vary across helicity channels. The leading effect in the opposite-helicity configurations is the initial-state dipole, which is Eq.~\eqref{eq:QI_diff_coh} evaluated on the $ZZ$ Born state: at $\sqrt s = 240\;\GeV$ one has $\beta\simeq0.12$, giving $\left\|\delta\hat\rho_R^{\rm NLP}\right\|\simeq5\times10^{-4}\,\left\|\partial\hat\rho_0/\partial\ln\sqrt s\right\|$ for $\Delta E=\Delta E_{\rm FCC}$. 

In the same-helicity initial-state configurations the Born amplitudes themselves are chirality-suppressed by $\mathcal O(m_e/\sqrt s)$, whereas bremsstrahlung from an initial leg supplies a chirality flip that converts the beam configuration into the unsuppressed opposite-helicity hard process. The corresponding splitting function is free of electron-mass singularities~\cite{Jadach:1988gb, Falk:1993tf}, so this contribution is governed by the unsuppressed rate rather than by the suppressed Born amplitude, and is enhanced relative to it by the inverse of the chirality suppression. It is of the same order in $\alpha$ as the virtual corrections we retain, so it is not removed by order counting; only the $\Delta E^2$ scaling suppresses it. Quantitatively, at $\Delta E=1\;\GeV$ it exceeds the same-helicity fixed-Born observables themselves by two to three orders of magnitude across the energies studied, and reaches parity with them only for $\Delta E\lesssim 10^{-1}\;\GeV$. The fixed-Born results for these configurations are therefore exclusive quantities valid at a photon threshold much smaller than the one at which the opposite-helicity results hold. We do not evaluate this contribution here.

Hard real photons with $\omega>\Delta E$ are also outside this fixed-Born setup. They either define genuine three-body events or, in the collinear limit, are absorbed into initial-state radiation (ISR) or final-state radiation (FSR) functions. The standard treatments are convolution with the QED structure function~\cite{Kuraev:1985hb,Altarelli:1977zs} for ISR\footnote{We note that the structure function is diagonal in beam helicity, so this absorption is complete only for the opposite-helicity configurations; recovering the same-helicity configurations requires a helicity-dependent structure function retaining the mass-suppressed off-diagonal channel~\cite{Falk:1993tf}.} and the soft-collinear effective theory (SCET)-based renormalization-group (RG) flow phase-flip channel~\cite{Gu:2025ijz} for FSR. These are measurement-level operations applied downstream of the partonic prediction. In the rest of this paper we therefore evaluate the quantum observables directly on the fixed-Born state of Eq.~\eqref{eq:rhohat_at_NLO}.

\subsection{Quantum Resources}
\label{sec:QI_resources}

\paragraph{Entanglement entropy.}
 
We now quantify these quantum resources in a pure state $\rho$ using standard density-matrix tools from quantum information theory. For bipartite pure states, entanglement is fully characterized by the von Neumann entropy of either reduced density matrix~\cite{Horodecki:2009zz}. 
Reduced density matrices are obtained by tracing out one subsystem,
\begin{equation}
\rho_A = \mathrm{Tr}_B(\rho),
\qquad
\rho_B = \mathrm{Tr}_A(\rho).
\end{equation}
The bipartite entanglement is fully characterized 
by the entanglement entropy, defined as the von Neumann entropy of 
either reduced state,
\begin{equation}
S_E(\rho)
=
S(\rho_A)
=
S(\rho_B),
\qquad
S(\rho) = -\mathrm{Tr}(\rho \log \rho).
\end{equation}
For separable states $S_E=0$, while $S_E>0$ signals entanglement 
between the two outgoing particles.

In practice, and particularly for analytic control of helicity amplitudes, 
it is convenient to employ the linear entropy,
\begin{equation}
E_L(\rho)
=
\frac{d}{d-1}
\left(1 - \mathrm{Tr}(\rho_A^2)\right),
\end{equation}
where $d$ is the dimension of the single-particle helicity space 
($d=2$ for fermions and $d=3$ for massive vectors). 
For pure states, $E_L$ provides a faithful entanglement witness 
and corresponds to the leading-order expansion of the entanglement 
entropy around purity. It ranges from $0$ for separable states 
to $1$ for maximally entangled states.

Regarding the density matrices that appear in this work, as already mentioned in Sec.~\ref{sec:scattering}, the fixed-Born partonic density matrices are pure. Consequently, the linear entropy is an adequate entanglement observable for the partonic states studied here. Moreover, away from unstable-particle poles, App.~\ref{app:pert_expansion} demonstrates that the NLO contribution to $E_L$ depends only on the real part of the non-universal electroweak one-loop matrix element. Near resonances, the complex-mass prescription modifies this simple real/imaginary separation through width effects, and we treat those regions directly in the numerical results. The same normalization argument also explains why purely multiplicative virtual corrections drop out of the other normalized quantum observables discussed below, including $M_2$.

\paragraph[Magic and stabilizer Renyi entropies.]{Magic and stabilizer R\'enyi entropies.}

Entanglement does not exhaust all non-classical features of quantum states. A particularly important additional resource is magic, or non-stabilizerness. Stabilizer states are those obtainable from computational-basis states by Clifford operations~\cite{Nielsen:2012yss}. Computations restricted to stabilizer states and Clifford gates can be efficiently simulated classically via the Gottesman--Knill framework~\cite{Gottesman:1998hu,Aaronson:2004xuh}. Non-stabilizer resources are therefore required for universal quantum computation~\cite{Emerson:2013zse}. To quantify magic explicitly we adopt the stabilizer R\'enyi entropy introduced in Ref.~\cite{Leone:2021rzd}; see also Ref.~\cite{Ohta:2025utz} for recent work on extremal magic states. 

We therefore quantify magic with respect to the 
helicity Hilbert space 
$\mathcal H_{\mathrm{hel}}$, 
where $d=2$ for fermion pairs and $d=3$ for vector-boson pairs. 
For a single $d$-level system we define the generalized Pauli operators 
via the shift and clock operators
$X\ket{j}=\ket{j+1\pmod d}$ and 
$Z\ket{j}=\omega^j\ket{j}$ with $\omega=e^{2\pi i/d}$.
Two-particle Pauli operators are tensor products 
$P=P_1\otimes P_2$ and form an orthogonal operator basis 
for matrices acting on $\mathcal H_{\mathrm{hel}}$. 
For a pure state $\ket{\psi}$ we define the Pauli spectrum
\begin{equation}
\Xi_P(\psi)
=
\frac{1}{d^2}
|\bra{\psi}P\ket{\psi}|^2,
\end{equation}
which satisfies $\sum_P\Xi_P(\psi)=1$.
The stabilizer $\alpha$-R\'enyi entropy is then
\begin{equation}
M_\alpha(\psi)
=
\frac{1}{1-\alpha}
\log\!\left(\sum_P \Xi_P(\psi)^\alpha\right)
-
\log{d^2}.
\label{eq:Malpha}
\end{equation}
The subtraction ensures $M_\alpha=0$ for stabilizer states. 
For mixed states $\rho$ one may use the extended form
\begin{equation}
\tilde M_\alpha(\rho)
=
\frac{1}{1-\alpha}
\log\!\left(
\frac{\sum_P |\mathrm{Tr}(\rho P)|^{2\alpha}}
{\sum_P |\mathrm{Tr}(\rho P)|^{2}}
\right),
\label{eq:Malpha_density}
\end{equation}
which reduces to the pure-state definition when $\rho=\ket{\psi}\bra{\psi}$. This construction is also invariant under Clifford unitaries, since Clifford conjugation permutes the Pauli basis. In this work we primarily use the second stabilizer R\'enyi entropy 
$M_2$ as our magic observable. 

\section{Tau-Pair Production}
\label{sec:results_tautau}

\begin{figure}[h]
\begin{subfigure}{.95\textwidth}
    \centering\includegraphics[width=1\linewidth]{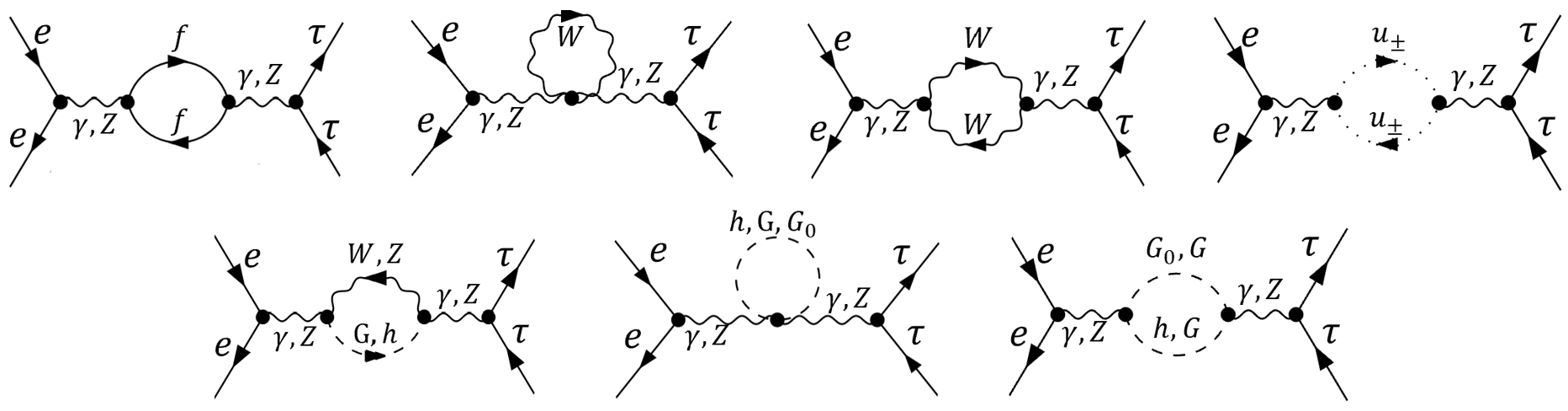}
    \caption{Self Energy Corrections to $e^- e^+ \rightarrow \tau^- \tau^+$ }
    \end{subfigure} 
    \begin{subfigure}{.95\textwidth}
    \centering
  \centering\includegraphics[width=.9\linewidth]{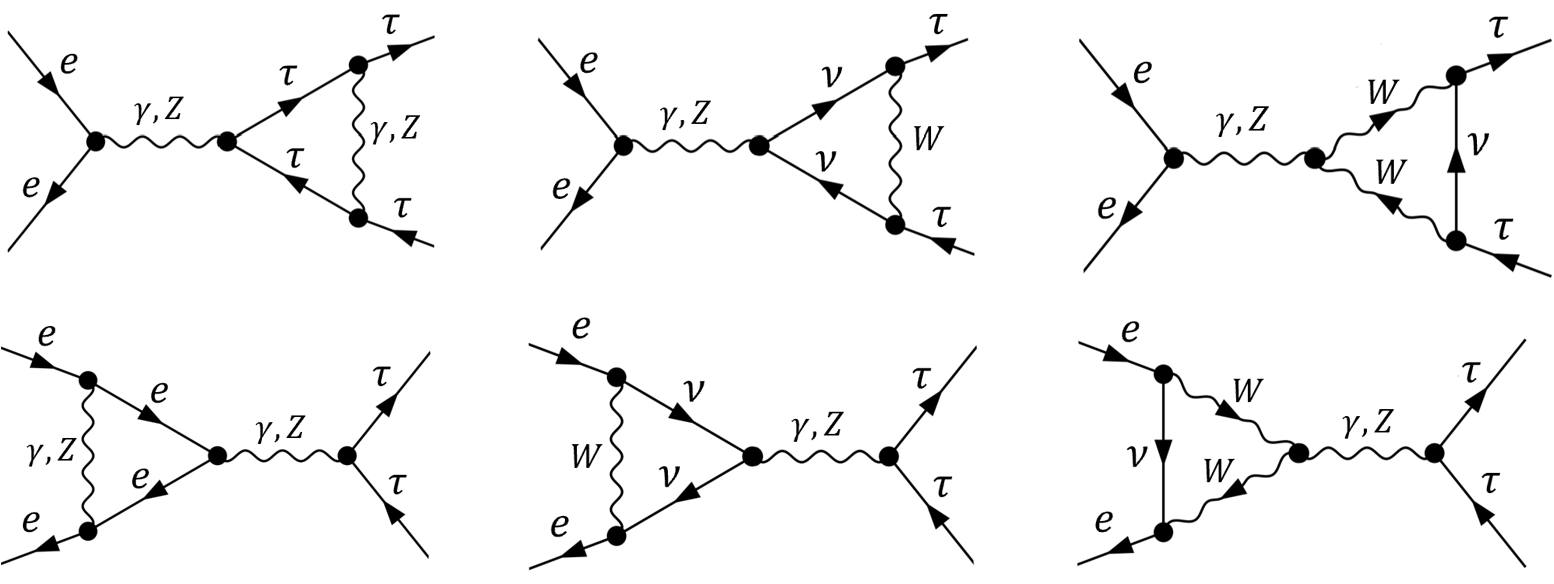}
    \caption{Vertex Corrections to $e^- e^+ \rightarrow \tau^- \tau^+$}
    \end{subfigure}
    \begin{subfigure}{.95\textwidth}
    \centering\includegraphics[width=1\linewidth]{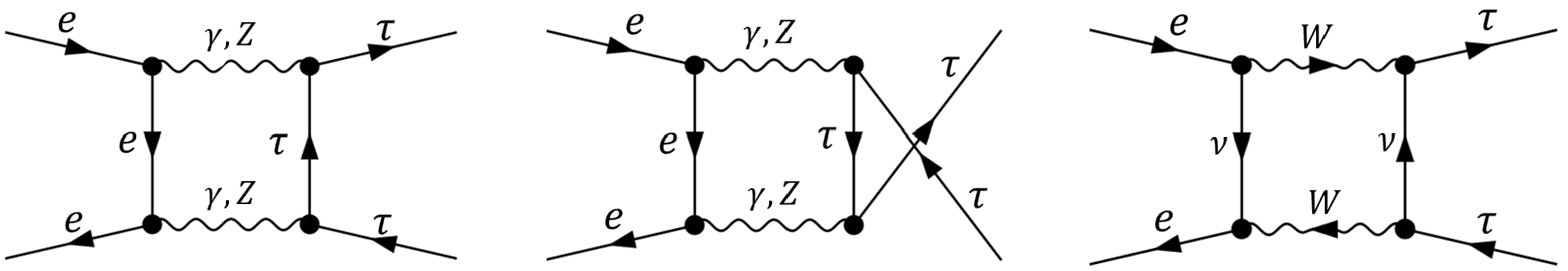}
    \caption{Box Diagram Corrections to $e^- e^+ \rightarrow \tau^-\tau^+ $}
    \end{subfigure}
\caption{Leading Feynman diagrams for $e^+ e^- \rightarrow \tau^+ \tau^-$ at NLO. Diagrams involving scalars are suppressed by the Yukawa couplings to the fermions.}
\label{fig:tau_feynDiag}
\end{figure}

We start with the $\tau^-\tau^+$ channel, which is the simplest two-qubit system among the processes in Eq.~\eqref{eq:processes} and has already been identified as a particularly promising target at future lepton colliders and at lower-energy $e^-e^+$ facilities~\cite{Gabrielli:2026tnl,Han:2025ewp}. The diagrams in Fig.~\ref{fig:tau_feynDiag} follow the standard QED/electroweak diagrammatic expansion introduced in Refs.~\cite{Feynman:1949zx,Dyson:1949ha}; the one-loop electroweak corrections to charged-lepton pair production were computed in Ref.~\cite{Passarino:1978jh}.

We note that the baseline FCC-ee design uses transverse beam polarization for energy calibration rather than sustained longitudinal polarization at the interaction point~\cite{FCC:2018evy}. Our helicity-resolved results should therefore be read as amplitude-level building blocks for a future collider-tomography analysis with realistic beam conditions, and as useful targets for future longitudinal-polarization studies such as those of Refs.~\cite{Altakach:2026fpl,Guo:2026yhz}.

In this section, as well as in Sec.~\ref{sec:results_zz}, we use the same presentation convention. The two-dimensional density plots display each quantum observable in the $(\cos\theta,\sqrt{s})$ plane, where $\theta$ is the scattering angle in the center-of-mass frame and the color scale encodes the value of $E_L$ or $M_2$. Fixed-energy slices show the same observable as a function of $\cos\theta$ at representative values of $\sqrt{s}$, while fixed-angle slices scan the energy dependence at a chosen scattering angle. When the NLO density plot is almost visually  indistinguishable from the LO one, we show only the NLO density plot in the main text and illustrate the radiative effects through these complementary LO--NLO slices. When the radiative effect is already visible in the two-dimensional plots, we show both LO and NLO density plots directly in addition to the slices.

The results reported at NLO are truncated to $\mathcal{O}(\alpha)$ in the quantum observables\footnote{We analytically continue this prescription near poles of resonant propagators  by dropping terms quadratic and higher in $\mathcal{M}_{\text{1-loop}}$, which is equivalent away from the poles to Taylor expanding in the gauge coupling. More details about our treatment of unstable particle singularities is found in Appendix \ref{app:loop_Comp}. }, while the leading order results are zeroth order in the gauge coupling away from resonant propagator singularities. Throughout this work, we estimate the sensitivity of our results to higher order corrections by computing the same observables without any perturbative truncation. We illustrate the deviation between the truncated and untruncated calculations in our fixed energy/ angle slice plots by light-colored bands around the respective truncated NLO curves. Since $E_L$ is a function of the density matrix squared, the untruncated results also contain contributions of the form $\mathcal{M}_{\text{1-loop}}^2\mathcal{M}_{\text{LO}}^2 \sim \mathcal{O}(\alpha^6)$,  $\mathcal{M}_{\text{1-loop}}^3\mathcal{M}_{\text{LO}} \sim \mathcal{O}(\alpha^7)$,  and $\mathcal{M}_{\text{1-loop}}^4 \sim \mathcal{O}(\alpha^8)$, with $M_2$ having even further sensitivity. The contributions from the preceding terms thus have additional corrections from NNLO through N$^4$LO for $E_L$ and through N$^8$LO for $M_2$, making the displayed bands a rough estimate. 

\subsection{Entanglement at NLO}

\begin{figure}[t]
\centering
\begin{subfigure}{.95\textwidth}
    \centering \includegraphics[width=0.6\linewidth]{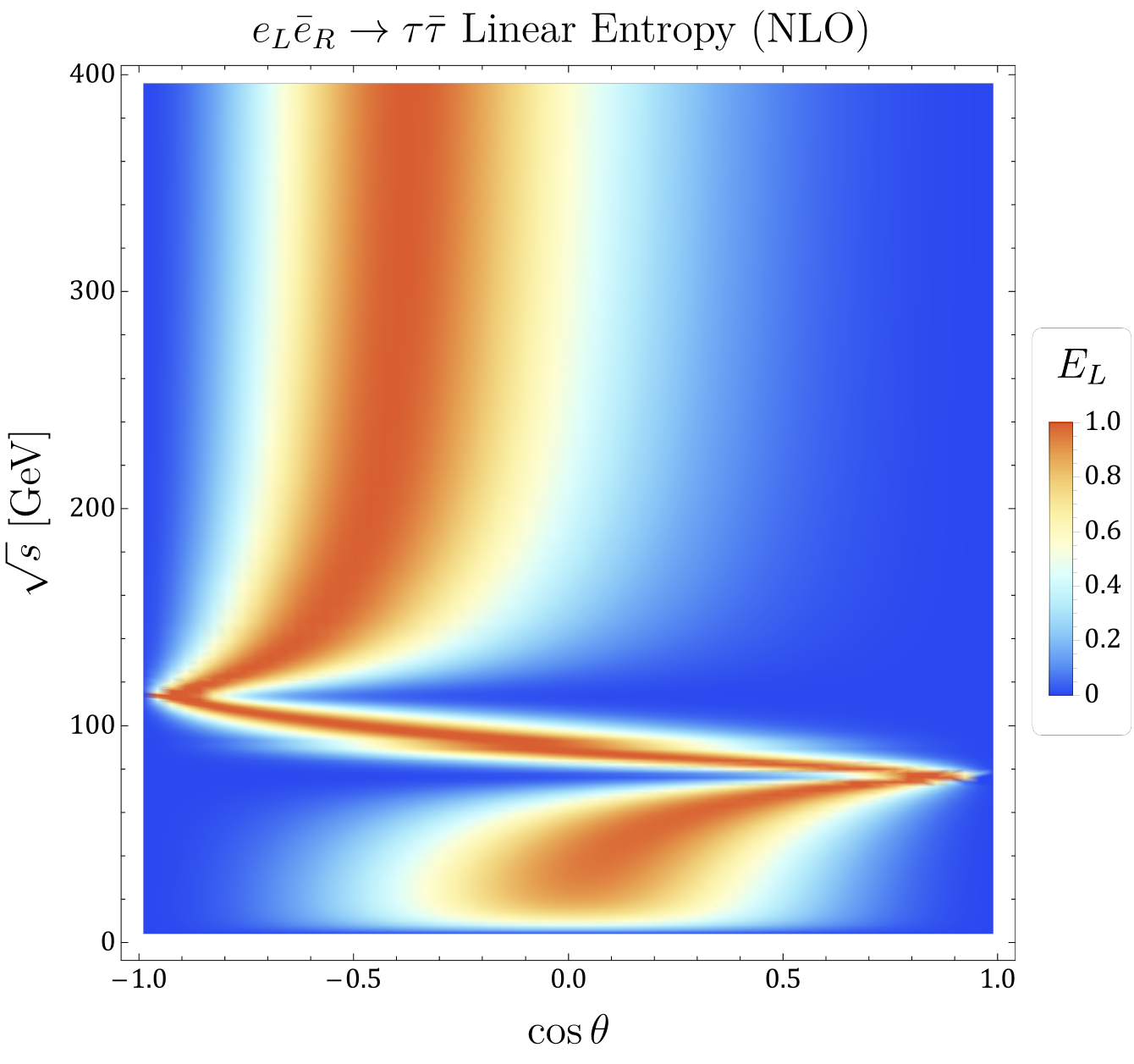}
    \caption{Density plot of linear entropy $E_L$ for $e^-e^+ \to \tau^-\tau^+$ with $LR$ initial states at NLO.}
    \label{fig:tau_ent_lr}
\end{subfigure}
\begin{subfigure}{.95\textwidth}
    \centering
    \includegraphics[width=0.32\linewidth]{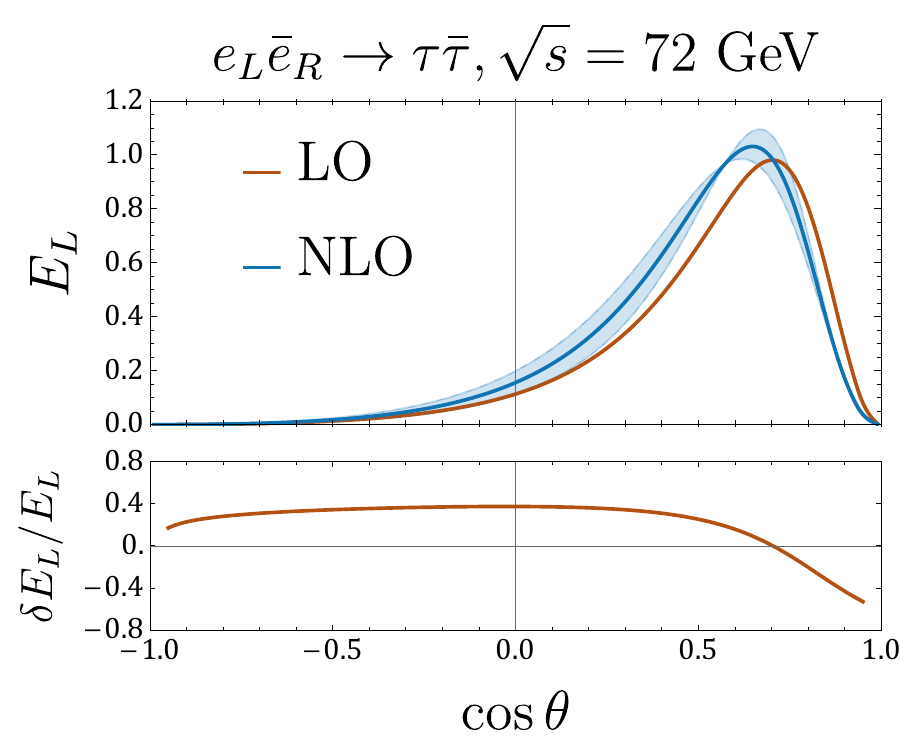}
    \includegraphics[width=0.32\linewidth]{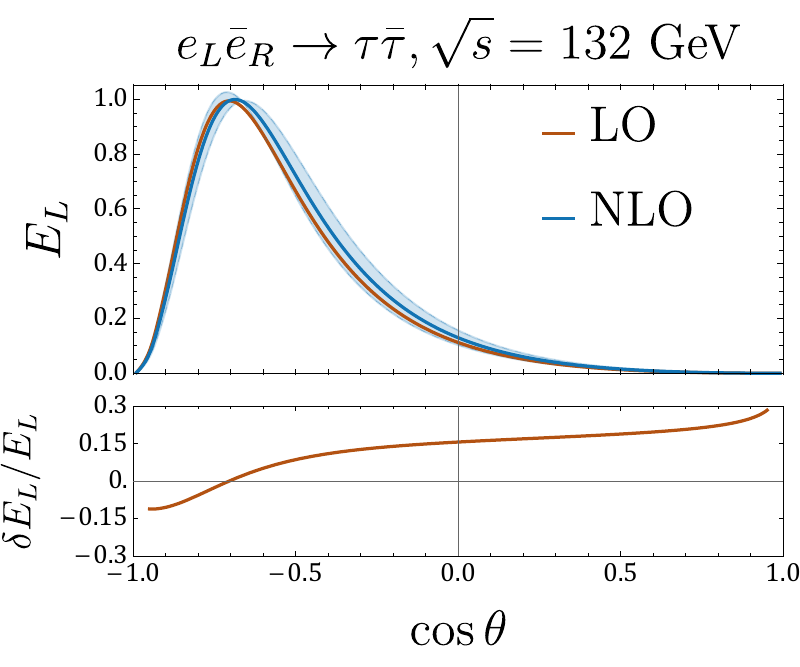}
    \includegraphics[width=0.32\linewidth]{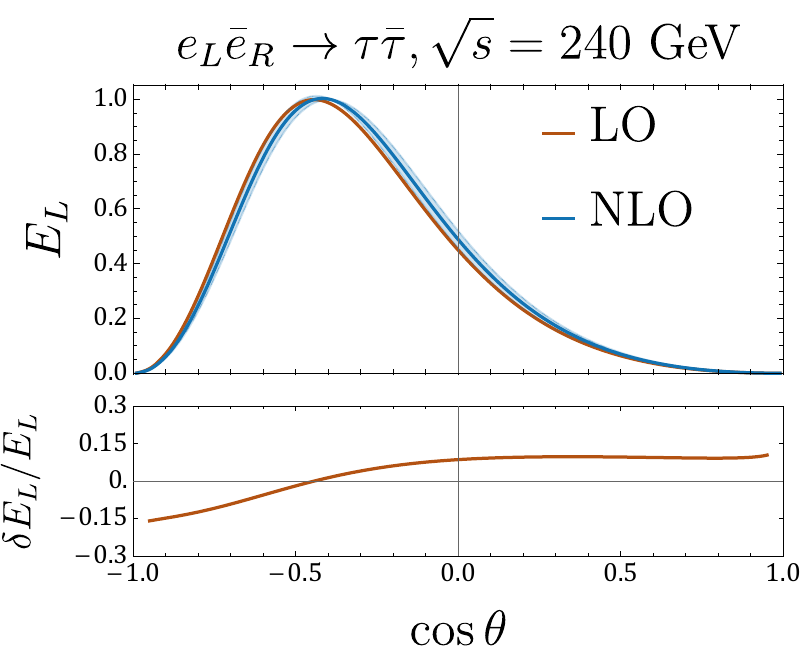}
    \caption{$\tau^-\tau^+$ linear-entropy slices for $LR$ initial states, comparing LO (red) and NLO (blue). Results for which $E_L >1 $ arise as an artifact of truncating the perturbative series to $\mathcal{O}(\alpha)$. The shaded regions correspond to the sensitivity of these calculations to NNLO contributions as estimated by the untruncated results (see Sec.~\ref{sec:results_tautau}).}
 \label{fig:tau_ent_slices_LR}
\end{subfigure}
 \caption{Linear entropy $E_L$ for $e^-e^+ \to \tau^-\tau^+$ with $LR$ initial states at NLO in the $(\cos\theta,\sqrt{s})$ plane. The LO density plot is visually similar, and the fixed-energy slices below make the LO--NLO comparison explicit. The $RL$ channel is qualitatively identical and is shown in App.~\ref{app:tau_slices}.}
\end{figure}

\paragraph{Opposite-helicity initial states.}
Figure~\ref{fig:tau_ent_lr} shows the two-dimensional density plot of the linear entropy $E_L$ for $LR$ initial states at NLO over the range $4 \leq \sqrt{s} \leq 400\;\GeV$. The $RL$ channel produces a qualitatively identical pattern and is collected in App.~\ref{app:tau_slices}.

% [tau ent RL heat map moved to appendix, see App.~\ref{app:tau_slices}]

Already at LO the opposite-helicity channels display a rich two-dimensional entanglement landscape, and the NLO corrections mostly deform this tree-level pattern rather than creating a new topology. High entanglement is concentrated along curved angular branches that emanate from the electroweak-scale region ($\sqrt{s}\sim 90$--$120\;\GeV$) and extend into the backward hemisphere over a broad energy range. The most striking feature is a prominent ridge of near-maximal entanglement ($E_L\simeq 1$) that sweeps from $\cos\theta\sim -0.7$ near the $Z$~pole to increasingly central angles at higher energies; a second, dimmer branch persists at forward angles below the electroweak scale.

The overall topology is qualitatively preserved between $LR$ and $RL$, with only a mild left--right asymmetry reflecting the different coupling strengths of the two helicity configurations to the $Z$ boson. Because $e^-e^+\to\tau^-\tau^+$ proceeds through the nearby $s$-channel $\gamma/Z$ system, the resource pattern is directly shaped by the $Z$ resonance and $\gamma$--$Z$ interference, which is why the strongest structures appear already around electroweak energies.

The global pattern of high-entanglement branches is therefore preserved, confirming that the dominant structure is set by tree-level interference of helicity amplitudes. The virtual corrections produce small angular displacements and mild changes in the branch gradients away from the pole. Exactly on the $Z$ resonance the difference is strongly reduced because the dominant NLO contributions are close to multiplicative corrections to the resonant amplitude, and such common rescalings cancel in the normalized density matrix. Resolving the residual off-pole distortions experimentally would require a dedicated tomography and luminosity study; we therefore treat the slices below as precision benchmarks rather than as immediate FCC-ee sensitivity projections.

These effects are made quantitative in the fixed-energy slices of Fig.~\ref{fig:tau_ent_slices_LR}  with values of $\sqrt{s}$  selected to roughly bookend the parameter space of significant virtual corrections. Substantial NLO effects first emerge at the weak scale, as reflected in the $\sqrt{s} = 72\;\GeV $ slice which features fractional corrections of $\sim 50\%$.  Slices at $132$ and $240 \text{ GeV}$ demonstrate the decreasing importance of NLO corrections with increasing center of mass energy, with the $\sim 10\%$ perturbations being characterized by subtle shifts to the angular location and sharpness of the entanglement peak while preserving the overall shape of $E_L$. This trend is not strictly monotonic however, as at $\sqrt{s}=92\;\GeV$, close to the $Z$ resonance, the LO and NLO curves nearly coincide, consistent with the cancellation of universal resonant corrections in the normalized density matrix. This $\sqrt{s}=92\;\GeV$ plot and other additional fixed-energy slices corresponding to future FCC-ee running points covering both helicity orderings  are collected in App.~\ref{app:tau_slices}.

\begin{figure}[H]
\begin{subfigure}{.95 \textwidth}
    \centering\includegraphics[width=0.49\linewidth]{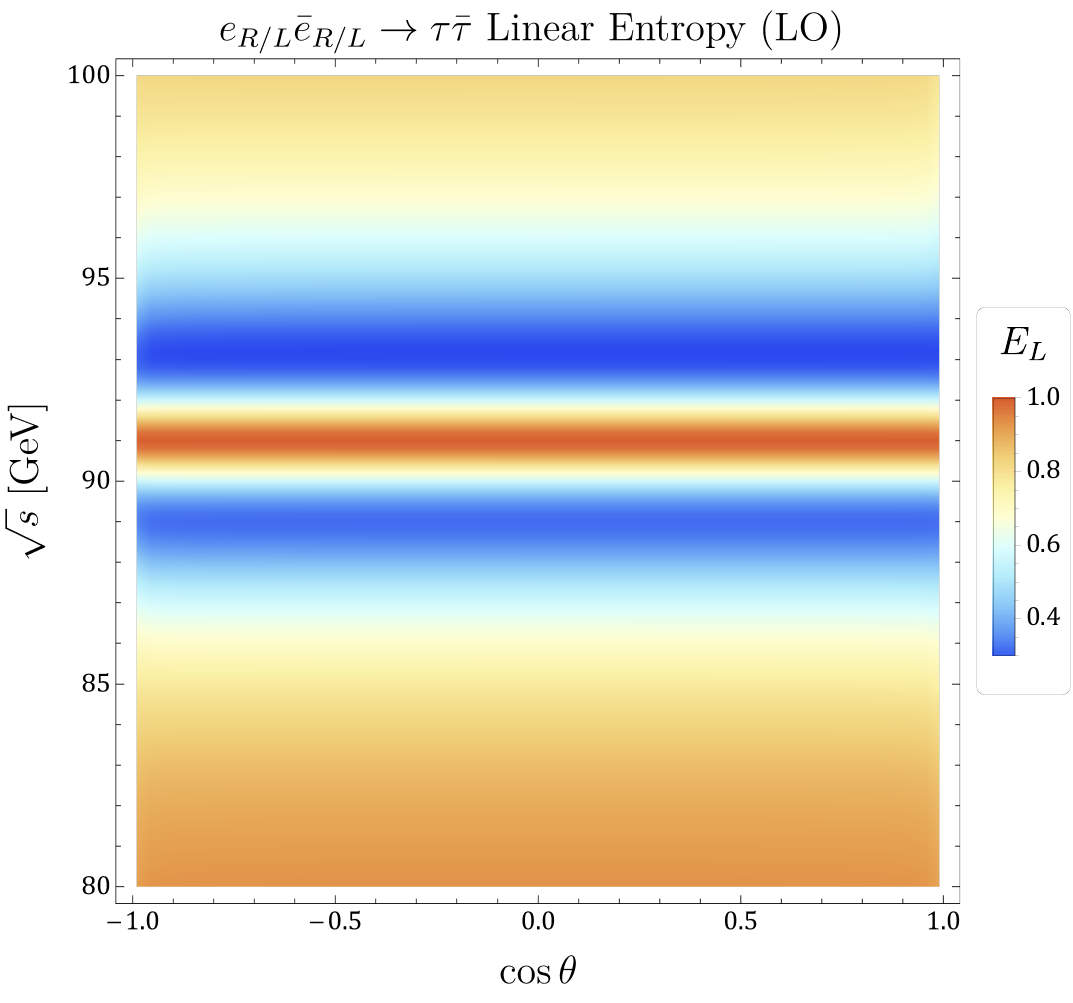}
    \includegraphics[width=0.49\linewidth]{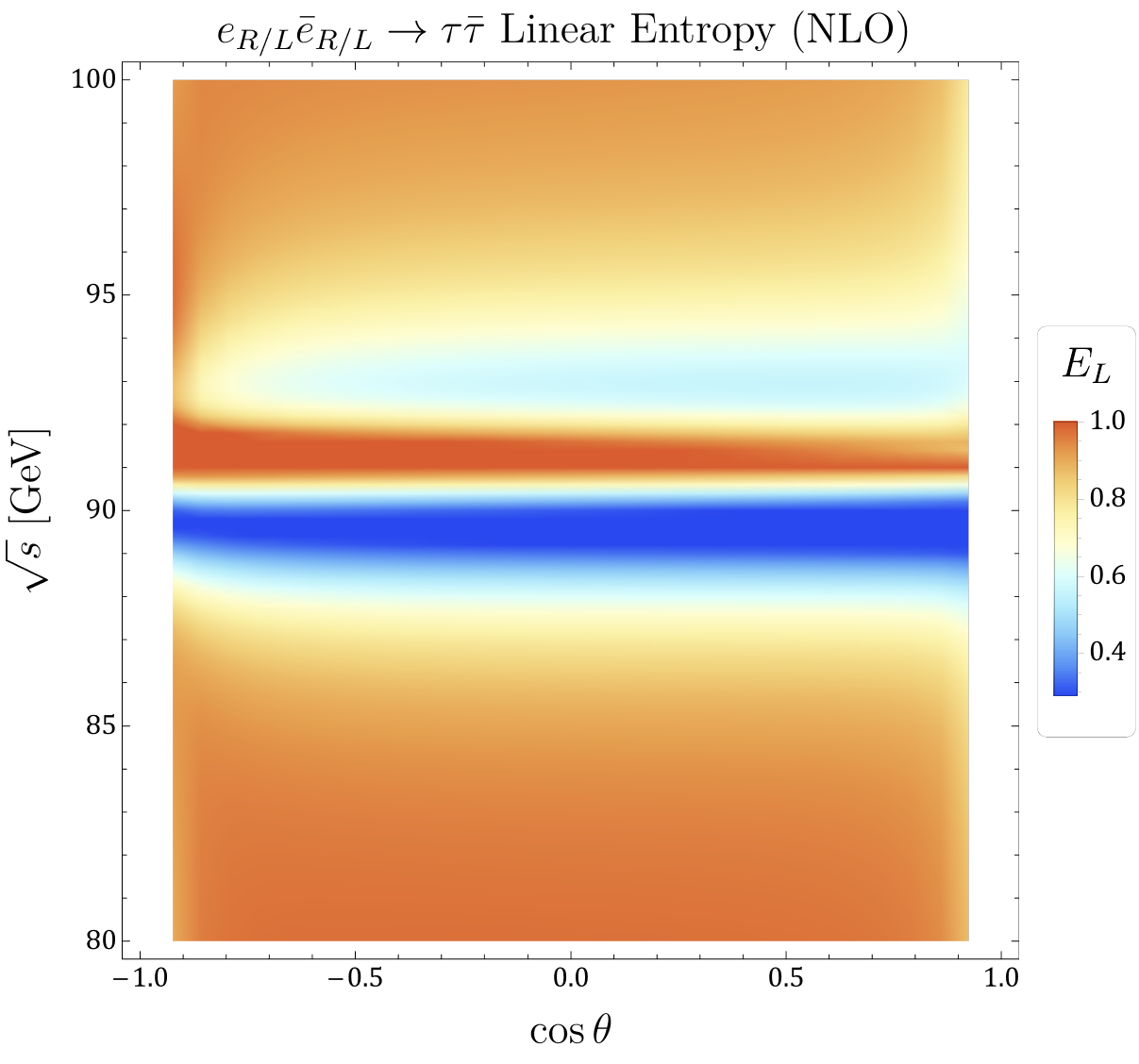}
    \caption{Density plots of linear entropy for $LL$ or $RR$ initial state scattering to $\tau^-\tau^+$, comparing LO (left) and NLO (right)}
\label{fig:tau_ent_ll}
\end{subfigure}
\begin{subfigure}{.95 \textwidth}
    \centering\includegraphics[width=0.95\linewidth]{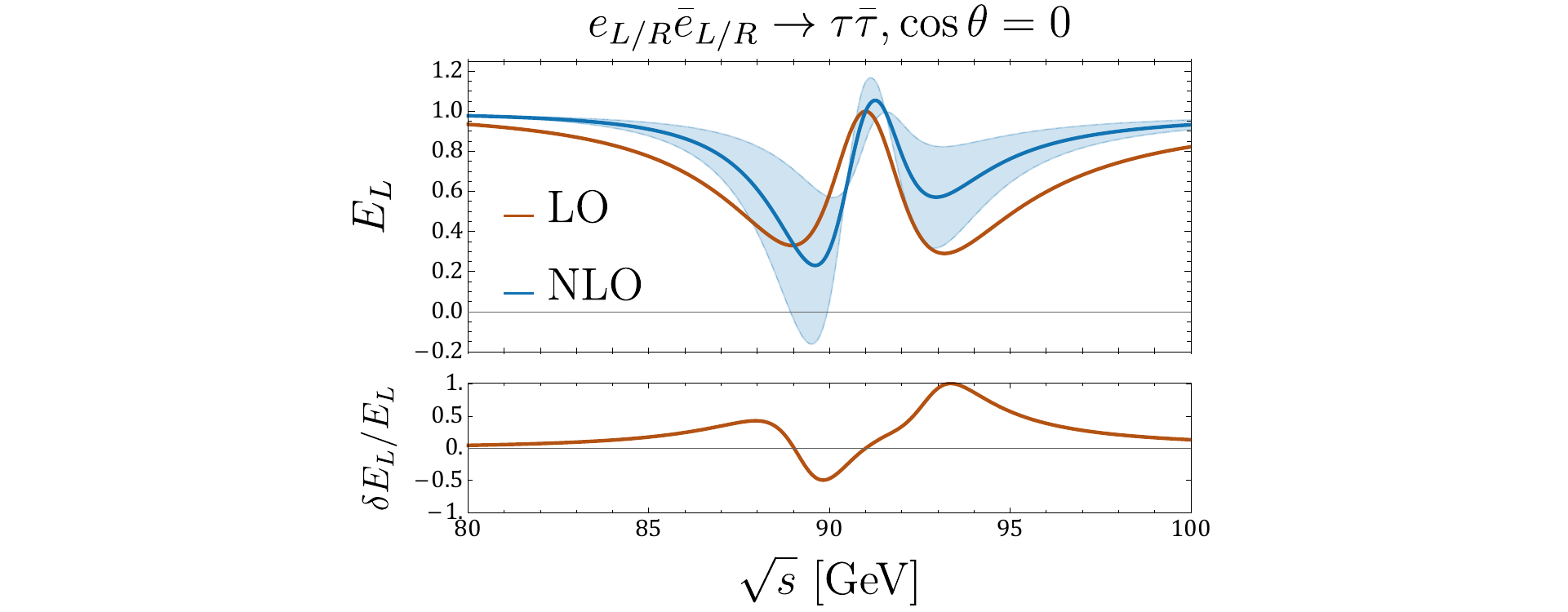}
\caption{Same-helicity $\tau^-\tau^+$ linear entropy at perpendicular scattering, $\cos\theta=0$, comparing LO (red) and NLO (blue) near the $Z$ pole. The shaded region corresponds to the uncertainty from NNLO contributions, which is estimated according to the prescription in Sec.~\ref{sec:results_tautau}.}
\label{fig:tau_ent_slices_LLRR}
\end{subfigure}
\caption{Linear entropy $E_L$ for $e^-e^+ \to \tau^-\tau^+$ with $LL$ or $RR$ initial states in the $Z$-pole region.}
\end{figure}

\paragraph{Same-helicity initial states.}
Figure~\ref{fig:tau_ent_ll} zooms into the $Z$-pole region ($\sqrt{s}=80$--$100\;\GeV$) for $LL$ or $RR$ initial states, where the entanglement structure is qualitatively different.
%
%
% [tau ent RR heat map moved to appendix, see App.~\ref{app:tau_slices}]
%
In contrast to the opposite-helicity channels, the same-helicity entanglement is driven predominantly by the collision energy, exhibiting nearly angle-independent horizontal bands at LO. A sharp maximum ($E_L\simeq 1$) appears as a narrow band at $\sqrt{s}\simeq 91\;\GeV$, coinciding with the $Z$~pole, flanked by local-minimum bands where the entanglement is strongly reduced but not driven all the way to zero once the finite resonance width is retained. Above and below this resonance the linear entropy rises again toward large values with only mild angular modulation. This band structure reflects the dominance of the $s$-channel $Z$ resonance for same-helicity configurations, where the angular dependence is suppressed compared to the $\gamma$--$Z$ interference that drives the opposite-helicity patterns.

The NLO corrections, in Figure~\ref{fig:tau_ent_ll} (right),  introduce visible yet controlled modifications which reweight the local extrema  observed at tree level\footnote{Results near $\sqrt{s} = M_Z$ are particularly sensitive to the power counting in $\alpha$ and the value of the $Z$-boson decay width used in the complex mass scheme. Using the values of the SM parameters in App.~\ref{app:loop_Comp}, we calculate $\Gamma_Z$ to $\mathcal{O}(\alpha)$, obtaining $\Gamma_Z = 2.3503 \text{ GeV}$. Proceeding analogously for the Higgs, we find $\Gamma_H = 0.00475 \text{ GeV}$ }.  As before, the linear entropy maximum remains tied to the Z pole, with a slight shift towards higher energies. The locations of the minima which flank this peak on either side are brought inward by roughly $2 \text{ GeV}$ while their heights are adjusted. The higher energy trough now sits closer to maximal entanglement while the lower energy extrema is now decreased relative to the tree level result. From the banded region displayed in Fig.~\ref{fig:tau_ent_slices_LLRR}, it is suggested that this behavior may be sensitive to contributions from NNLO and beyond. Below $80 \text{ GeV}$ and above $110 \text{ GeV}$, the NLO modifications (and the NNLO sensitivity) decrease below the $1\%$ level and become vanishingly small as one ventures out further. Fig.~\ref{fig:tau_ent_ll} (right) reveals the emergence of a subtle angular dependence beyond leading order near $\sqrt{s} = M_Z$. At tree level, the dominant helicity amplitudes carry a common $\sin\theta$ dependence that is removed by the density-matrix normalization in much of the same-helicity region. Loop corrections break this simple factorization through non-universal chiral electroweak box and vertex contributions, generating the residual angular structure.

\subsection{Magic at NLO}

\begin{figure}[t]
\begin{subfigure}{.95\textwidth}
    \centering\includegraphics[width=0.6\linewidth]{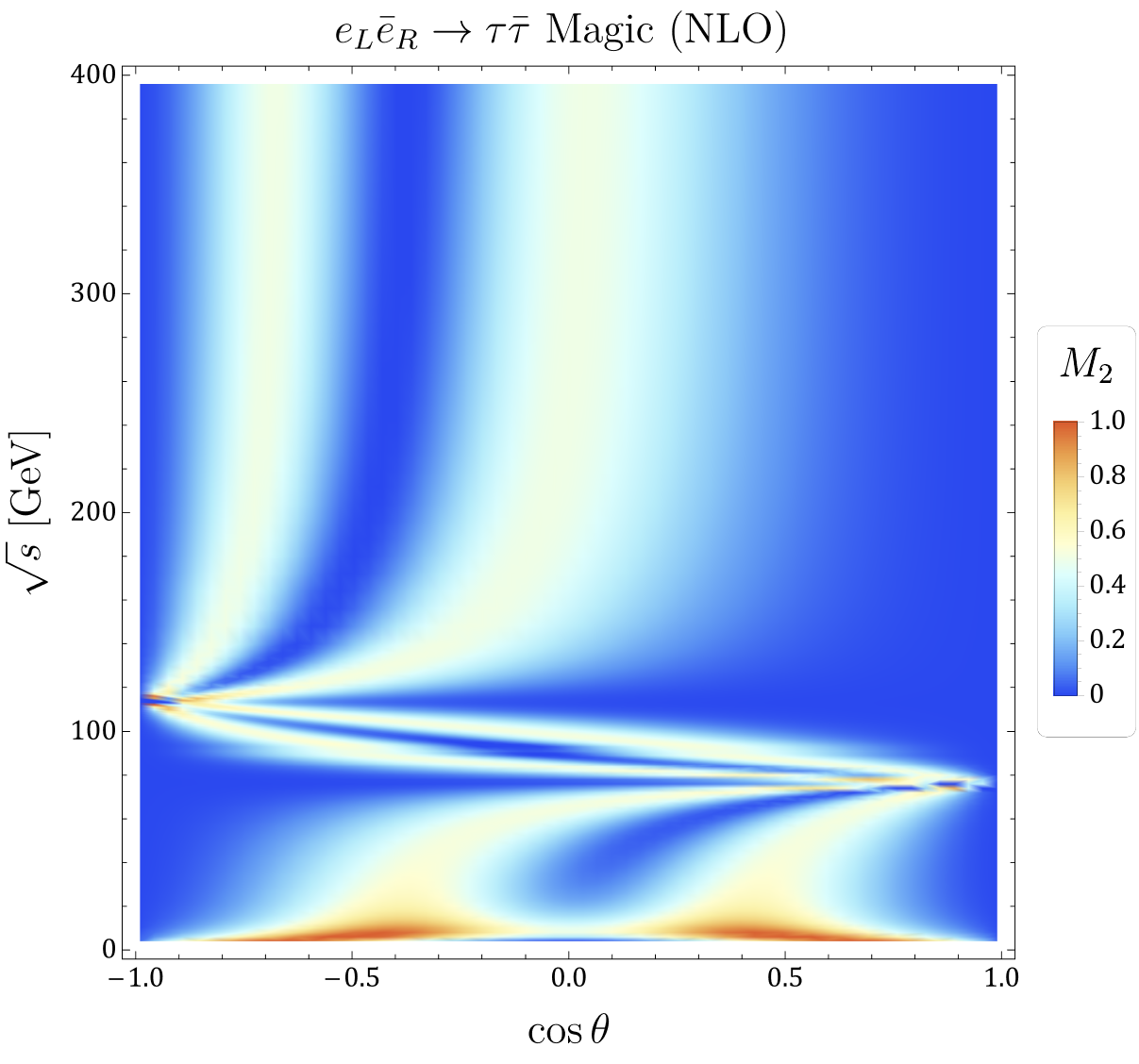}
    \caption{Density plot of stabilizer second  R\'enyi entropy $M_2$ for $e^-e^+ \to \tau^-\tau^+$ with $LR$ initial states at NLO}
    \label{fig:tau_mag_lr}
\end{subfigure}
\begin{subfigure}{.95\textwidth}
    \centering
    \includegraphics[width=0.32\linewidth]{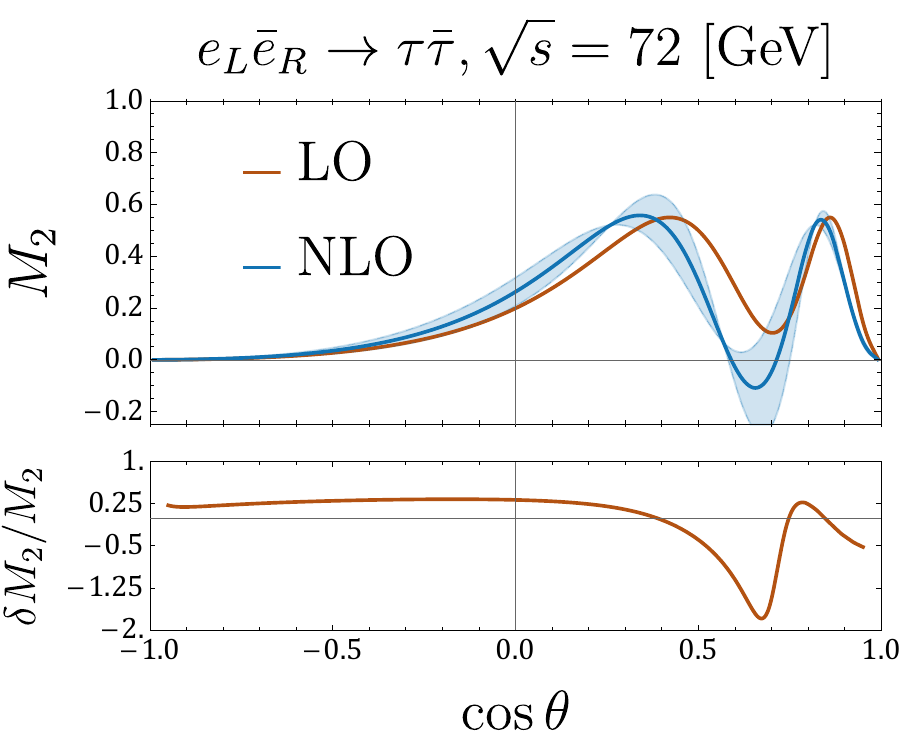}
    \includegraphics[width=0.32\linewidth]{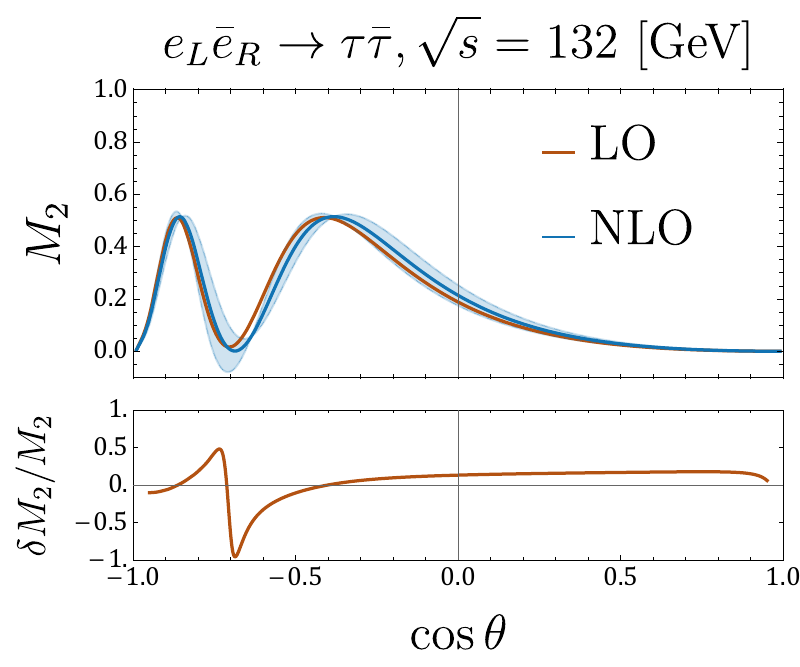}
    \includegraphics[width=0.32\linewidth]{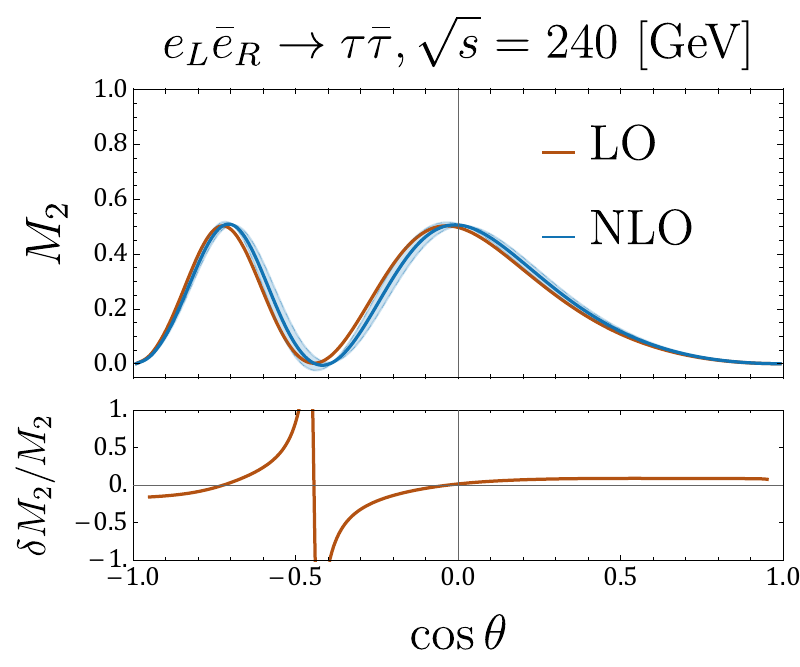}
    \caption{Representative fixed-energy slices of the $\tau^-\tau^+$ stabilizer second R\'enyi entropy $M_2$ at LO and NLO. The NLO corrections shift the peak positions and reshape the minima between them while preserving the non-monotonic relation between entanglement and magic. Results for which $M_2 < 0$ arise as an artifact of truncating the perturbative series to $\mathcal{O}(\alpha)$. The shaded regions correspond to the sensitivity of these calculations to NNLO contributions, as estimated by the untruncated results (see Sec.~\ref{sec:results_tautau}).}
    \label{fig:tau_mag_slices_lr}
\end{subfigure}
\caption{Stabilizer second R\'enyi entropy $M_2$ for $e^-e^+ \to \tau^-\tau^+$ with $LR$ initial states at NLO in the $(\cos\theta,\sqrt{s})$ plane. The LO density plot is visually similar, and the $RL$ channel is shown in App.~\ref{app:tau_slices}.}

\end{figure}

\paragraph{Opposite-helicity initial states.}
Figure~\ref{fig:tau_mag_lr} shows the two-dimensional density plot of the stabilizer second R\'enyi entropy $M_2$ for $LR$ initial states at NLO. The $RL$ channel is collected in App.~\ref{app:tau_slices} because it exhibits the same qualitative branch structure, with only modest left--right shifts inherited from the chiral electroweak couplings.

The magic maps share the same broad branch-like topology as the entanglement, but the local maxima of the two resources are distributed differently in the $(\cos\theta,\,\sqrt{s})$ plane. The bright $M_2$ ridges tend to flank, rather than coincide with, the regions of maximal entanglement: the wider high-resource branches are correlated, but where $E_L$ is closest to its maximum $M_2$ is often reduced. This complementarity is already manifest at LO and persists at NLO, confirming that the most entangled states in this process are close to stabilizer states, while the largest non-stabilizerness is generated in intermediate-entanglement regions. The pattern is consistent with the general findings of Refs.~\cite{Guo:2026yhz,Altakach:2026fpl} and is visible at the amplitude level without invoking EFT deformations or beam-polarization scans.

The LO--NLO comparison is therefore best seen in the fixed-energy slices in Fig.~\ref{fig:tau_mag_slices_lr}. As for entanglement, the magic branches undergo small angular shifts in the electroweak resonance region which decrease as $\sqrt{s}$ tends towards higher energies.

\paragraph{Same-helicity initial states.}

\begin{figure}[t]
\begin{subfigure}{.95\textwidth}
\centering
\includegraphics[width=0.49\linewidth]{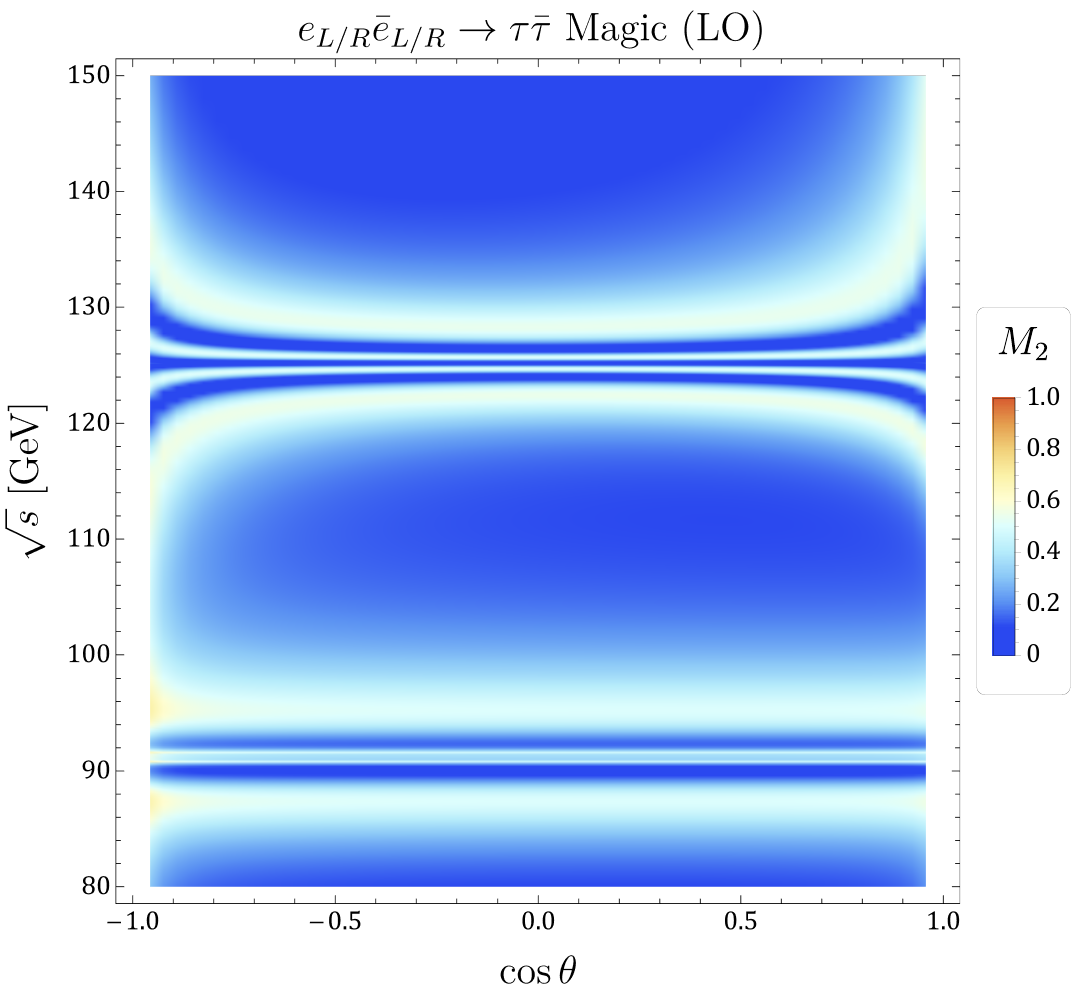}
\includegraphics[width=0.49\linewidth]{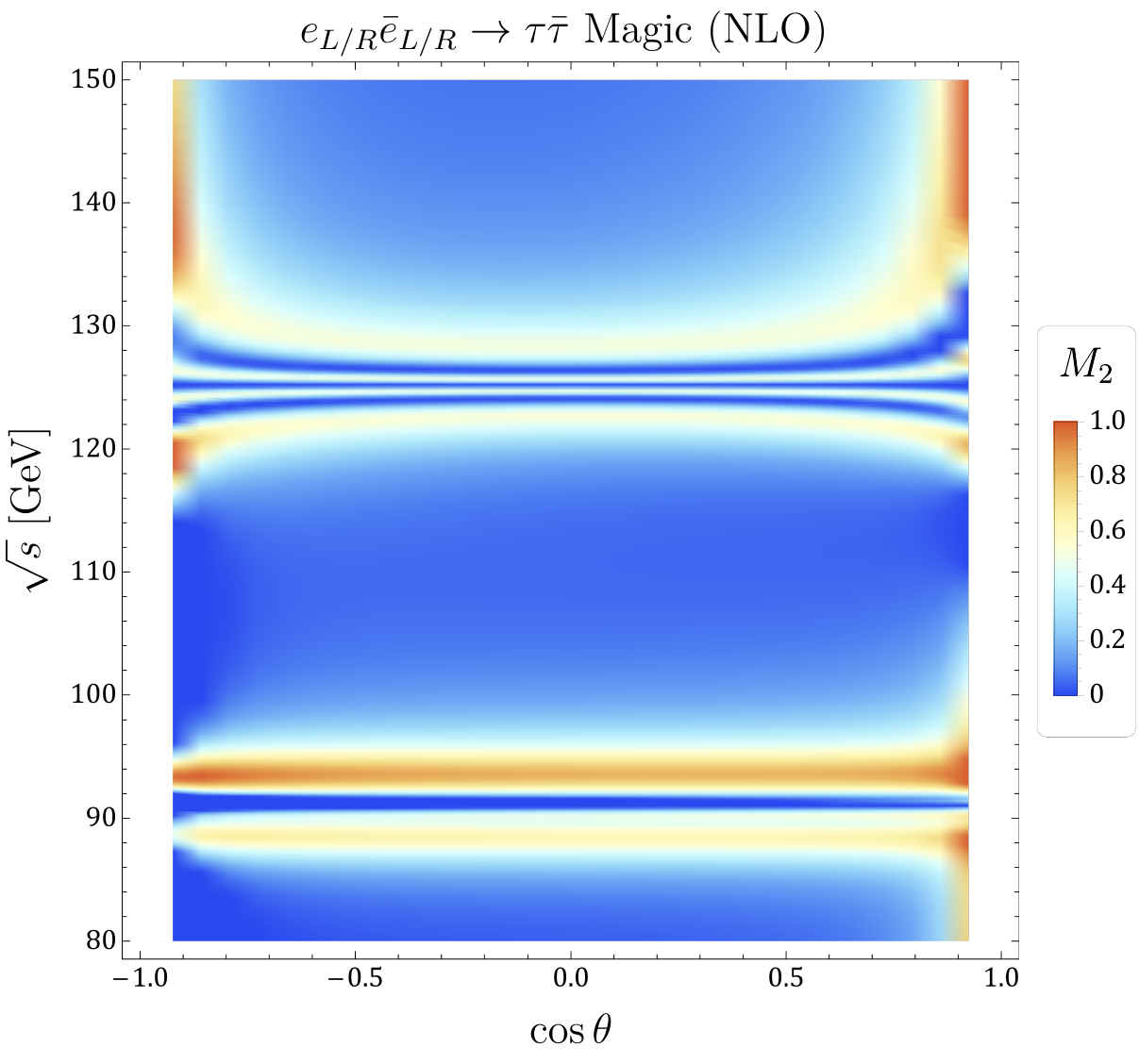}
\caption{Density plots of stabilizer second R\'enyi entropy $M_2$ for $e^-e^+ \to \tau^-\tau^+$ with same-helicity initial states in the electroweak region, comparing LO (left) and NLO (right).}
\label{fig:tau_mag_ll}
\end{subfigure}
\begin{subfigure}{.95 \textwidth}
\centering\includegraphics[width=0.99\linewidth]{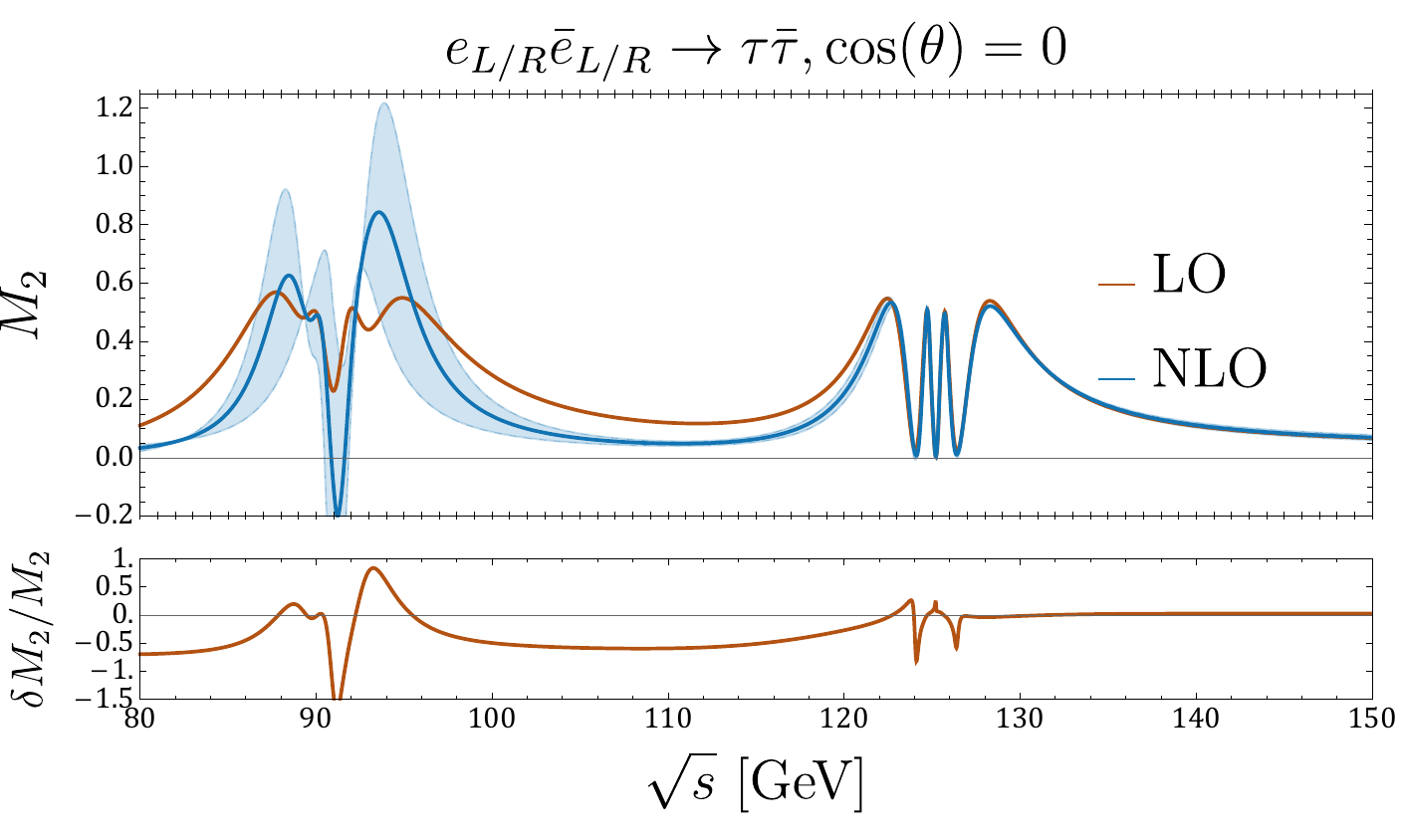}
\caption{Same-helicity $\tau^-\tau^+$ magic at perpendicular scattering, $\cos\theta=0$, comparing LO and NLO near the $Z$ and Higgs resonance regions. The shaded region corresponds to the uncertainty from NNLO contributions, which is estimated according to the prescription in Sec.~\ref{sec:results_tautau}.}
\label{fig:tau_mag_energy_scans}
\end{subfigure}
\caption{Stabilizer second R\'enyi entropy $M_2$ for $e^-e^+ \to \tau^-\tau^+$ with same-helicity initial states in the electroweak region.}
\end{figure}

Figure~\ref{fig:tau_mag_ll} displays $M_2$ for the same-helicity channels in the range $80 \leq \sqrt{s} \leq 150\;\GeV$. The magic exhibits a distinctive horizontal band structure with prominent features in two energy windows. A broad region of suppressed $M_2$ appears between $\sqrt{s}\simeq 95$ and $120\;\GeV$, separated from a second low-magic region below $\sqrt{s}\simeq 85\;\GeV$ by a turbulent zone of relatively larger $M_2$ with oscillatory features centered at the $Z$ pole. Notably, the $Z$-pole band that maximizes entanglement (Fig.~\ref{fig:tau_ent_ll}) sits precisely in the local minima of the magic oscillations at $\sqrt{s} \sim M_Z$, reinforcing the complementarity between the two resources. A similar oscillating band structure is found around the Higgs mass (from roughly $120 \text{ GeV}\leq \sqrt{s} \leq 130 \text{ GeV}$), above which Magic is again relatively suppressed. 
The angular dependence of $M_2$ can be understood from the helicity hierarchy of the amplitudes. Where the opposite-helicity final-state components dominate, their common leading $\sin\theta$ dependence is largely removed by normalization, leaving only weak angular structure. Stronger angular dependence emerges where the same-helicity components can no longer be ignored, notably at extreme scattering angles and near the Higgs-pole region, where chirality-flip amplitudes can become important for the normalized state even though the absolute rate remains tiny.

At NLO the two-band structure is preserved, but the magic content is redistributed. The intermediate region ($\sqrt{s}\sim 100$--$120\;\GeV$) becomes less prominent, with the $M_2$ values decreasing by roughly $50\%$ compared to LO, with similar corrections being realized below the $Z$ pole. Above the oscillating behavior at $\sqrt{s} \sim M_H$, the NLO corrections fade to $\mathcal{O}(1\%)$. The oscillations between the bands sharpen and develop additional fine structure at NLO  near the $Z$-pole, indicating that the loop corrections introduce new interference features in the helicity amplitudes that selectively enhance or suppress magic at specific energies. These same-helicity magic density plots show the largest radiative reshuffling in the $\tau^-\tau^+$ channel and make $M_2$ the most sensitive resource measure for NLO effects in this sector.

The energy scan at perpendicular scattering, shown in Fig.~\ref{fig:tau_mag_energy_scans}, resolves these structures more sharply. Near the $Z$ pole the NLO corrections shrink the oscillatory region inward by ${\cal O}(1\;\GeV)$ and increase the oscillatory amplitude. The two peaks at higher $\sqrt{s}$ also merge into a single local max at NLO. However, the shaded region suggests that these features in the lower oscillatory zone are sensitive to higher order corrections. More subtle effects are observed in the Higgs-region window, where no resolvable phase shift occurs. Besides the artificially large fractional corrections occurring where the $\text{LO}$ result is approximately zero, the dominant effect is a re-weighting of the outermost extrema, superimposed on a smooth electroweak background. These same-helicity electron-beam channels are, however, chirality-flip suppressed by the tiny electron mass, so even FCC-ee luminosities are unlikely to make them directly accessible; the analogous muon-collider channels would be much less suppressed, making them more promising. 

\section{$ZZ$ Production}
\label{sec:results_zz}
\begin{figure}[h]
\centering
\begin{subfigure}{1\textwidth}
\centering
\includegraphics[width=.475\linewidth]{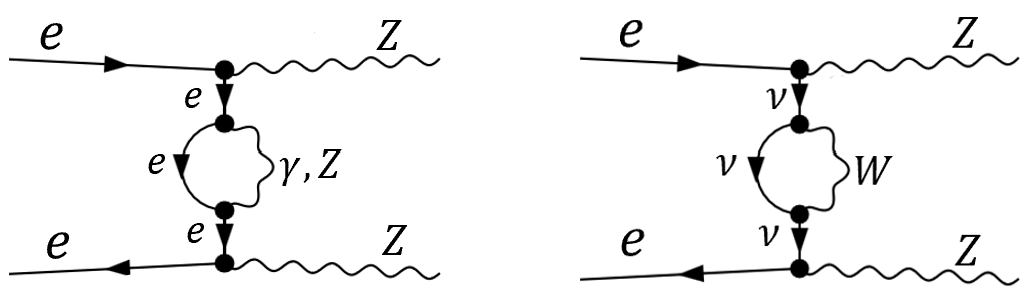}
\caption{Self-energy corrections to $e^-e^+ \rightarrow ZZ$}
\end{subfigure}
\begin{subfigure}{1\textwidth}
\centering
\includegraphics[width=0.95\linewidth]{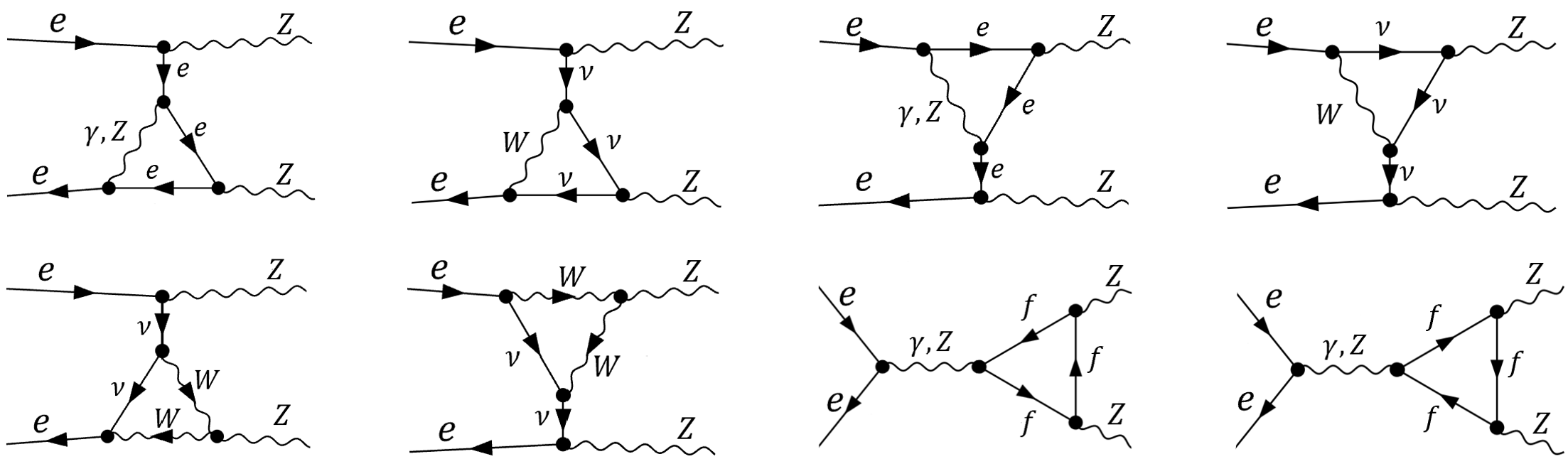}
\caption{Vertex corrections to $e^+ e^- \rightarrow ZZ$}
\end{subfigure}
\begin{subfigure}{1\textwidth}
\centering
\includegraphics[width=0.95\linewidth]{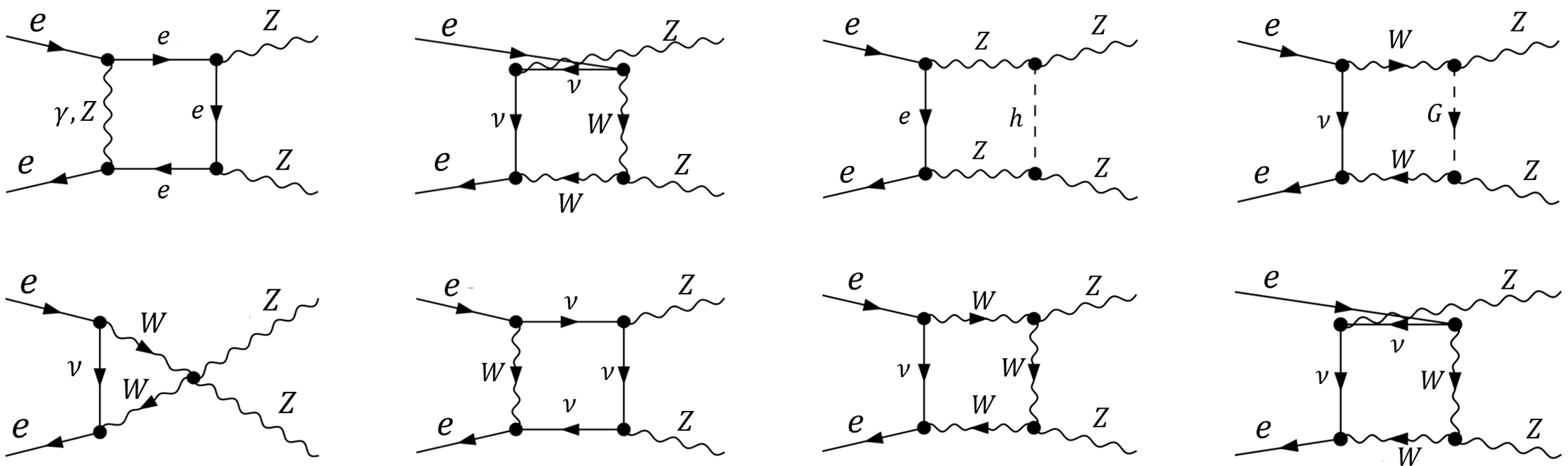}
\caption{Box diagram corrections to $e^-e^+ \rightarrow ZZ$}
\end{subfigure}
\caption{Leading Feynman diagrams for $e^+ e^- \rightarrow ZZ$ at NLO, up to rearrangements of the external legs.}
\label{fig:ZZ_feynDiag}
\end{figure}

We now turn to $e^-e^+\to ZZ$, where the outgoing particles are massive spin-1 bosons and the helicity Hilbert space is nine-dimensional (qutrit$\,\otimes\,$qutrit). Production requires $\sqrt{s}\geq 2M_Z\simeq 182\;\GeV$, so the electroweak resonance region is replaced by a threshold region whose structure is governed by the interplay of $t$- and $u$-channel electron-exchange diagrams and their one-loop electroweak completions, shown schematically in Fig.~\ref{fig:ZZ_feynDiag}. The original one-loop calculation for this channel was given in Ref.~\cite{Denner:1988tv}, building on the standard diagrammatic QED framework of Refs.~\cite{Feynman:1949zx,Dyson:1949ha}.

We present results for $2M_Z \leq \sqrt{s} \leq 1\;\TeV$. This interval spans the neutral-diboson threshold region together with the standard FCC-ee benchmark energies used for the Higgs-factory and top-threshold stages~\cite{FCC:2018evy}; in particular, the fixed-energy slice at $\sqrt{s}=365\;\GeV$ corresponds to the $t\bar t$-run operating point, and the density plots cover the Higgs-factory energy $\sqrt{s}=240\;\GeV$ within their range. The $\sqrt{s}=750\;\GeV$ and $1\;\TeV$ slices then provide a first benchmark for a higher-energy lepton-collider regime, including future muon-collider stages~\cite{InternationalMuonCollider:2024jyv,AlAli:2021let}. The estimated sensitivity to NNLO effects is determined as in the $\tau^- \tau^+$ channels, and is again represented by the shaded regions in the below slices. For all helicity channels studied below, the hard $ZZ$ amplitudes are the same for electron and muon beams up to negligible lepton-mass effects in this energy range, so the same partonic results provide useful input for the fixed-order component of muon-collider studies, where vector-boson-fusion descriptions and electroweak parton-distribution effects become increasingly important at multi-TeV energies~\cite{AlAli:2021let,Costantini:2020stv,Han:2020uid,Garosi:2023bvq} and where gauge-boson entanglement has already been investigated~\cite{Ding:2025mzj}.

\subsection{Entanglement at NLO}

\paragraph{Opposite-helicity initial states.}

\begin{figure}[t]
\begin{subfigure}{.99\textwidth}
    \centering\includegraphics[width=0.6\linewidth]{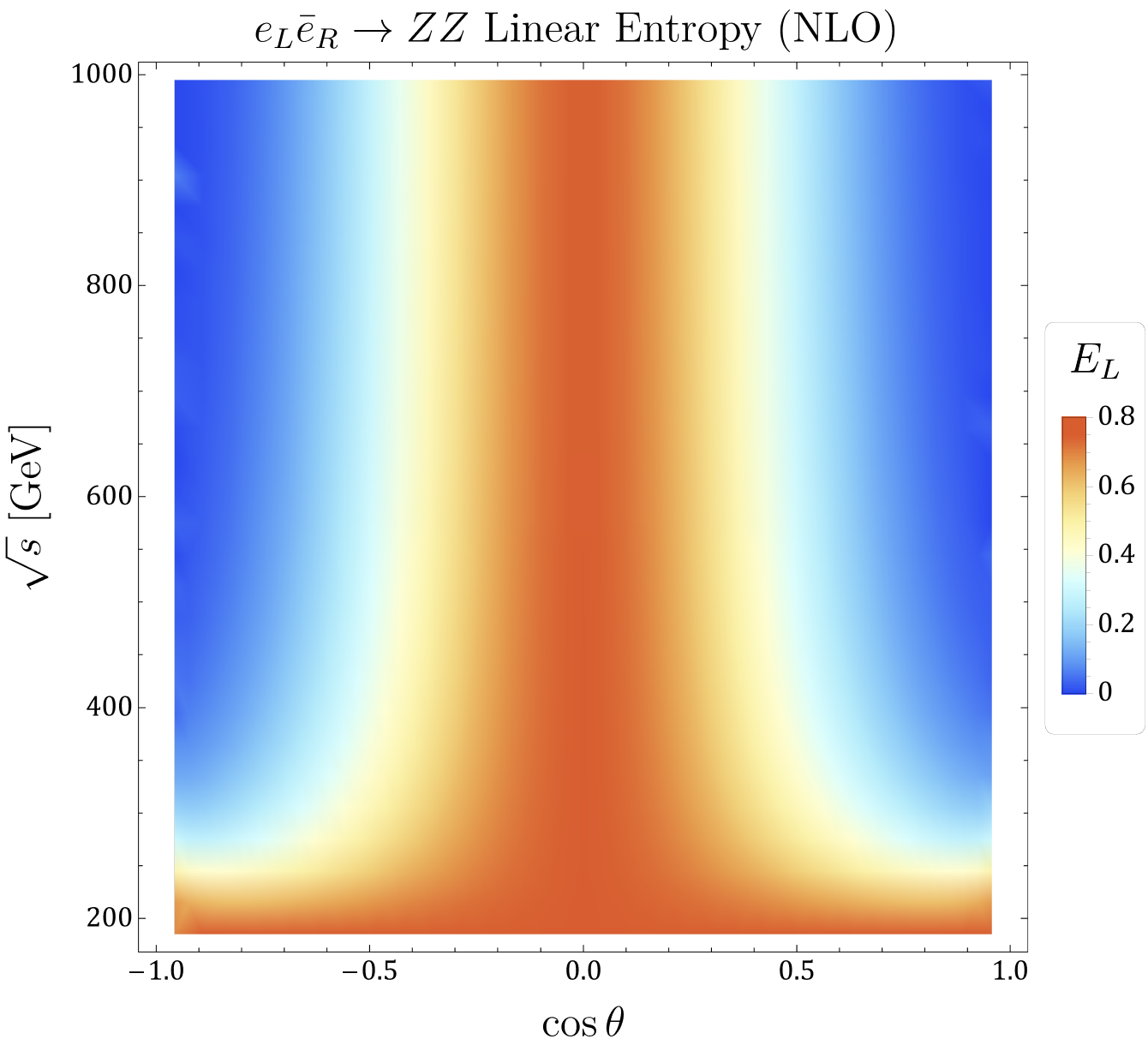}
    \caption{Density plot of linear entropy for  $e^-e^+ \to ZZ$ with $LR$ initial states at NLO. The LO density plot is visually similar, and the $RL$ channel is shown in App.~\ref{app:tau_slices}. }
    \label{fig:zz_ent_lr}
\end{subfigure}
\begin{subfigure}{.99\textwidth}
    \centering
    \includegraphics[width=0.32\linewidth]{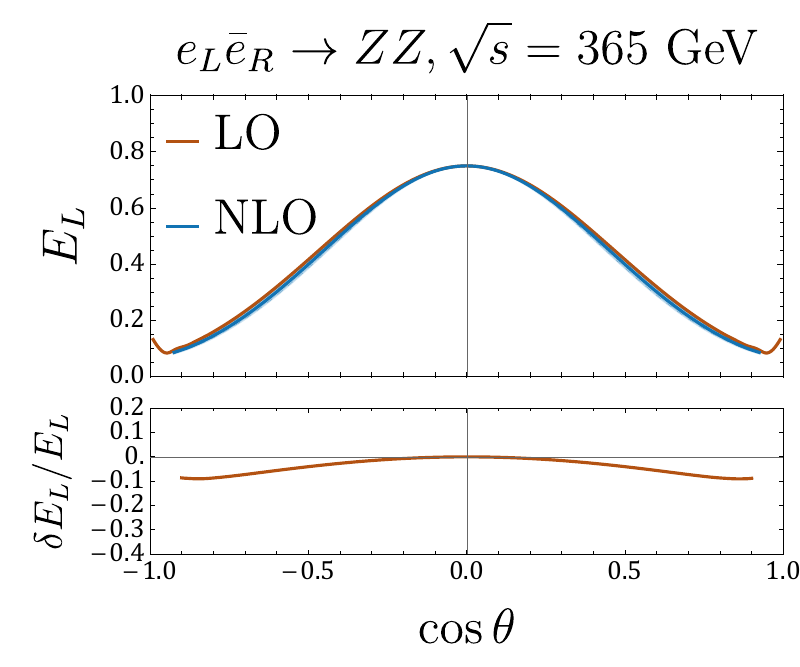}
    \includegraphics[width=0.32\linewidth]{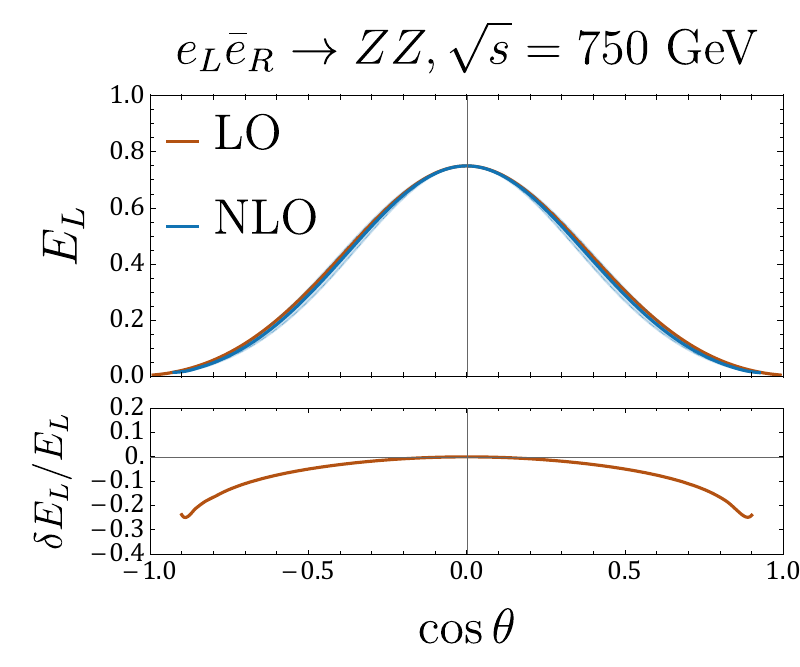}
    \includegraphics[width=0.32\linewidth]{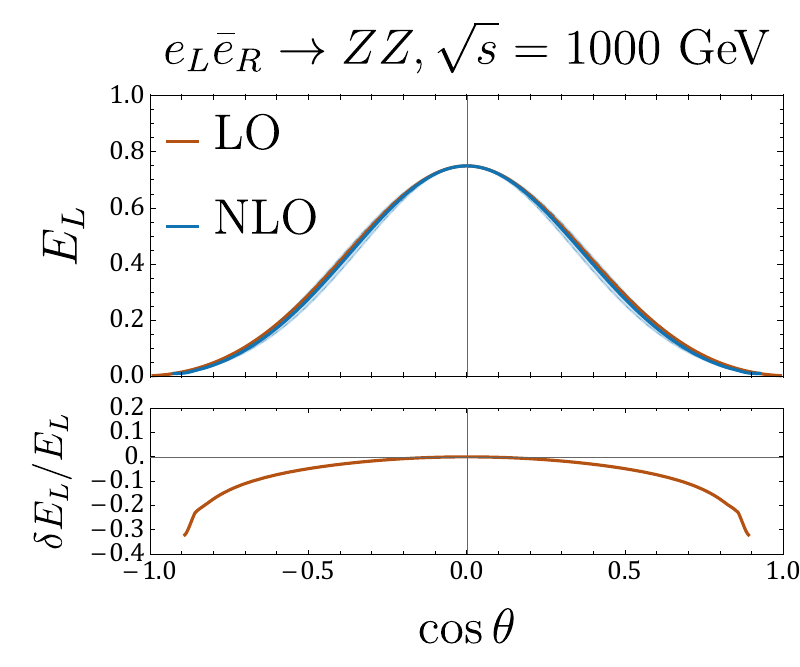}
    \caption{$ZZ$ linear-entropy slices for $LR$ initial states, comparing LO (red) and NLO (blue). At $\sqrt{s}=1\;\TeV$, the absolute LO and NLO curves approach each other near the angular endpoints, although the fractional difference grows there. The $365 \text{ GeV}$ slice corresponds to the FCC-ee $t \bar{t}$ running point, while those at higher energies convey the trend of growing NLO effects with increasing $\sqrt{s}$. The shaded regions correspond to the sensitivity of these calculations to NNLO contributions as estimated by the untruncated results (see Sec.~\ref{sec:results_tautau}).}
\label{fig:zz_ent_slices_lr}
\end{subfigure}
\caption{Linear entropy $E_L$ for $e^-e^+ \to ZZ$ with $LR$ initial states.}

\end{figure}

Figure~\ref{fig:zz_ent_lr} shows the two-dimensional density plot of the linear entropy $E_L$ for $LR$ initial states at NLO. The $RL$ channel is collected in App.~\ref{app:tau_slices}. For opposite-helicity initial states the linear-entropy density plots are remarkably stable under virtual NLO corrections. Both $LR$ and $RL$ channels retain the same threshold-to-high-energy topology seen at LO, and the differences are hard to resolve at the level of the two-dimensional color-density plots. This behavior is consistent with the general density-matrix argument in the infrared discussion above: loop contributions that are close to common rescalings of the dominant helicity amplitudes largely cancel after normalization, leaving only the non-universal helicity-dependent pieces to affect $E_L$.

The subtle NLO corrections grow with energy and are made visible via the slices in Figs.~\ref{fig:zz_ent_slices_lr} and~\ref{fig:zz_ent_slices_rl}. In localized regions of the $LR$ channel the relative loop effect can reach the order-10-percent level, while the absolute changes in $E_L$ remain small and the corresponding $RL$ modifications are much weaker. This behavior can be understood from the chiral couplings of the SM and from the diagrams that first appear at NLO. Specifically, the $RL$ (right-chiral) initial state does not couple to diagrams with internal $W$ lines, while the $LR$ (left-chiral) state does. Charged-current corrections to the differential cross section can be as large as $\sim 30\%$ relative to the tree result~\cite{Denner:1988tv}; since internal $W$ lines do not appear in the tree-level graphs, their contribution is non-factorizable with the leading result and naturally affects normalized resource measures through non-universal helicity reshuffling.

The generally weaker NLO behavior of the $ZZ$ final states contrasts with the $\tau^-\tau^+$ channel for a simple physical reason: there is no nearby $s$-channel $Z$ resonance or $\gamma$--$Z$ interference pattern shaping the $ZZ$ production amplitudes. Instead, the tree-level process starts only at threshold and is governed mainly by $t$ and $u$-channel exchange, so the resource structure varies smoothly through the electroweak region; visibly different loop-sensitive features appear only at high energies, where several tree-level helicity amplitudes become power-suppressed and loop-induced contributions can compete with them. 

The fixed-energy slices in Fig.~\ref{fig:zz_ent_slices_lr} confirm this pattern: the LO and NLO curves are close at the FCC-ee $t\bar t$-run benchmark $\sqrt{s}=365\;\GeV$ and remain similar even at $\sqrt{s}=1\;\TeV$, although localized relative distortions can become order-10-percent where the absolute entropy difference is small. The persistence of small but nonzero high-energy effects in the vector channel is relevant for future lepton colliders and, at higher energies, for the fixed-order matching of vector-boson-fusion descriptions~\cite{AlAli:2021let,Costantini:2020stv,Garosi:2023bvq}.

\paragraph{Same-helicity initial states.}

\begin{figure}[t]
\begin{subfigure}{.99\textwidth}
    \centering
    \includegraphics[width=0.49\linewidth]{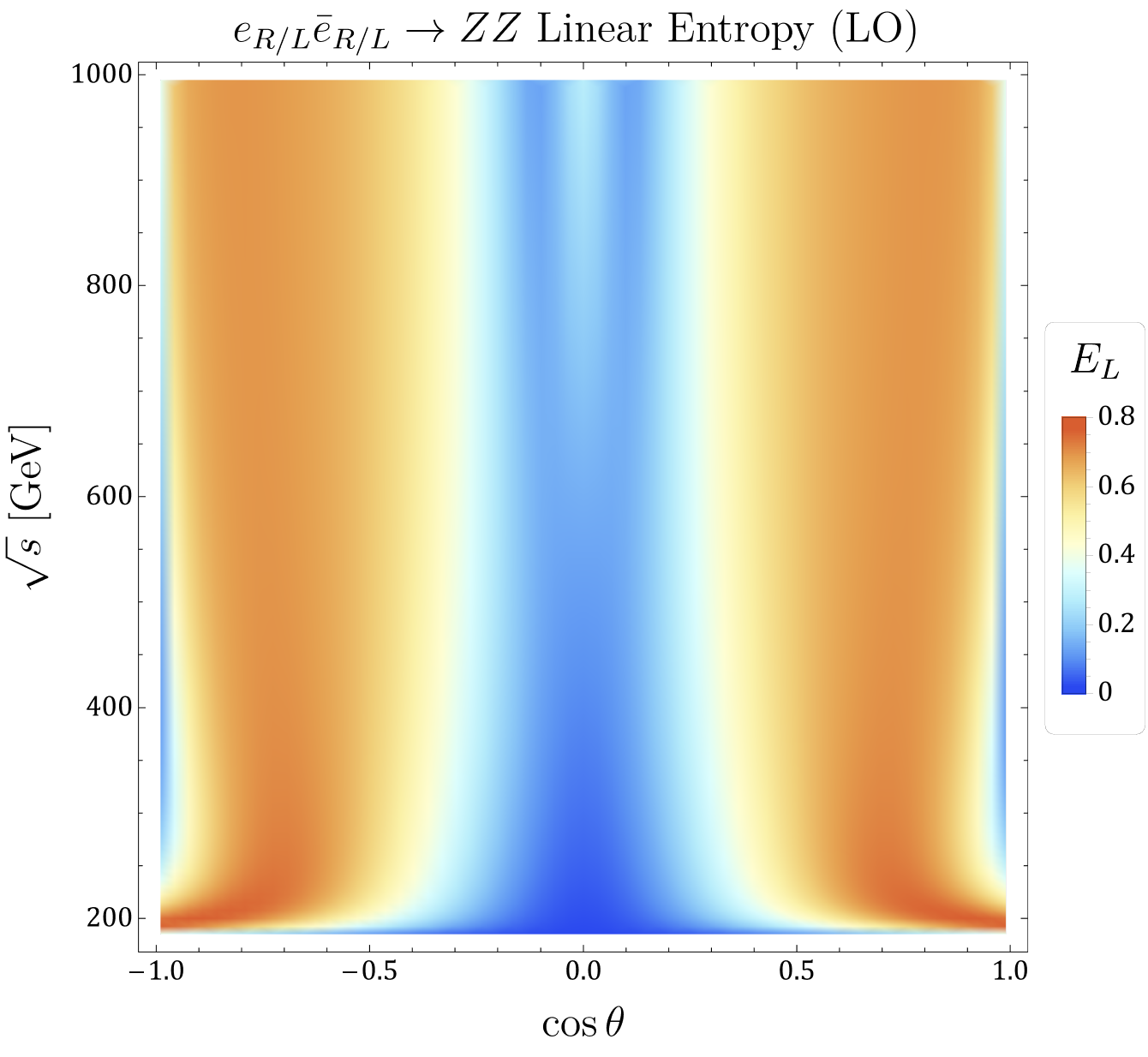}
    \includegraphics[width=0.49\linewidth]{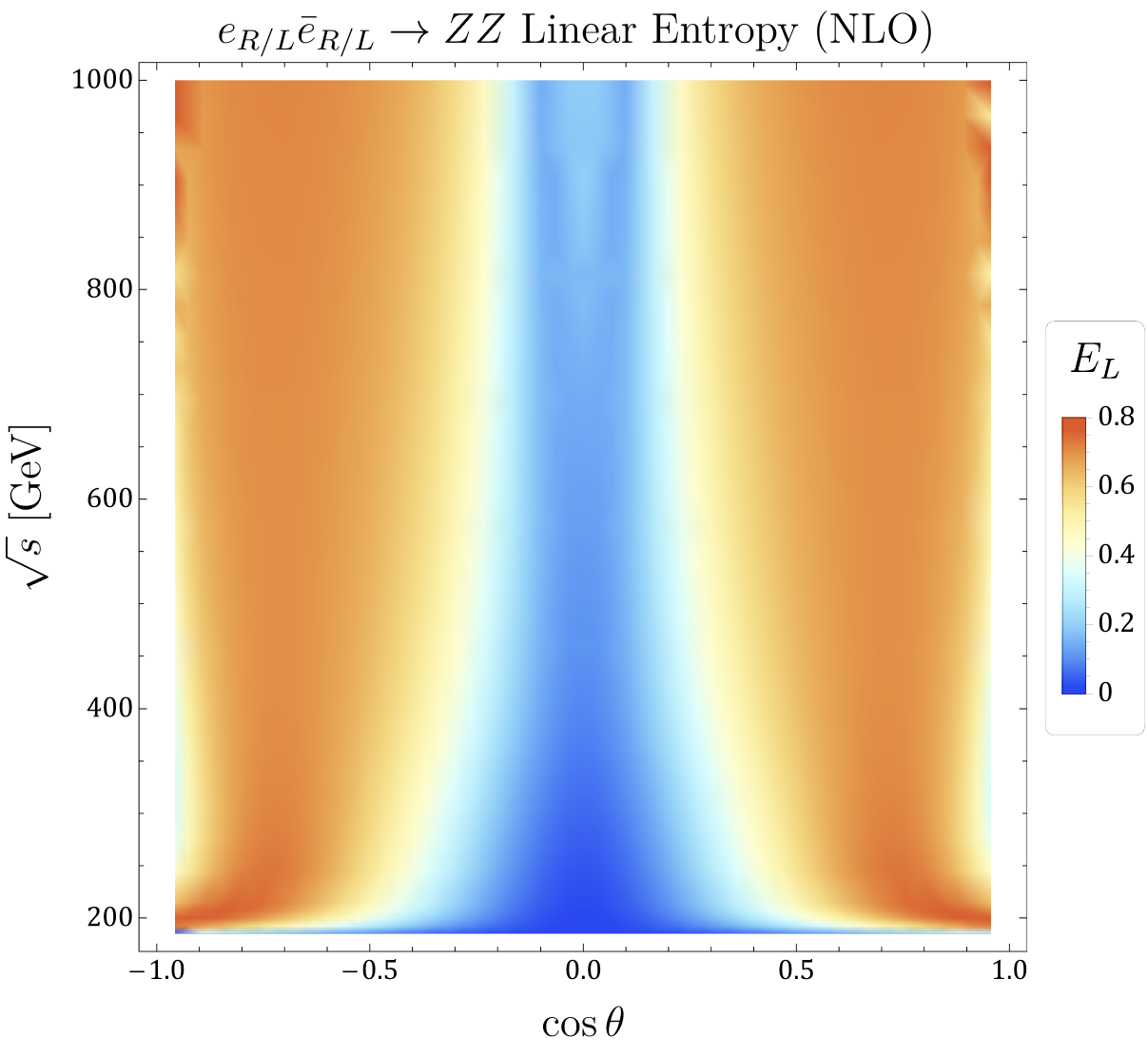}
    \caption{Density plots of linear entropy for same-helicity initial states scattering to $ZZ$ final states, shown at LO (left) and NLO (right). }
    \label{fig:zz_ent_rr}
\end{subfigure}
\begin{subfigure}{.99\textwidth}
    \centering
    \includegraphics[width=0.32\linewidth]{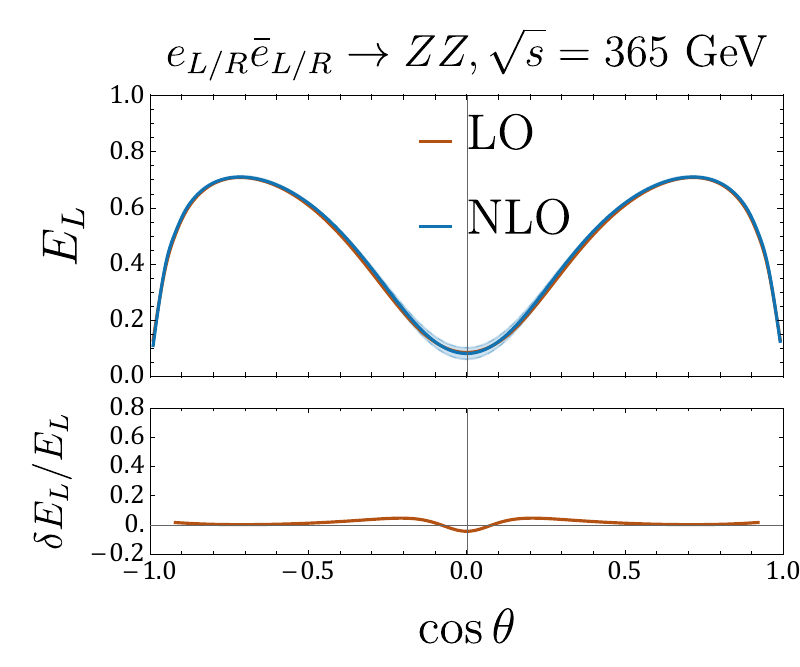}
    \includegraphics[width=0.32\linewidth]{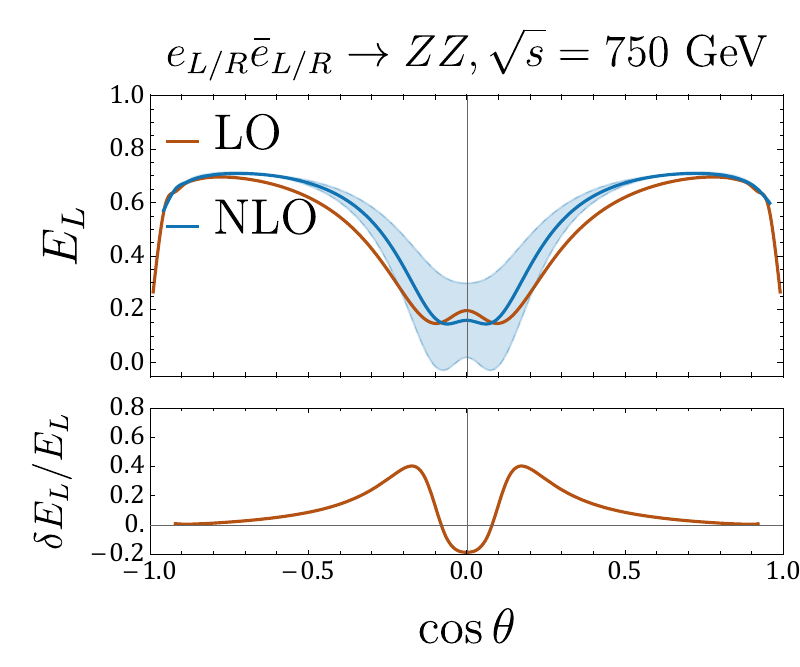}
    \includegraphics[width=0.32\linewidth]{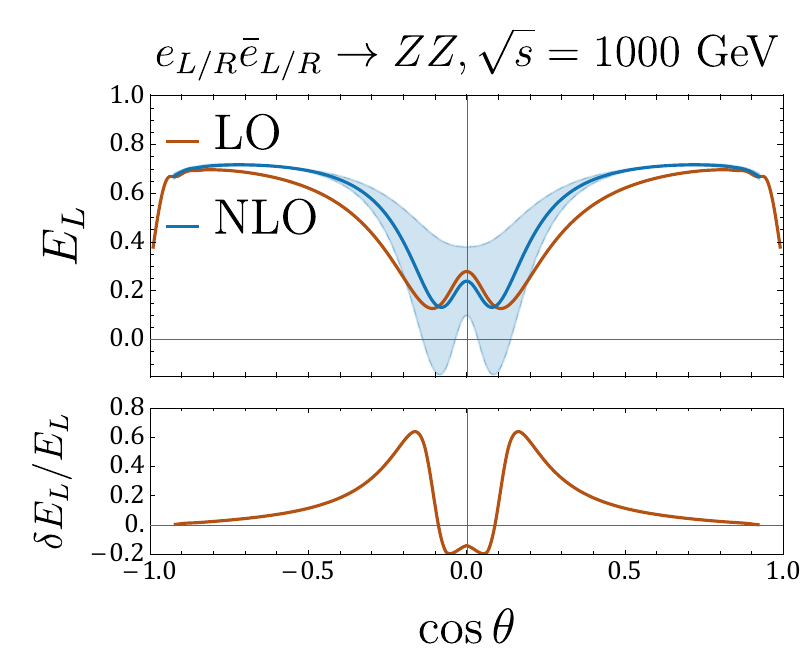}
    \caption{$ZZ$ linear-entropy slices for $LL$ and $RR$ initial states, comparing LO (red) and NLO (blue). The shaded regions highlight the sensitivity of these calculations to NNLO contributions as estimated by the untruncated results (see Sec.~\ref{sec:results_tautau}).}
\label{fig:zz_ent_slices_ll}
\end{subfigure}
\caption{Linear entropy $E_L$ for $e^-e^+ \to ZZ$ with same-helicity initial states.}

\end{figure}

Figure~\ref{fig:zz_ent_rr} shows the linear entropy for same-helicity initial states. This sector is more featureful than the opposite-helicity one: the LO density plot already contains a pronounced central low-entanglement valley bracketed by high-entanglement side bands, and the NLO correction reshapes their relative intensity and introduces visible angular dependence away from threshold. The fixed-energy slices in Fig.~\ref{fig:zz_ent_slices_ll} show this effect more explicitly, with the central entropy valley sharpening at the TeV benchmark. Since vector-boson production by same-helicity initial states is suppressed or symmetry-protected in many tree-level limits, it is a natural place for loop effects and higher-dimensional neutral-gauge or Higgs-current operators to compete. Operators such as $H^\dagger H\,Z_{\mu\nu}Z^{\mu\nu}$, $H^\dagger H\,Z_{\mu\nu}\tilde Z^{\mu\nu}$, and derivative Higgs-current structures that modify neutral-current helicity amplitudes would therefore be expected to leave especially clear imprints in these QI maps at the partonic level~\cite{Fabbrichesi:2023jep,Bernal:2023ruk,Bernal:2024xhm,Aoude:2023hxv}. The slices also suggest interesting SM effects may be found at higher orders of perturbation theory. Fig.~\ref{fig:zz_ent_slices_ll} displays growing uncertainty towards higher energies, thus NNLO effects could impact the oscillatory behavior near perpendicular scattering.

\subsection{Magic at NLO}

\begin{figure}[t]
\begin{subfigure}{.95 \textwidth}
    \centering\includegraphics[width=0.6\linewidth]{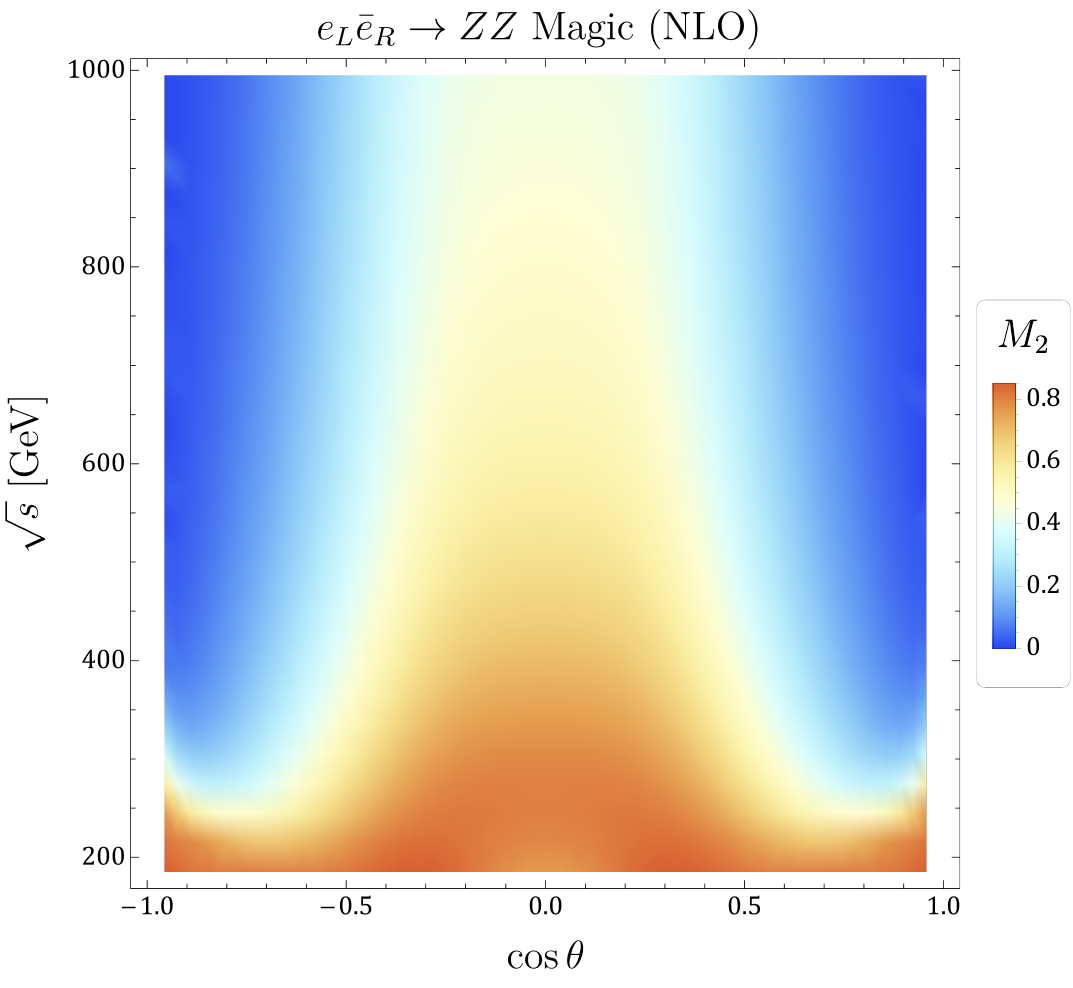}
    \caption{Density plot of stabilizer second R\'enyi entropy $M_2$ for $e^-e^+ \to ZZ$ with $LR$ initial states at NLO. The LO density plot is visually similar, and the $RL$ channel is shown in App.~\ref{app:tau_slices}. }
\label{fig:zz_mag_lr}
\end{subfigure}
\begin{subfigure}{.95 \textwidth}
    \centering
    \includegraphics[width=0.32\linewidth]{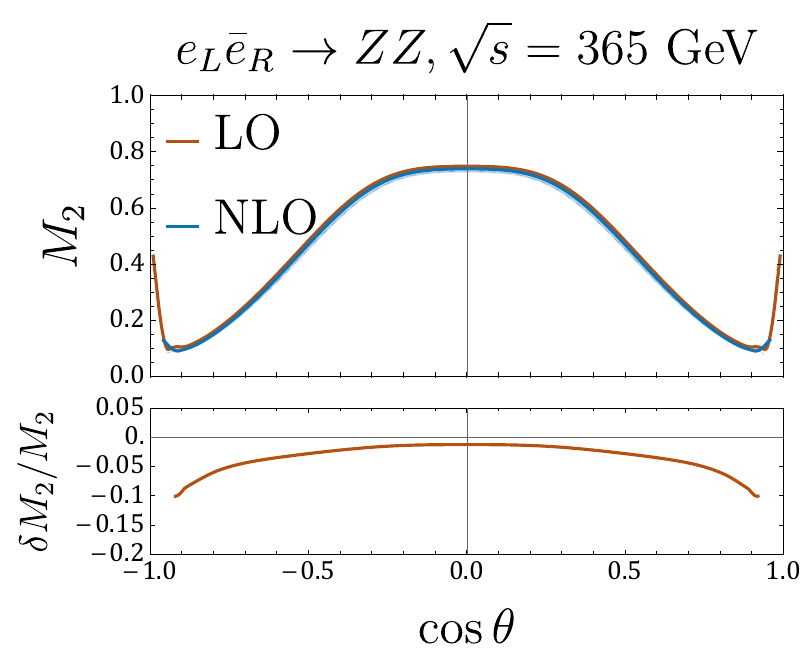}
    \includegraphics[width=0.32\linewidth]{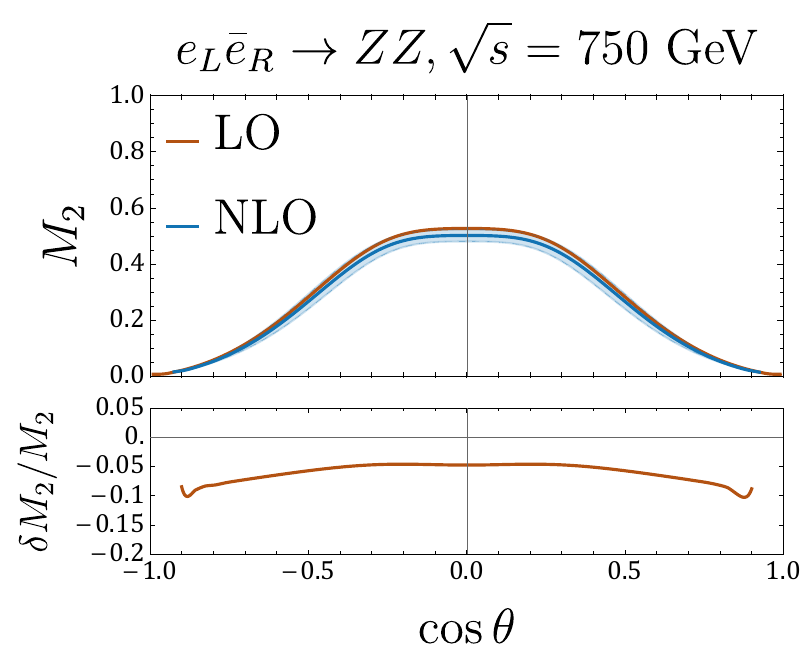}
    \includegraphics[width=0.32\linewidth]{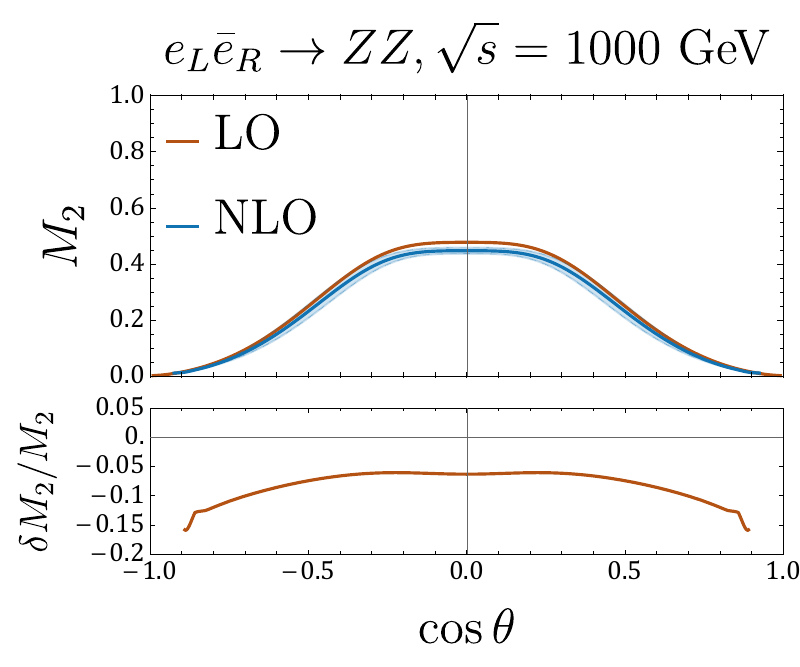}
    \caption{$ZZ$ stabilizer second R\'enyi entropy slices for $LR$ initial states, comparing LO (red) and NLO (blue). The shaded regions highlight the sensitivity of these calculations to NNLO contributions as estimated by the untruncated results (see Sec.~\ref{sec:results_tautau}).}
\label{fig:zz_mag_slices_lr}
\end{subfigure}
\caption{Stabilizer second R\'enyi entropy $M_2$ for $e^-e^+ \to ZZ$ with $LR$ initial states. }
\end{figure}

\paragraph{Opposite-helicity initial states.}
Figure~\ref{fig:zz_mag_lr} shows the stabilizer second R\'enyi entropy $M_2$ for $LR$ initial states at NLO. The $RL$ channel is collected in App.~\ref{app:tau_slices}.

The opposite-helicity magic density plots are somewhat more sensitive to the non-universal parts of the loop corrections than the corresponding entanglement plots. The broad ridge structure remains the same, but the fixed-energy slices reveal shifts in the local extrema which grow with $\sqrt{s}$. This mild separation between entanglement and magic mirrors the $\tau^-\tau^+$ channel: normalization removes corrections that act as common amplitude rescalings, while magic remains sensitive to helicity-dependent changes in the relative composition of the outgoing qutrit state. The $\sqrt{s}=365\;\GeV$ slice in Fig.~\ref{fig:zz_mag_slices_lr} is included as the FCC-ee $t \bar{t}$-run benchmark, while the higher-energy slices show the loop-induced shifts more clearly.
 
\paragraph{Same-helicity initial states.}

\begin{figure}[t]
\begin{subfigure}{.95 \textwidth}
    \centering
    \includegraphics[width=0.49\linewidth]{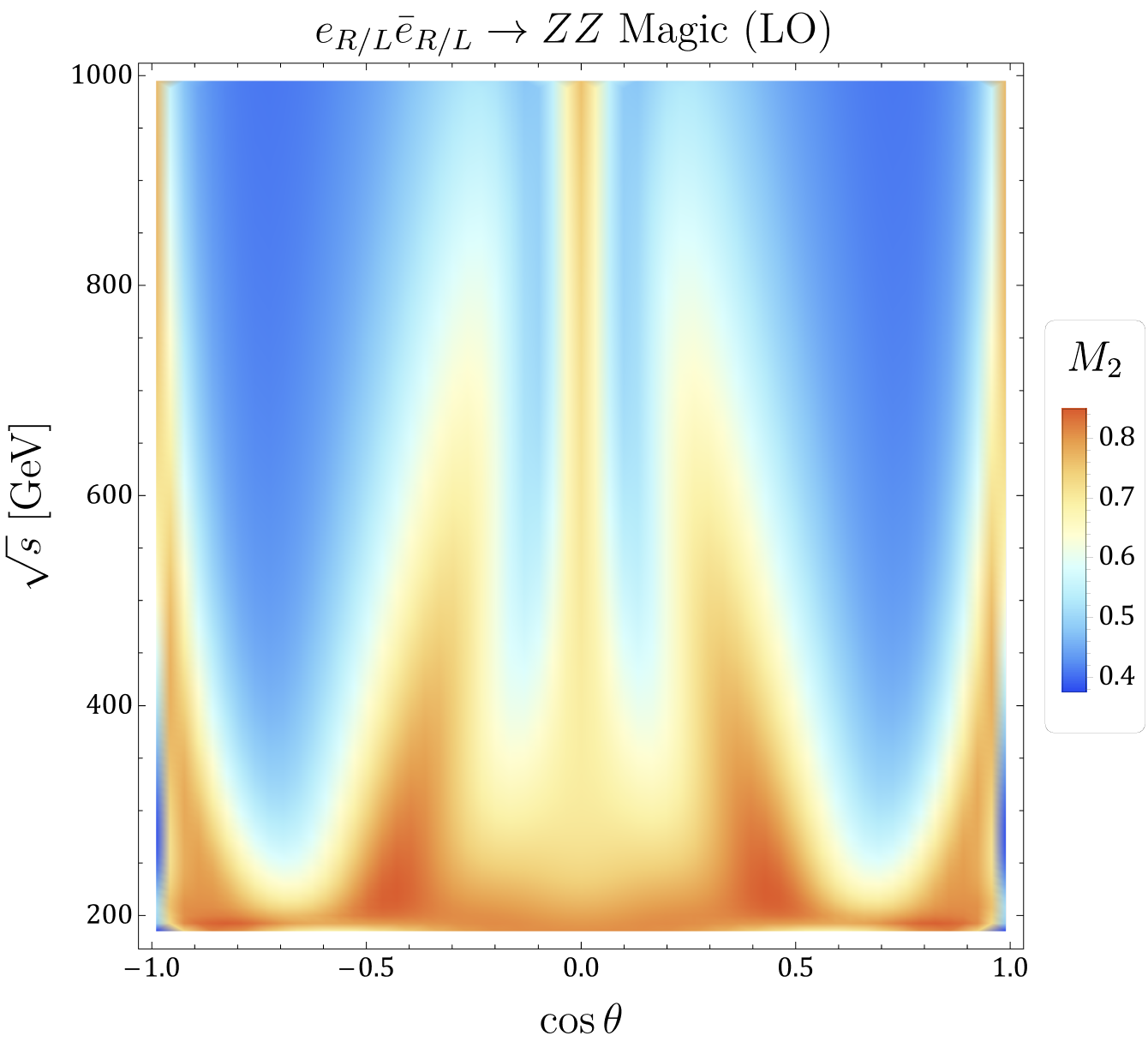}
    \includegraphics[width=0.49\linewidth]{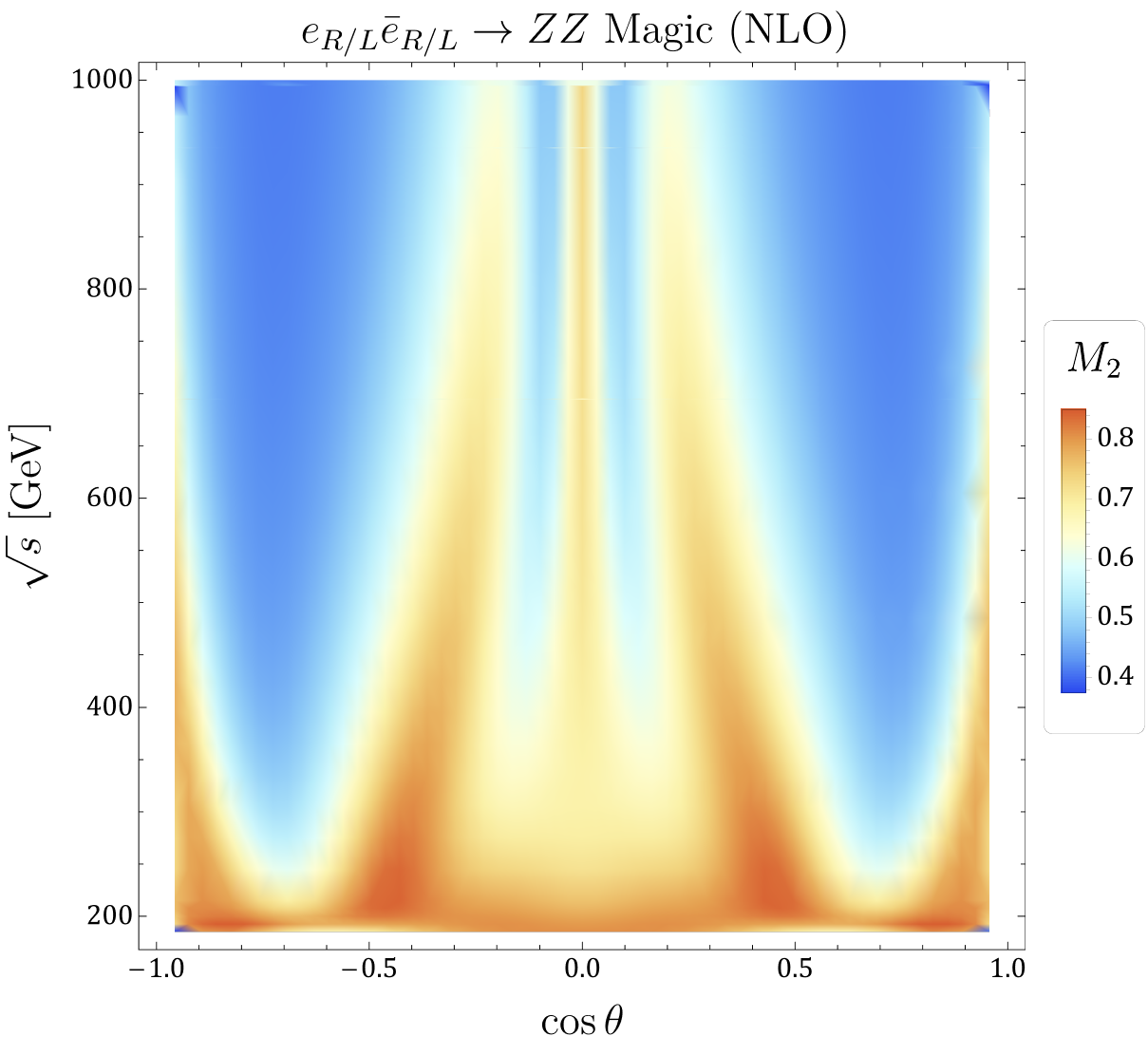}
    \caption{Density plots at LO (left) and NLO (right) of stabilizer second R\'enyi entropy for ZZ production from $RR$ or $LL$ initial states.}
\end{subfigure}
\begin{subfigure}{.95\textwidth}
    \centering
    \includegraphics[width=0.32\linewidth]{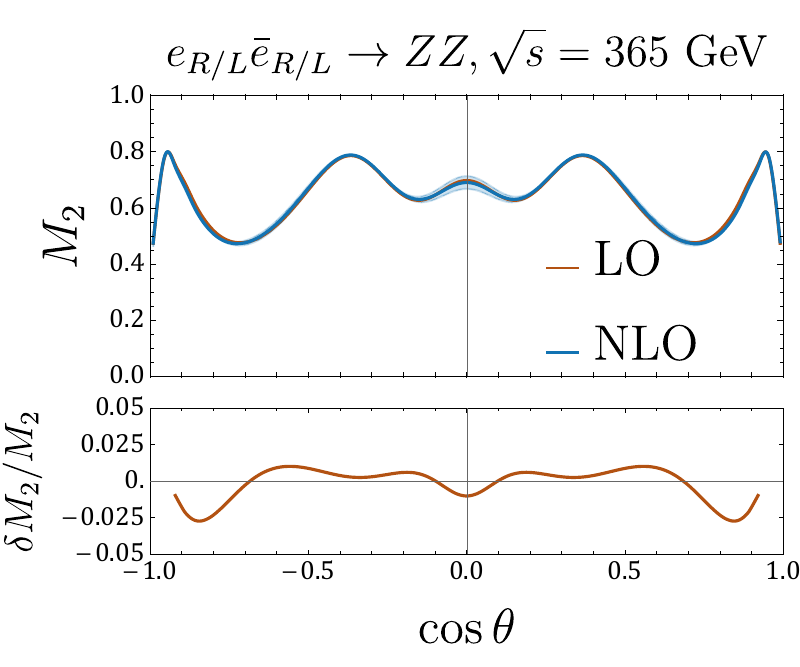}
    \includegraphics[width=0.32\linewidth]{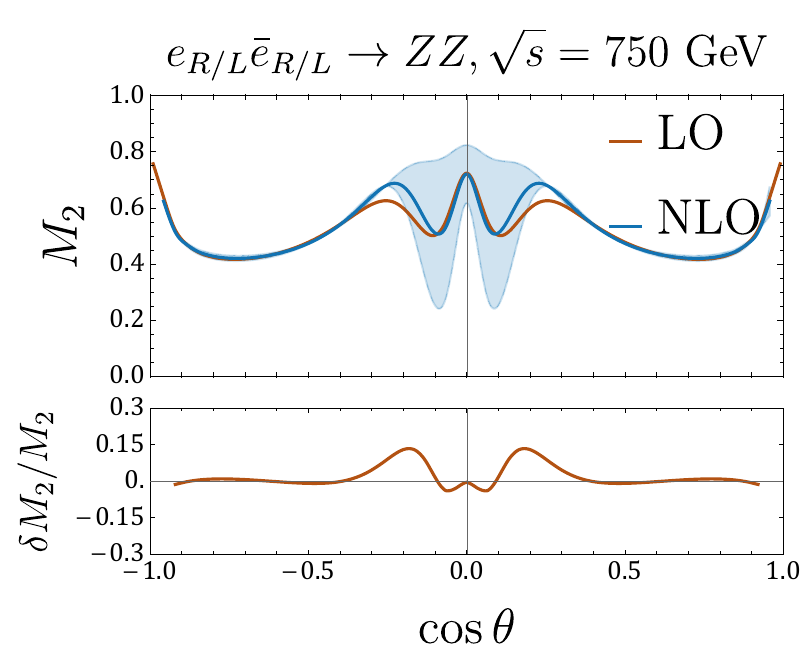}
    \includegraphics[width=0.32\linewidth]{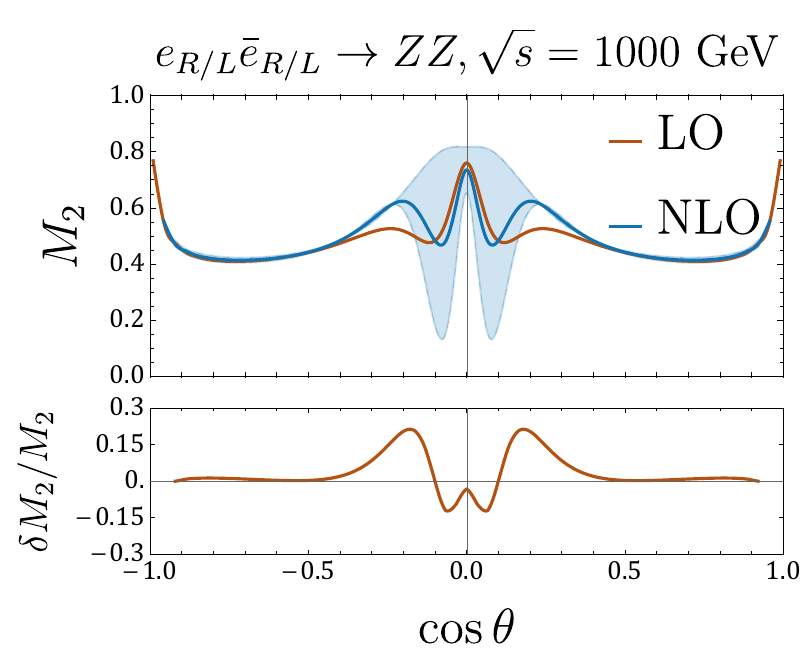}
\caption{$ZZ$ stabilizer second R\'enyi entropy $M_2$ for $LL$ and $RR$ initial states, comparing LO (red) and NLO (blue). The shaded regions highlight the sensitivity of these calculations to NNLO contributions as estimated by the untruncated results (see Sec.~\ref{sec:results_tautau}).}
\end{subfigure}
\caption{Stabilizer second R\'enyi entropy $M_2$ for $e^-e^+ \to ZZ$ with same-helicity initial states, shown at LO (left) and NLO (right).}
\label{fig:zz_mag_rr}
\end{figure}

As shown in Fig.~\ref{fig:zz_mag_rr}, the same-helicity magic observable shows the clearest NLO reshuffling in the $ZZ$ channel, though NNLO contributions could be significant in the regions where the quantum observable varies most rapidly (namely, near perpendicular scattering). Up to these truncated results, $M_2$ responds more sharply (compared to $E_L$ for the same channel) to loop-induced changes in the relative weights and phases of the vector-boson helicity amplitudes, especially away from the threshold region. Same-helicity $ZZ$ production is therefore where new-physics deformations of the neutral-current vector-boson amplitudes would register most strongly in these observables. Whether this sensitivity is experimentally accessible is a separate question: the chirality-suppressed Born rate makes these configurations exclusive quantities, requiring photon thresholds and beam-polarization purities well beyond realistic FCC-ee conditions (Sec.~\ref{sec:IR_structure}), and we therefore regard the same-helicity maps as theoretical diagnostics of the loop structure rather than as projected measurements. CP-even and CP-odd contact structures such as $H^\dagger H\,Z_{\mu\nu}Z^{\mu\nu}$ and $H^\dagger H\,Z_{\mu\nu}\tilde Z^{\mu\nu}$, as well as higher-derivative neutral-gauge operators, can populate helicity configurations that are loop- or interference-suppressed in the SM and would therefore be naturally amplified in magic-based diagnostics~\cite{Fabbrichesi:2023jep,Bernal:2023ruk,Bernal:2024xhm,Aoude:2023hxv}.

\section{Conclusions}
\label{sec:conclusions}

We have performed the first systematic study of how virtual NLO electroweak corrections reshape the quantum resources of entanglement and magic in leptonic scattering processes at lepton colliders, covering the qubit channel $e^-e^+\to\tau^-\tau^+$ and the qutrit channel $e^-e^+\to ZZ$. 

In $\tau^-\tau^+$ production, the NLO corrections are strongest at the electroweak scale, where they shift the angular positions of both entanglement and magic peaks, while preserving the qualitative branch topology. For same-helicity states, the entanglement and magic band structure exhibits novel angular dependence at NLO, with the most dramatic effects occurring near the $Z$ boson resonance. In this region, the oscillatory behavior of the magic content is significantly enhanced; within the idealized exclusive setting that same-helicity beams require (Sec.~\ref{sec:IR_structure}), magic is thus the more responsive of the two observables to radiative corrections in this sector.

In $ZZ$ production, the opposite-helicity linear-entropy density plots are much more stable: LO and NLO remain nearly indistinguishable at energies accessible to FCC-ee, with localized order-10-percent relative distortions appearing first in the TeV regime where the absolute entropy difference remains small. This contrasts with the $\tau^-\tau^+$ channel, where the nearby $\gamma/Z$ resonant structure and chiral interference already amplify loop-induced reshuffling around the electroweak scale; in neutral diboson production, by contrast, the tree-level helicity pattern evolves more smoothly from threshold, so the loop-sensitive distortions only become prominent once the high-energy helicity structure takes over. The most pronounced changes in $ZZ$ production appear in magic and in same-helicity configurations, where loop-induced helicity reshuffling can compete with suppressed tree-level structures; these configurations are, however, exclusive quantities in the sense of Sec.~\ref{sec:IR_structure}, and we present them as theoretical diagnostics rather than collider projections.

Our results establish the NLO electroweak baseline for the quantum observables, a prerequisite for any quantitative use of these resources at future lepton colliders.
More broadly, NLO effects reshape the geometry of the outgoing quantum state in spin space, shifting the regions of phase space where each resource is maximized or minimized compared to the tree-level expectation. They are necessary input for extending the analysis to realistic beam-polarization effects, collider-specific tomography strategies, and systematic studies of sensitivity to higher-dimensional operators.

Finally, beyond the specific processes studied here, the tools developed in this work for calculating higher-order corrections to quantum observables should be relevant more broadly to the line of work in which such observables are used as guides towards organizing principles underlying the structure of effective field theories. This includes the use of extremization of entanglement or magic to diagnose enhanced symmetry, symmetry breaking, and parameter structure in low-energy hadronic scattering~\cite{Beane:2018oxh,Beane:2021zvo,Low:2021ufv,Liu:2022grf,Miller:2023ujx,Low:2024hvn,Low:2024mrk,Hu:2024hex,Hu:2025lua,Cavallin:2025kjn,McGinnis:2025iab,McGinnis:2025xgt,Sone:2026jmo}, in Higgs-sector models~\cite{Carena:2023vjc,Carena:2025wyh,Chang:2024wrx,Kowalska:2024kbs,Liu:2025iwh,Busoni:2025dns}, in the flavor and electroweak sectors of the SM~\cite{Thaler:2024anb,Chernyshev:2024pqy,Liu:2025bgw} as well as in Yang--Mills theory and gravity~\cite{Cervera-Lierta:2017tdt,Gargalionis:2025iqs,Nunez:2025xds}. These studies have so far been formulated predominantly at leading order, and it is therefore important to understand which of the resulting extrema of interest are stable under radiative corrections.

\section*{Acknowledgements}
We thank Tao Han, Fabio Maltoni, Giovanni Pelliccioli, Jesse Thaler, and Lian-Tao Wang for useful discussions.
S.C.~and T.D.~are supported in part by the DOE under grant DE-SC0011640. S.T. was supported by the Swiss National Science Foundation project number P5R5PT\_222350. S.T. is also supported in part by the U.S. Department of Energy (DOE) under Contracts DE-SC0023522, DE-SC0010143, and No. 89243024CSC000002 (QuantISED Program). We acknowledge CERN TH Department for hospitality while this research was being carried out. The authors used Claude Fable 5 for independent cross-checks of selected algebraic and numerical calculations; responsibility for all results and conclusions rests with the authors.

\clearpage
\raggedbottom
\appendix

\section{Supplementary Plots}
\label{app:tau_slices}

This appendix collects supplementary plots that support the main discussion but are not essential to its flow. We keep them here to document the broader pattern across helicity channels and benchmark energies, while the main text focuses on the representative examples that carry the central physics interpretation.

\subsection{Tau-pair Supplementary Plots}

\begin{figure}[H]
\centering
\includegraphics[width=0.48\linewidth]{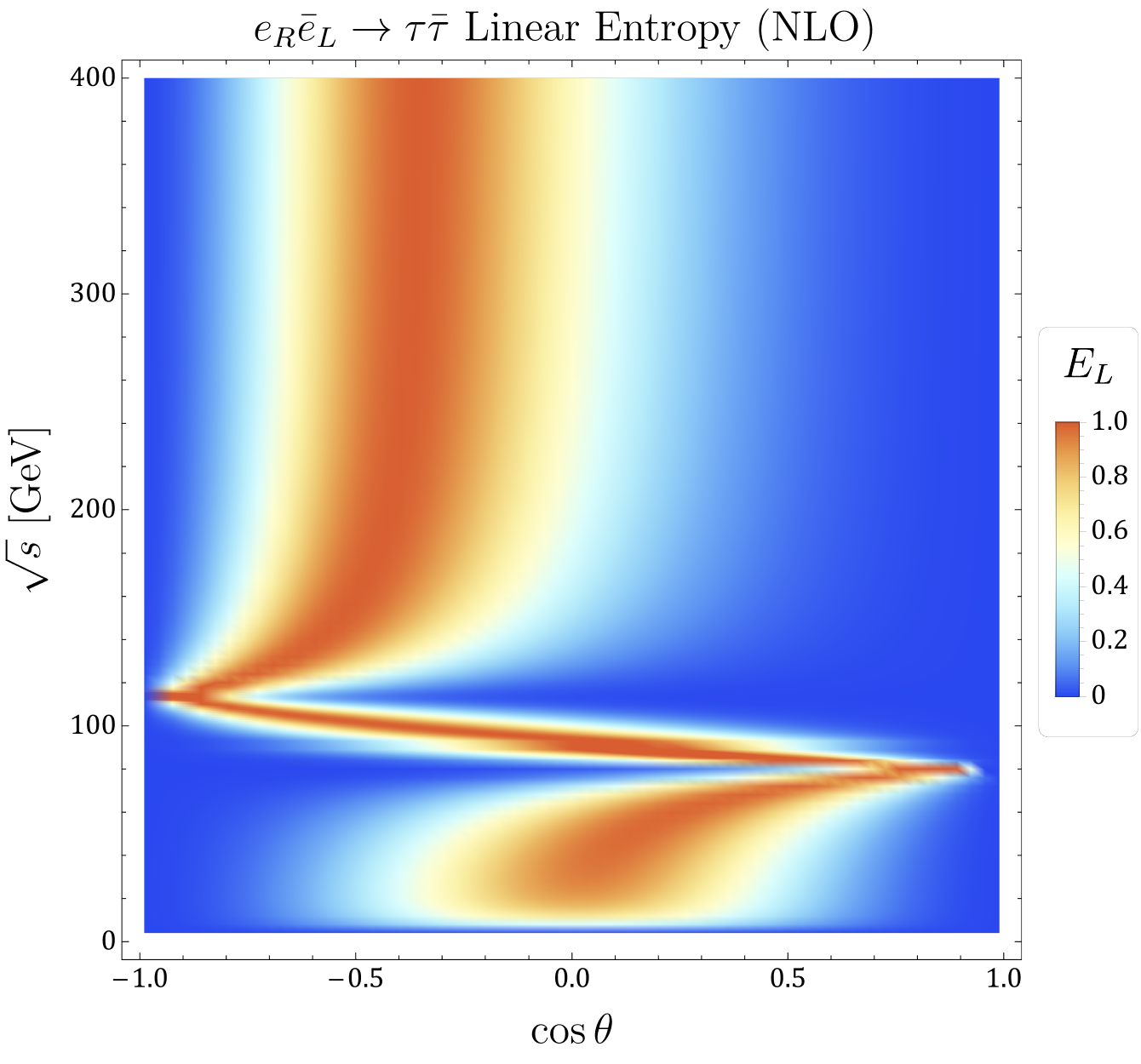}\hfill
\includegraphics[width=0.48\linewidth]{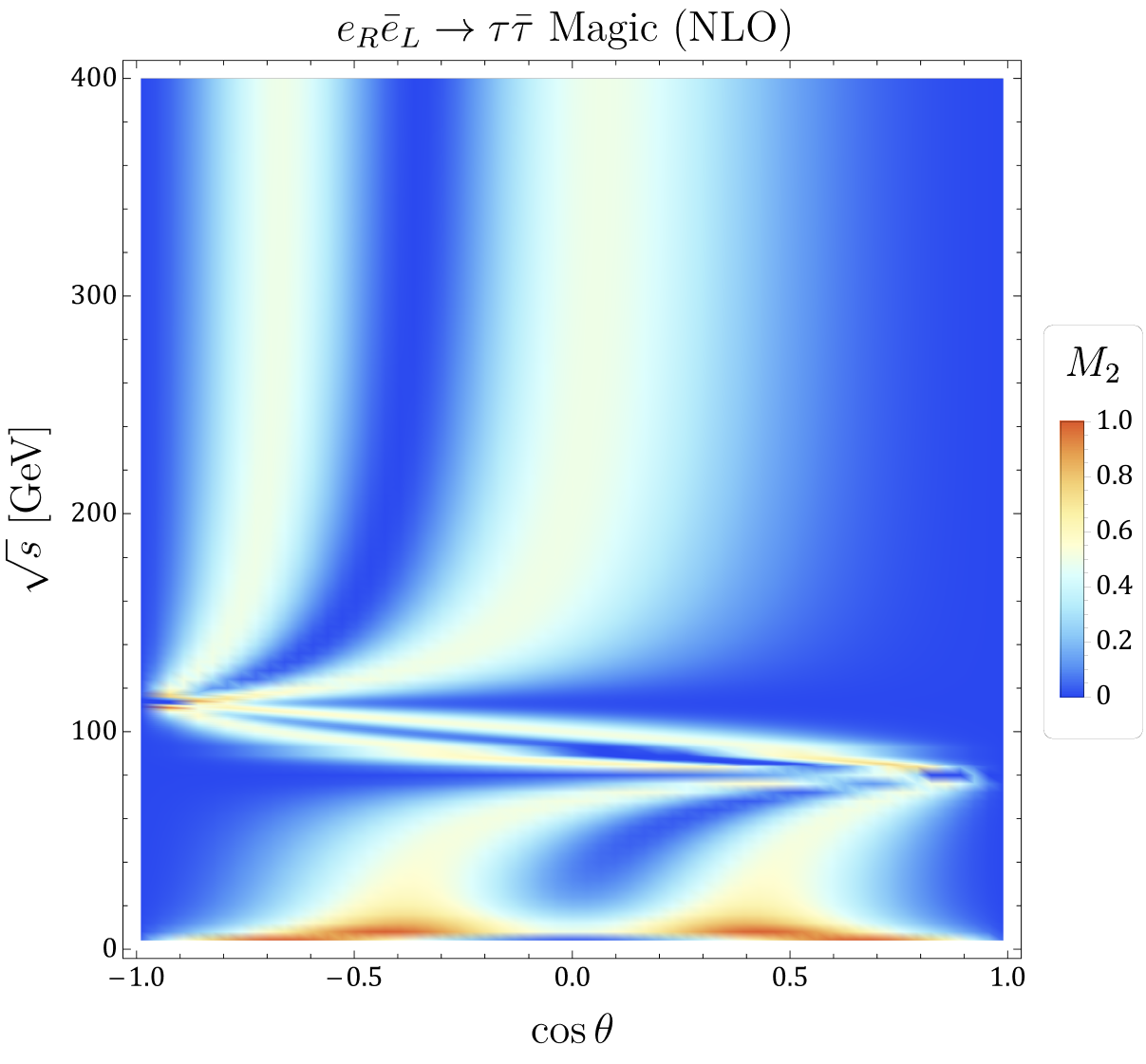}
\caption{$e^-e^+ \to \tau^-\tau^+$ with $RL$ initial states at NLO in the $(\cos\theta,\sqrt{s})$ plane, showing the linear entropy $E_L$ (left) and the stabilizer second R\'enyi entropy $M_2$ (right). Compare with the corresponding $LR$ channels in Figs.~\ref{fig:tau_ent_lr} and~\ref{fig:tau_mag_lr}.}
\label{fig:tau_ent_rl}
\label{fig:tau_mag_rl}
\end{figure}

\begin{figure}[H]
\centering
\includegraphics[width=0.32\linewidth]{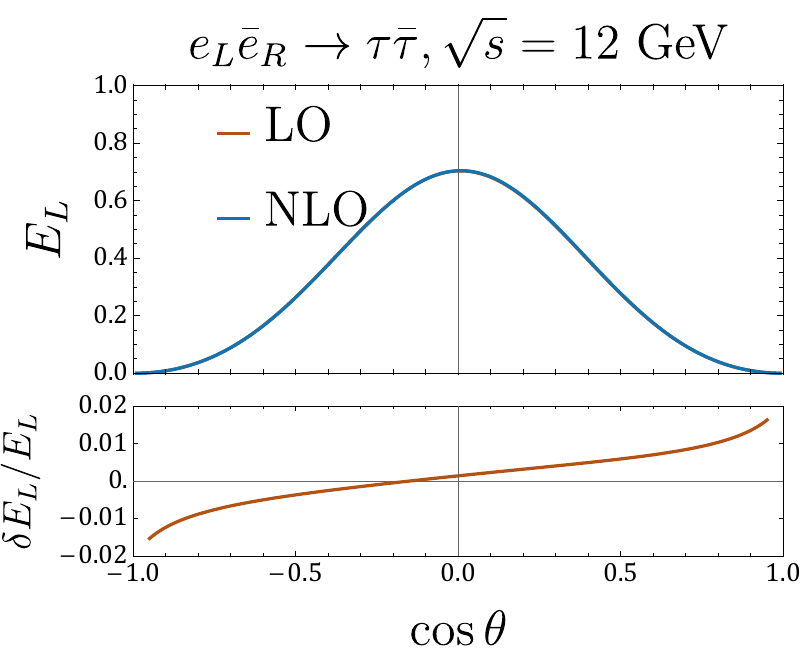}
\includegraphics[width=0.32\linewidth]{LinEntPlotsDraft/eetautau/Slices_w_Error/LR72e.pdf}
\includegraphics[width=0.32\linewidth]{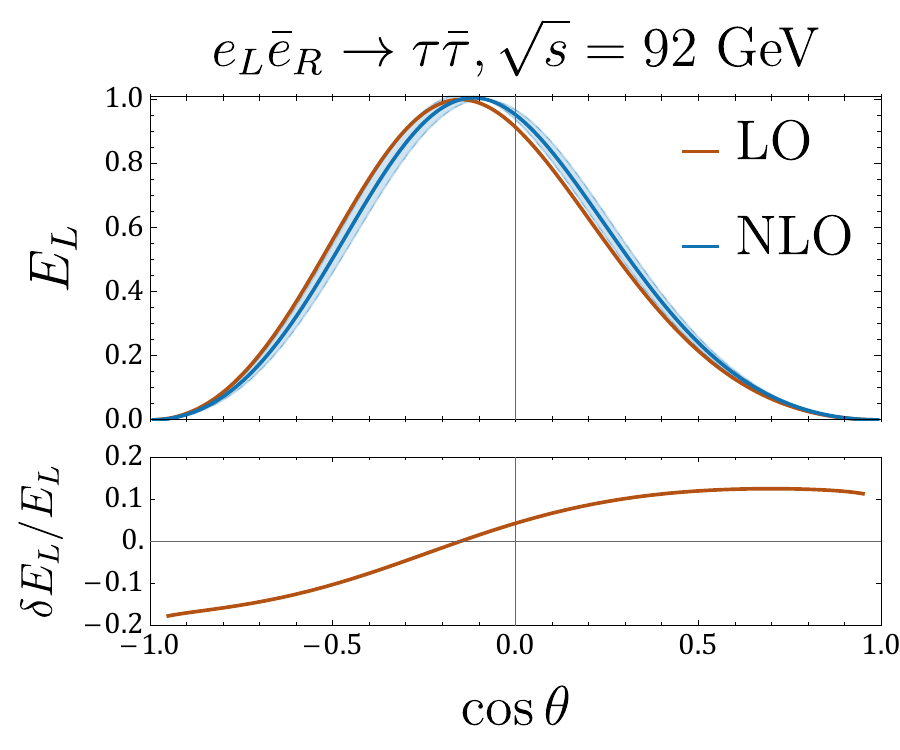}\\[3pt]
\includegraphics[width=0.32\linewidth]{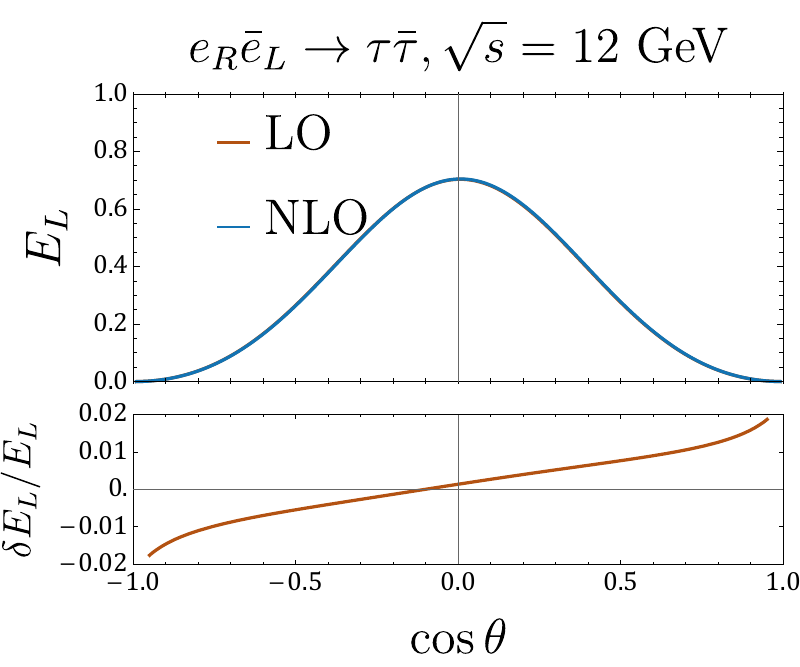}
\includegraphics[width=0.32\linewidth]{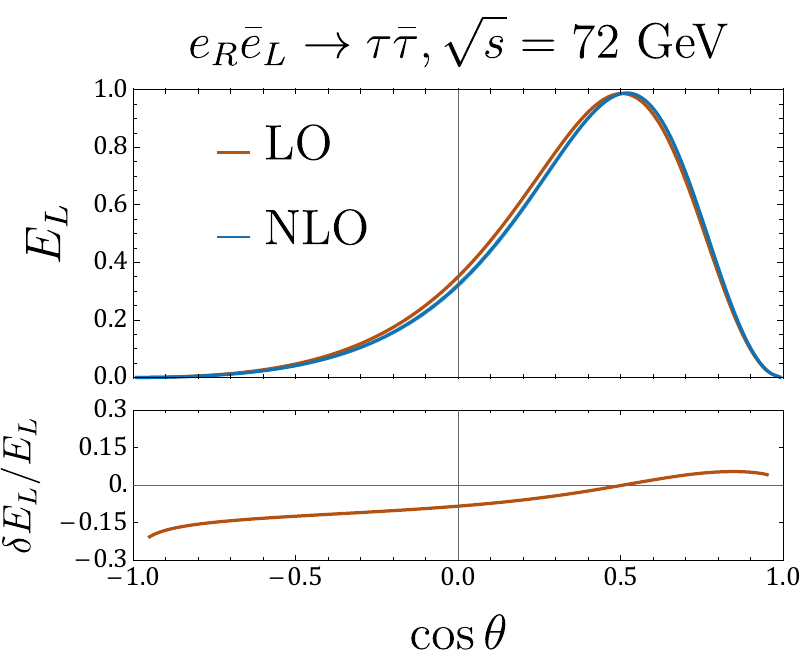}
\includegraphics[width=0.32\linewidth]{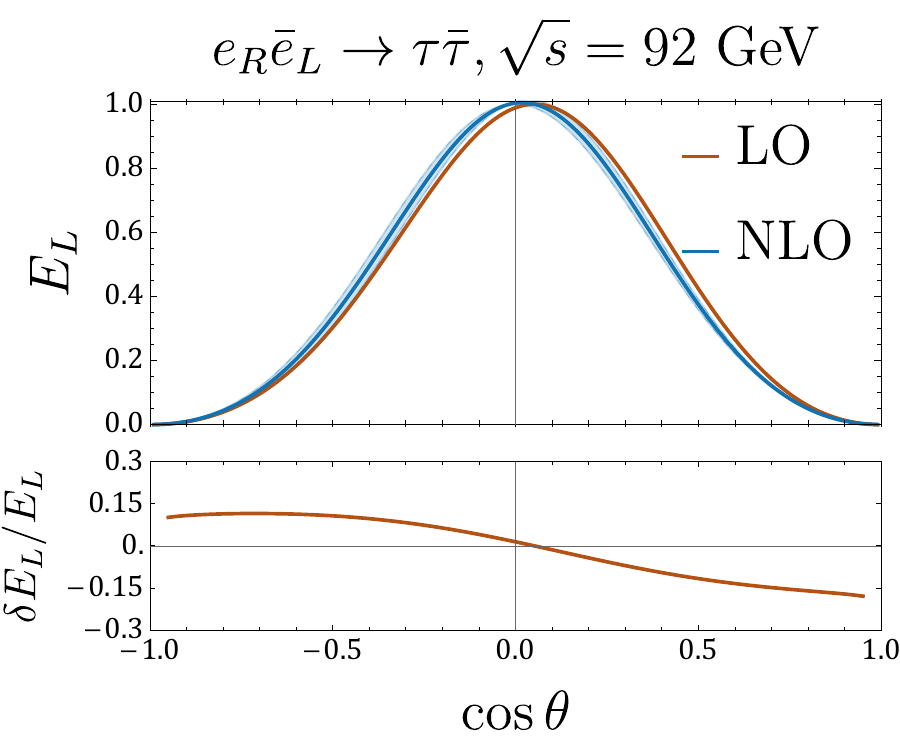}
\caption{Representative fixed-energy slices of the $\tau^-\tau^+$ linear entropy at LO and NLO, displaying the transition from $1\%$ level corrections at low energies where QED dominates, to the $\sim50\%$ corrections realized at energies near the weak boson masses. Results for which $E_L > 1$ arise as an artifact of truncating the perturbative series to $\mathcal{O}(\alpha)$. The $12 \text{ GeV}$ slice was selected to correspond with the energy scale of Belle II, while the $92 \text{ GeV}$ slice serves to benchmark the proposed FCC-ee running mode at the Z-pole. Meanwhile, the $72 \text{ GeV}$ slice captures the energy scale at which virtual corrections first become substantial. The shaded regions highlight the sensitivity of these calculations to NNLO contributions as estimated by the untruncated results (see Sec.~\ref{sec:results_tautau}).  }
\end{figure}

\begin{figure}[H]
\centering
\includegraphics[width=0.32\linewidth]{LinEntPlotsDraft/eetautau/Slices_w_Error/LR132e.pdf}
\includegraphics[width=0.32\linewidth]{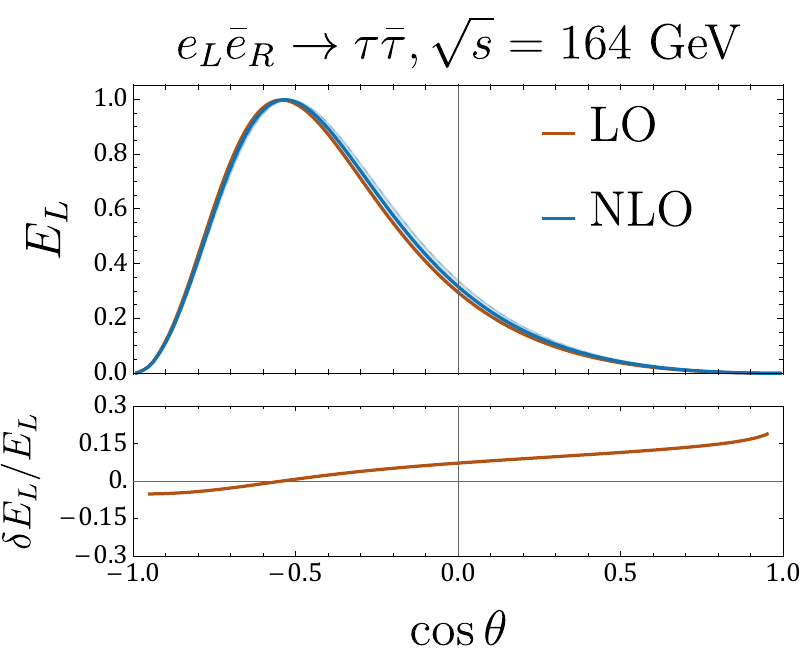}
\includegraphics[width=0.32\linewidth]{LinEntPlotsDraft/eetautau/Slices_w_Error/LR240e.pdf}\\[3pt]
\includegraphics[width=0.32\linewidth]{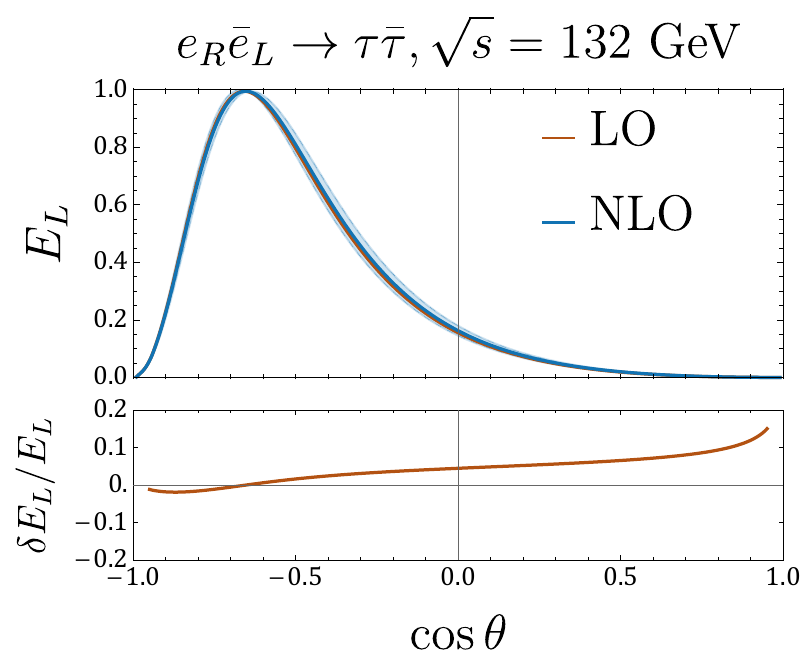}
\includegraphics[width=0.32\linewidth]{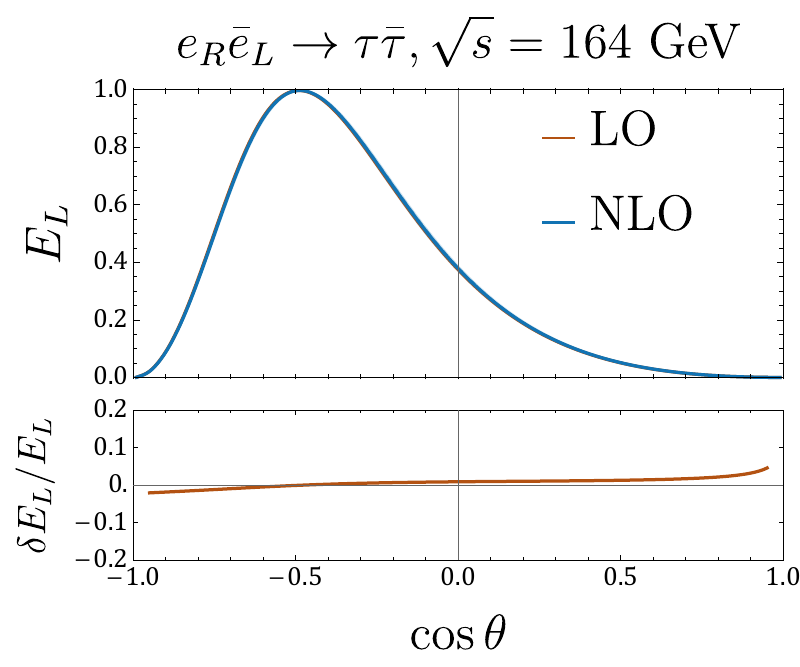}
\includegraphics[width=0.32\linewidth]{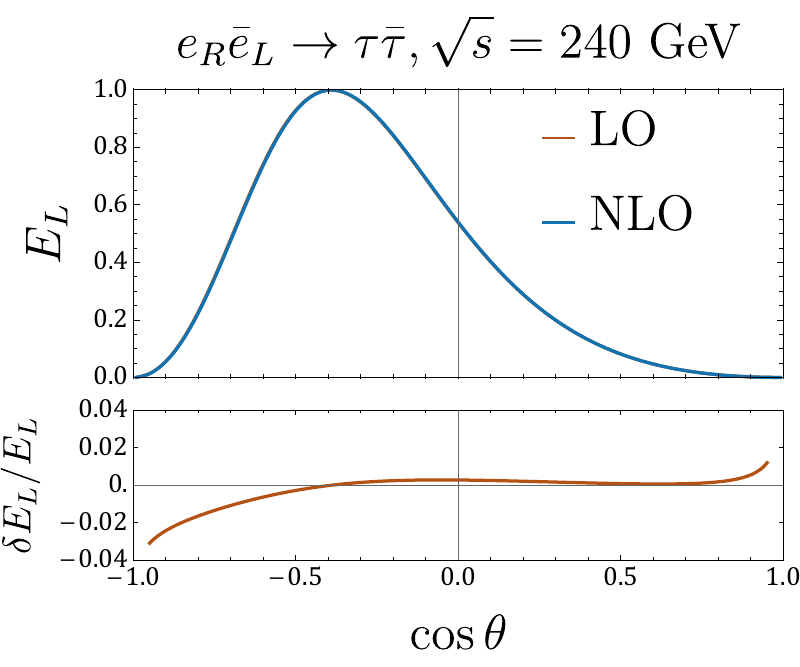}
\caption{ Additional fixed-energy slices of the $\tau^-\tau^+$ linear entropy at LO and NLO. The virtual corrections shift the peak positions while preserving the qualitative angular structure. Slices at $164 \text{ and } 240 \text{ GeV}$ correspond to the $WW$ threshold and $ZH$ proposed FCC-ee running points. The $132 \text{ GeV }$slice is included to help articulate the trend of generally decreasing NLO contributions  as energies increase beyond $M_Z$. The shaded regions highlight the sensitivity of these calculations to NNLO contributions as estimated by the untruncated results (see Sec.~\ref{sec:results_tautau}).}
\end{figure}

\begin{figure}[H]
\centering
\includegraphics[width=0.32\linewidth]{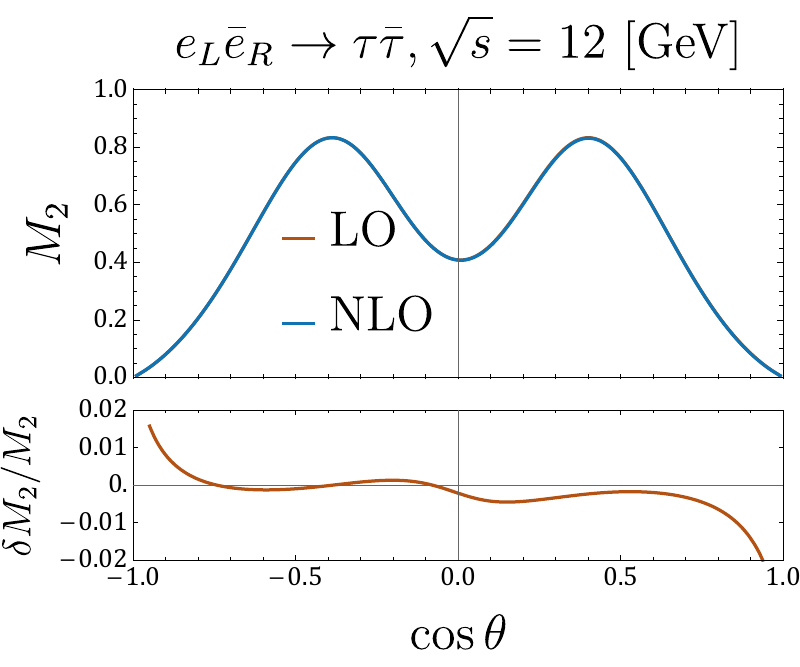}
\includegraphics[width=0.32\linewidth]{MagPlotsDraft/eetautau/Slices_w_Error/LR72M.pdf}
\includegraphics[width=0.32\linewidth]{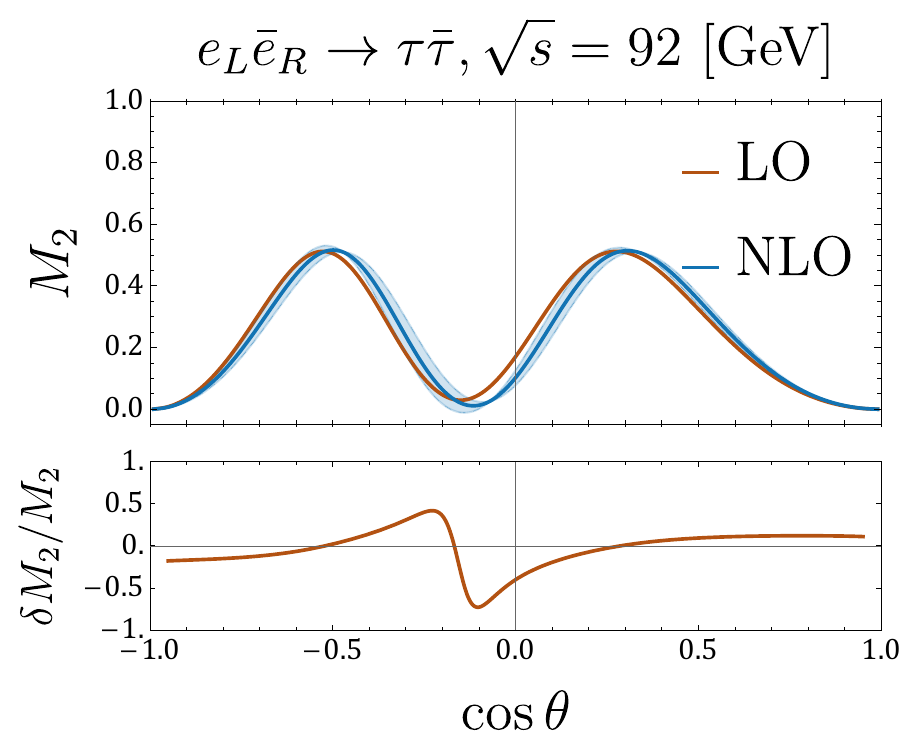}\\[3pt]
\includegraphics[width=0.32\linewidth]{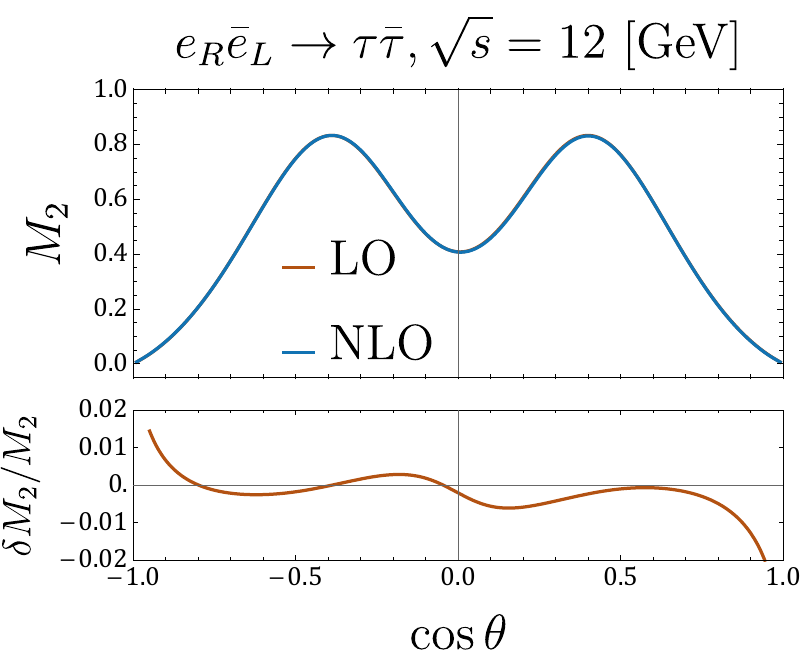}
\includegraphics[width=0.32\linewidth]{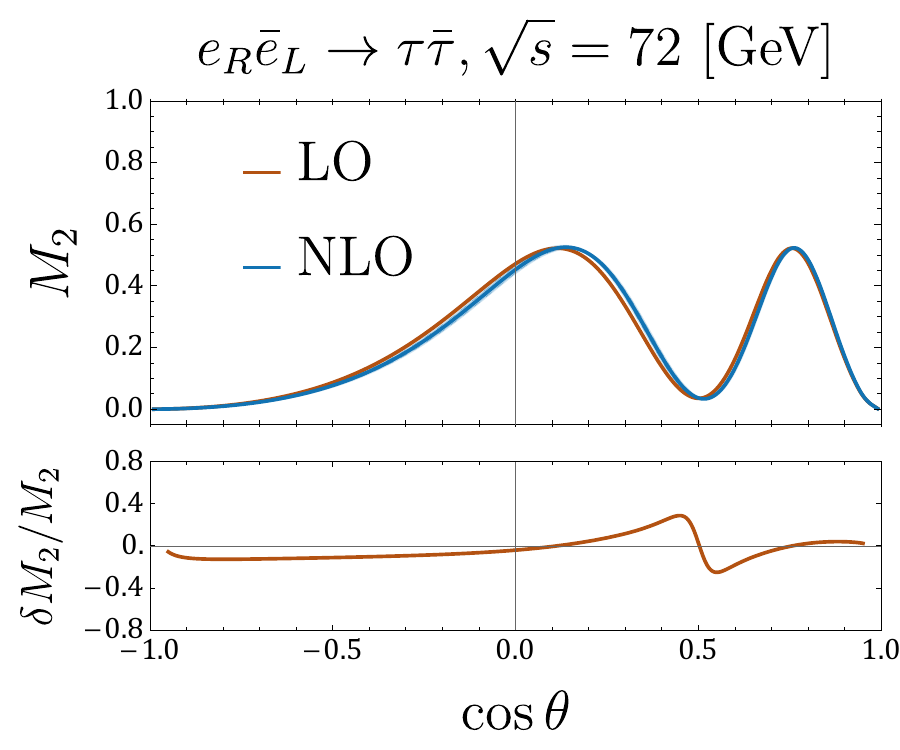}
\includegraphics[width=0.32\linewidth]{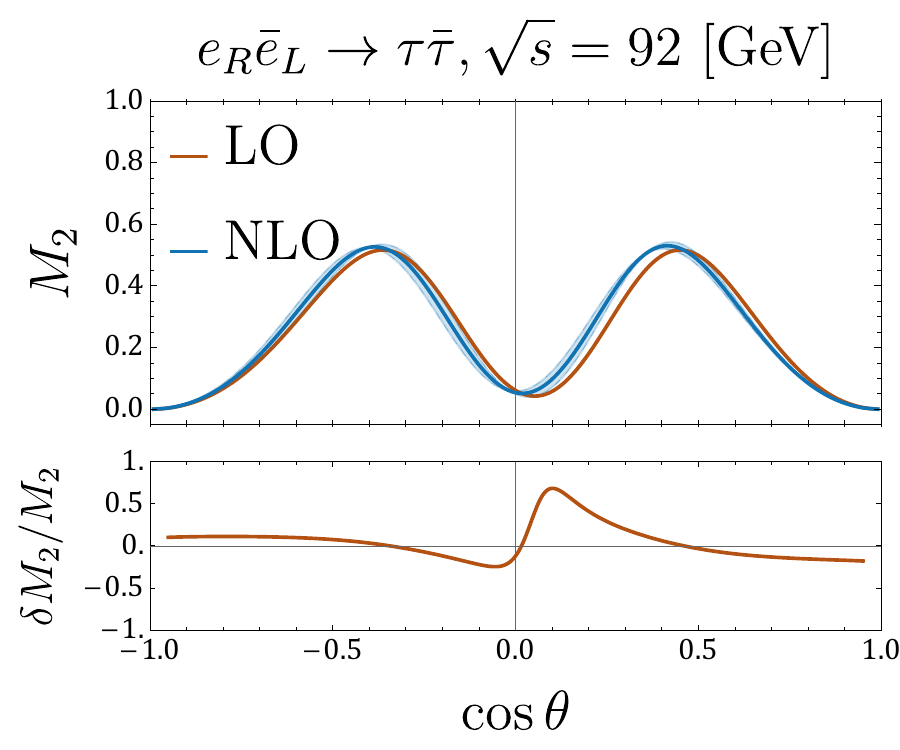}
\caption{Representative fixed-energy slices of the $\tau^-\tau^+$ stabilizer R\'enyi entropy $M_2$ at LO and NLO. As with $E_L$, the corrections are small $(\sim1\%)$ until energies near $M_W, M_Z$ are reached. Results for which $M_2 < 0$ arise as an artifact of truncating the perturbative series to $\mathcal{O}(\alpha)$. The shaded regions highlight the sensitivity of these calculations to NNLO contributions as estimated by the untruncated results (see Sec.~\ref{sec:results_tautau}).}
\end{figure}

\begin{figure}[H]
\centering
\includegraphics[width=0.32\linewidth]{MagPlotsDraft/eetautau/Slices_w_Error/LR132M.pdf}
\includegraphics[width=0.32\linewidth]{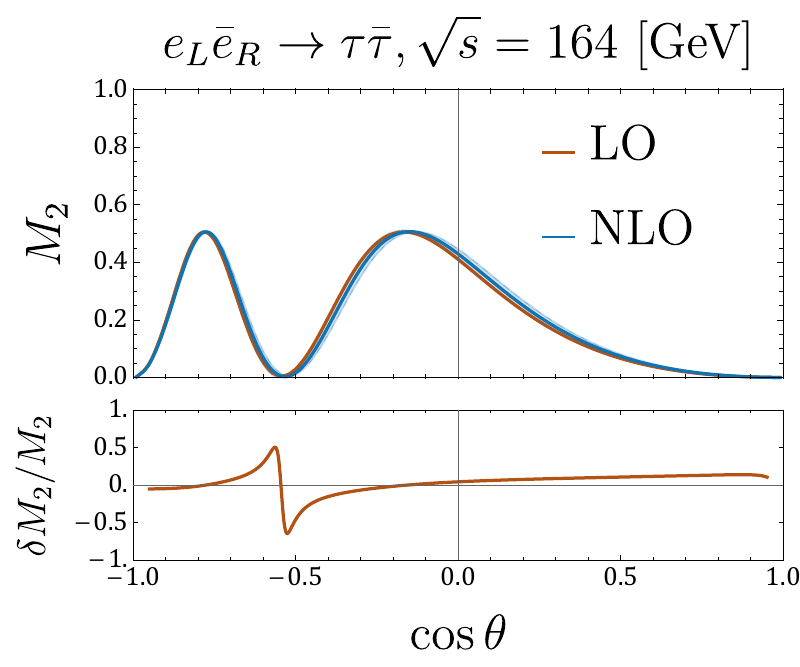}
\includegraphics[width=0.32\linewidth]{MagPlotsDraft/eetautau/Slices_w_Error/LR240M.pdf}\\[3pt]
\includegraphics[width=0.32\linewidth]{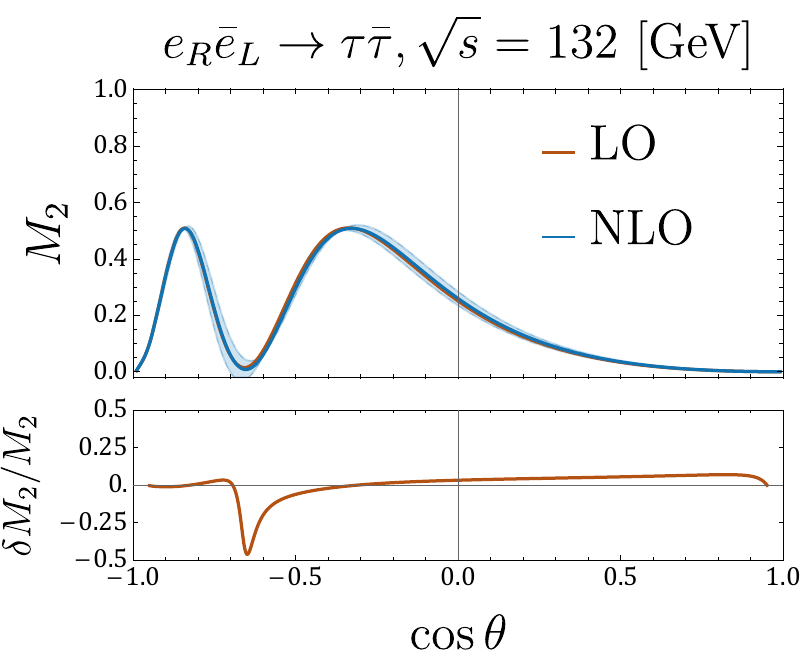}
\includegraphics[width=0.32\linewidth]{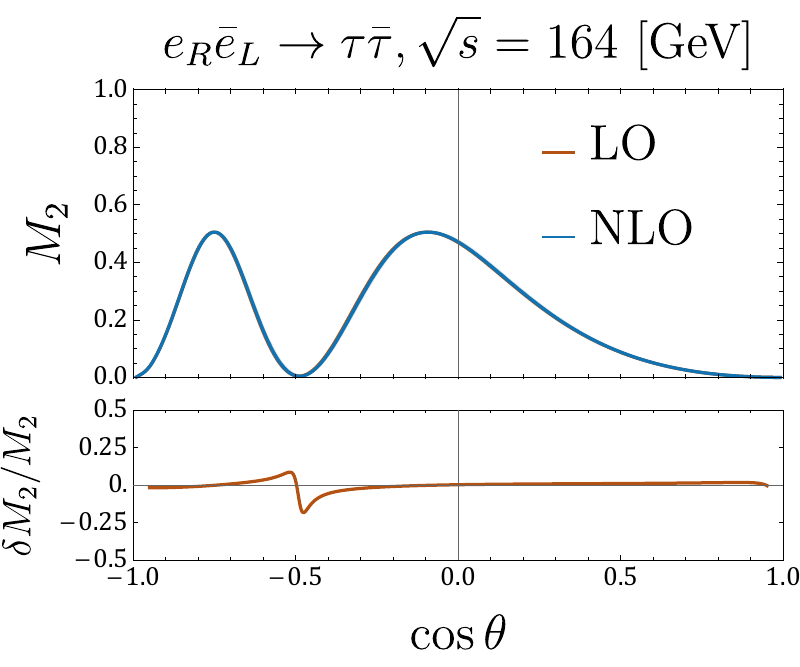}
\includegraphics[width=0.32\linewidth]{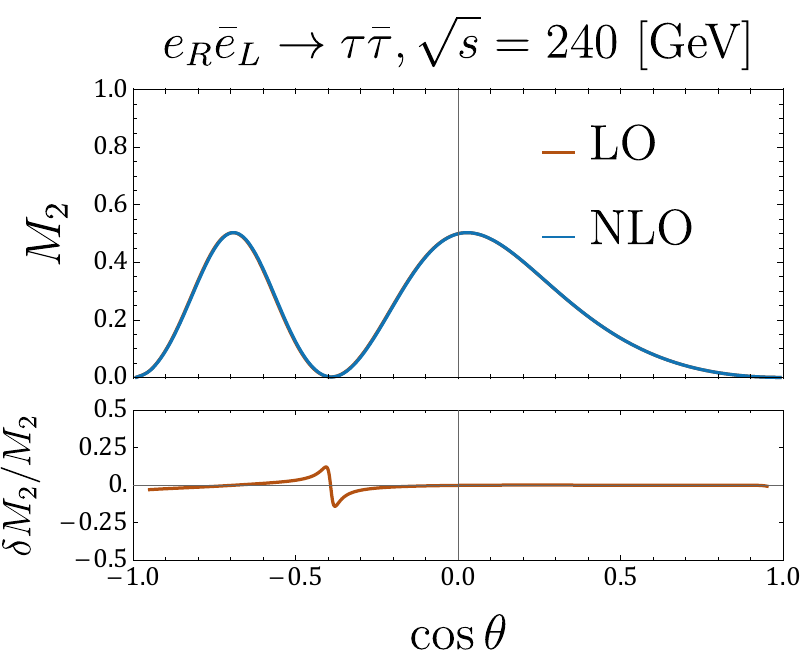}
\caption{Additional fixed-energy slices of the $\tau^-\tau^+$ stabilizer R\'enyi entropy $M_2$ at LO and NLO. Increasing $\sqrt{s}$ beyond $M_Z$ leads to generically decreasing NLO corrections. The shaded regions highlight the sensitivity of these calculations to NNLO contributions as estimated by the untruncated results (see Sec.~\ref{sec:results_tautau})}
\label{fig:tau_magic_slices}
\end{figure}

\subsection{$ZZ$ Supplementary Plots}

\begin{figure}[H]
\begin{subfigure}{.95\textwidth}
    \centering
    \includegraphics[width=0.48\linewidth]{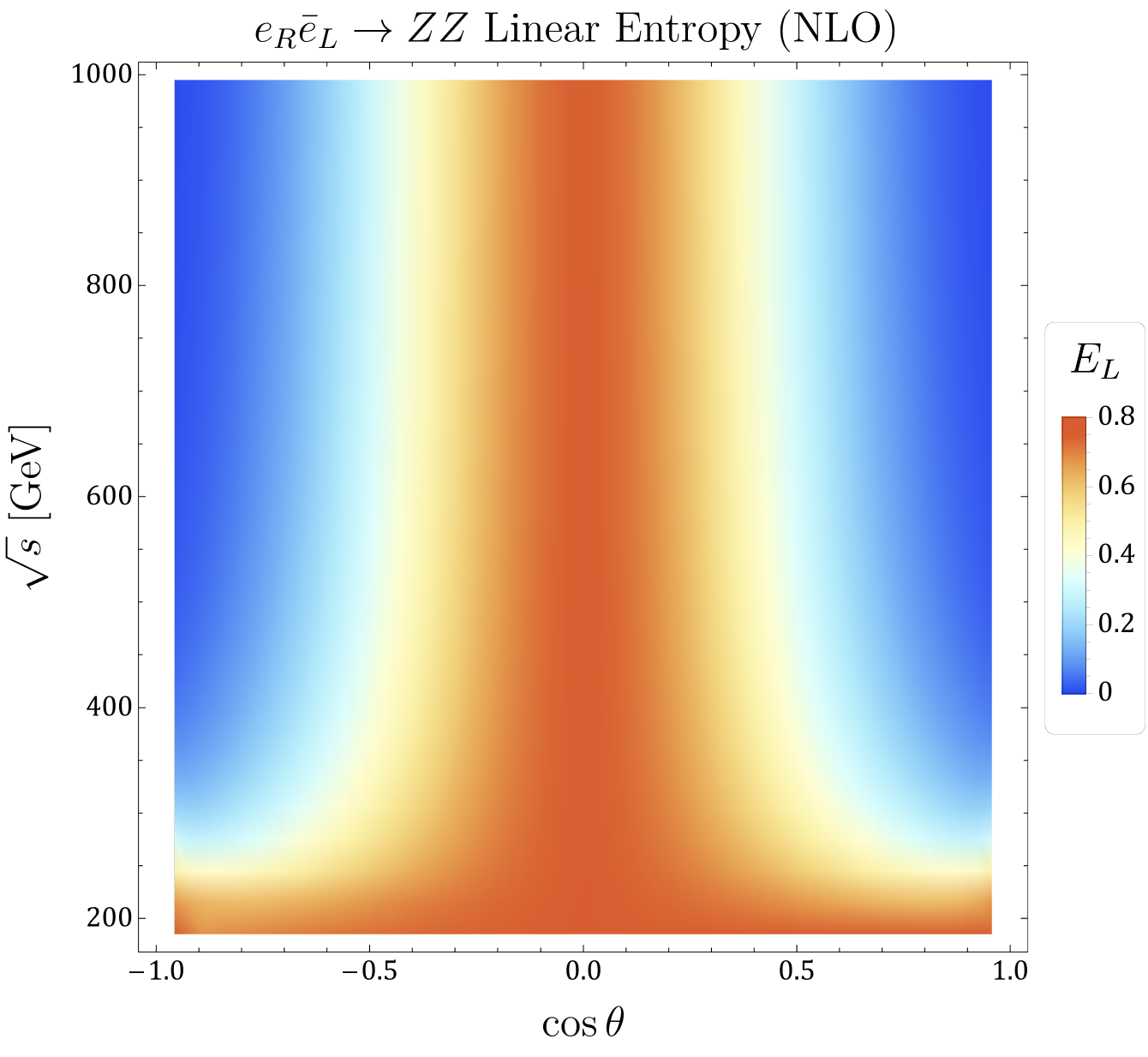}\hfill
    \includegraphics[width=0.48\linewidth]{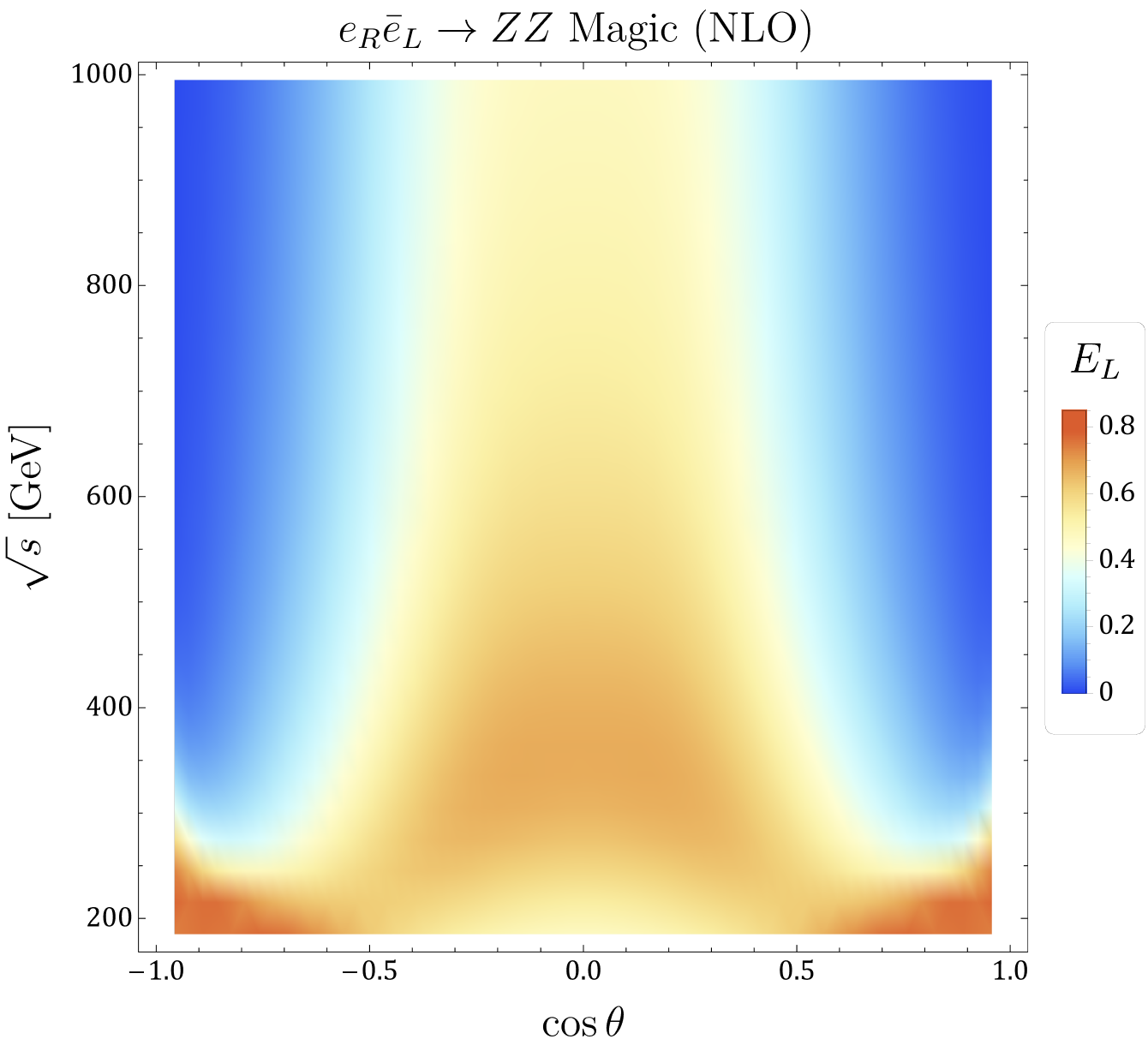}
    \caption{Density plots of linear entropy and stabilizer second R\'enyi entropy for $ZZ$ production from $RL$ initial states. }
\end{subfigure}
\begin{subfigure}{.95 \textwidth}
    \centering
    \includegraphics[width=0.32\linewidth]{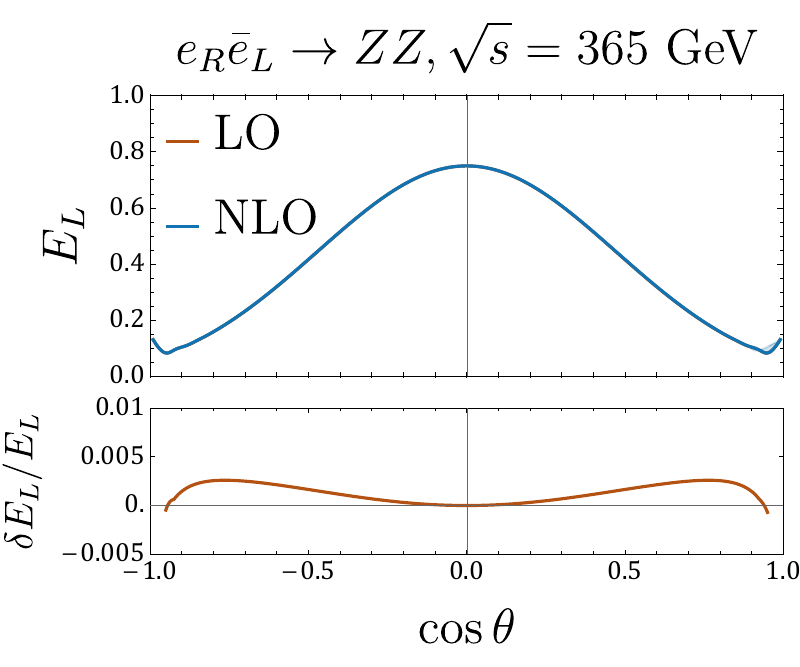}
    \includegraphics[width=0.32\linewidth]{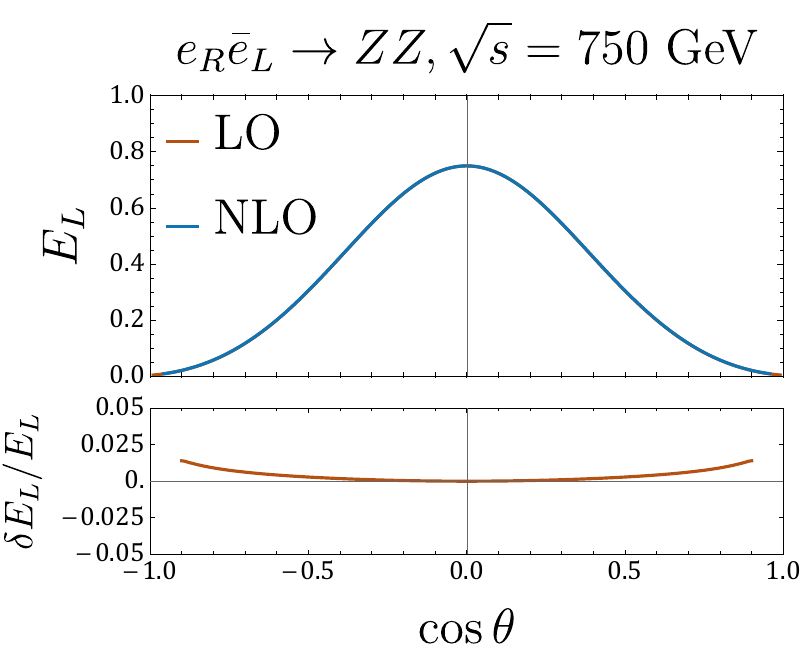}
    \includegraphics[width=0.32\linewidth]{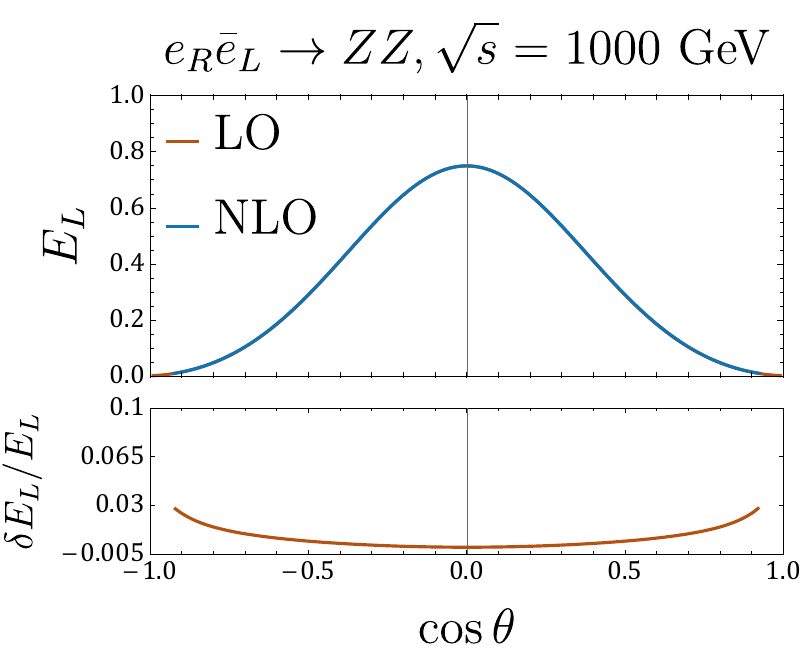}
    \includegraphics[width=0.32\linewidth]{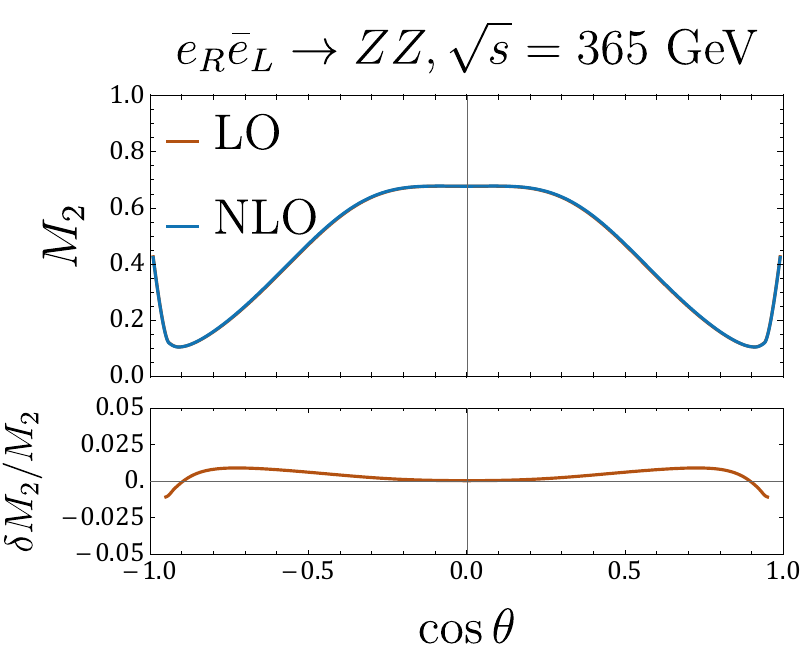}
    \includegraphics[width=0.32\linewidth]{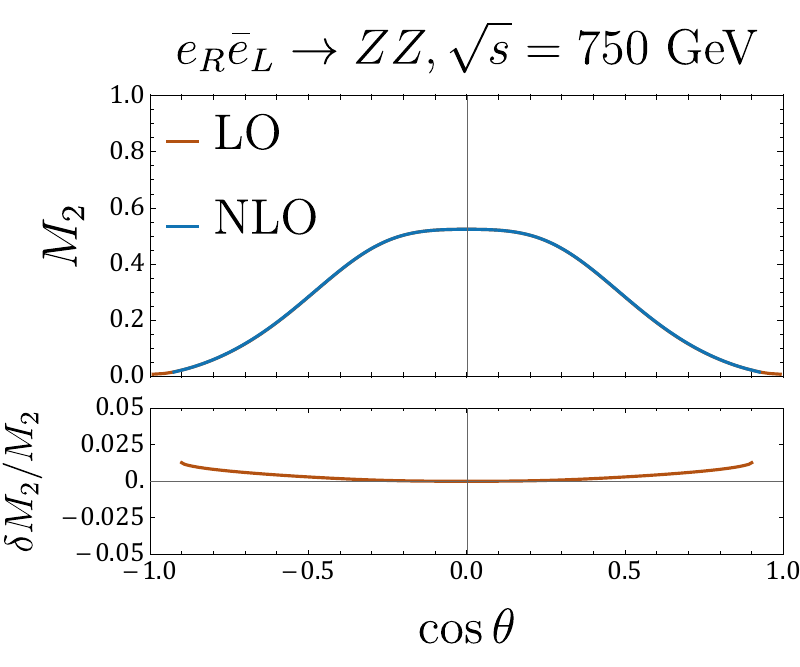}
    \includegraphics[width=0.32\linewidth]{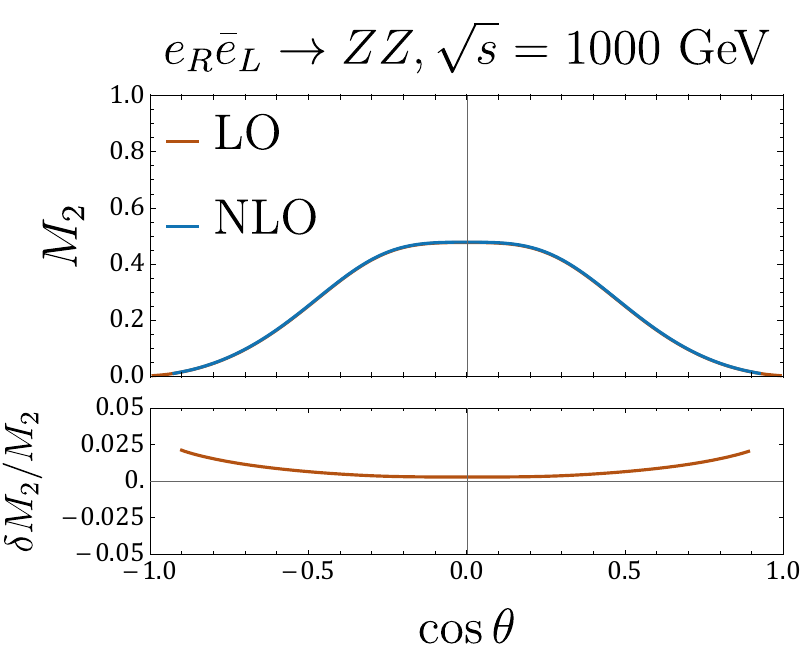}
    \caption{$ZZ$ linear-entropy and stabilizer second R\'enyi entropy slices for $RL$ initial states, comparing LO (red) and NLO (blue). The curves remain nearly indistinguishable, confirming that only small non-universal virtual corrections survive density-matrix normalization in this channel. The shaded regions (which fail to extend beyond the width of the NLO line) highlight the sensitivity of these calculations to NNLO contributions as estimated by the untruncated results (see Sec.~\ref{sec:results_tautau}).}
    \label{fig:zz_ent_slices_rl}
\end{subfigure}
\caption{ QI density plots and fixed energy slices for $e^-e^+ \to ZZ$ with $RL$ initial states at NLO. Compare with the corresponding $LR$ channels in Figs.~\ref{fig:zz_ent_lr} and~\ref{fig:zz_mag_lr}.}
\label{fig:zz_ent_rl}
\label{fig:zz_mag_rl}
\end{figure}

\section{Perturbative Expansion of the Normalized Quantum Observables}
\label{app:pert_expansion}

We show here that quantum observables computed from the normalized density matrix, away from unstable-particle poles, depend only on the real parts of the non-universal one-loop helicity amplitudes at NLO. This is the analytic counterpart of the cancellation argument discussed in Sec.~\ref{sec:IR_structure}.
\subsection{General NLO Expansion of Linear Entropy}
For a fixed initial-state helicity channel $ij$, we write the amplitude as \footnote{Unlike in section \ref{sec:IR_structure}, we separate out the factors of $\alpha$ from $\mathcal{M}$ and $\rho$ here to illustrate the power counting in expressions containing products of $\rho$ and $\mathcal{M}$. } 
\begin{equation}
\mathcal{M}_{kl,ij}
=
\alpha \mathcal{M}^{\text{LO}}_{kl,ij}
+ \alpha^2\mathcal{M}^{\text{1-loop}}_{kl,ij}
+ \mathcal{O}(\alpha^3),
\end{equation}
where $\alpha$ denotes the fine structure constant used for power counting and $k,l$ label the final-state helicities. The unnormalized two-body density matrix then has components
\begin{align}
[\rho]_{kl,k'l'} &= \alpha^2\,\mathcal{M}^{\text{LO}}_{kl}\,\overline{\mathcal{M}}^{\text{LO}}_{k'l'}
+ \alpha^3\big(\mathcal{M}^{\text{1-loop}}_{kl}\,\overline{\mathcal{M}}^{\text{LO}}_{k'l'} + \overline{\mathcal{M}}^{\text{1-loop}}_{k'l'}\,\mathcal{M}^{\text{LO}}_{kl}\big)
+ \mathcal{O}(\alpha^4)\,,
\label{eq:rhomatrix}
\end{align}
where the initial-state indices are suppressed from now on. Away from the poles of unstable-particle propagators, the tree-level amplitudes can be chosen real for the channels studied here, and the trace becomes
\begin{equation}
\mathrm{Tr}(\rho) = \alpha^2\,\mathrm{Tr}(\rho^{\text{LO}}) + 2 \alpha^3\sum_{k,l}\mathcal{M}^{\text{LO}}_{kl}\,\mathrm{Re}\!\big(\mathcal{M}^{\text{1-loop}}_{kl}\big) + \mathcal{O}(\alpha^4)\,,
\end{equation}
which depends only on $\mathrm{Re}\!\big(\mathcal{M}^{\text{1-loop}}\big)$.

The reduced density matrix for sector A $[\rho_A]_{k,k'} = \sum_l [\rho]_{kl,k'l}$ and its squared trace evaluate to
\begin{align}
\mathrm{Tr}\!\big((\rho_A)^2\big) &= \alpha^4\,\mathrm{Tr}\!\big((\rho_A^{\text{LO}})^2\big)
+ 4\alpha^5\sum_{k,l,m} [\rho_A^{\text{LO}}]_{km}\,\mathcal{M}^{\text{LO}}_{kl}\,\mathrm{Re}\!\big(\mathcal{M}^{\text{1-loop}}_{ml}\big)
+ \mathcal{O}(\alpha^6)\,,
\end{align}
which again involves only the real part of the one-loop amplitude. After normalization, the purity of the reduced state becomes
\begin{align}
\mathrm{Tr}\!\big((\hat\rho_A)^2\big)
&=
\frac{\mathrm{Tr}\!\big((\rho_A^{\text{LO}})^2\big)}{\mathrm{Tr}(\rho^{\text{LO}})^2}
+ \frac{4\alpha}{\mathrm{Tr}(\rho^{\text{LO}})^2}
\Bigg[
\sum_{k,l,m}[\rho_A^{\text{LO}}]_{km}\,\mathcal{M}^{\text{LO}}_{kl}\,
\mathrm{Re}\!\big(\mathcal{M}^{\text{1-loop}}_{ml}\big) \notag\\
&\qquad\qquad
- \frac{\mathrm{Tr}\!\big((\rho_A^{\text{LO}})^2\big)}{\mathrm{Tr}(\rho^{\text{LO}})}
\sum_{k,l}\mathcal{M}^{\text{LO}}_{kl}\,
\mathrm{Re}\!\big(\mathcal{M}^{\text{1-loop}}_{kl}\big)
\Bigg]
+ \mathcal{O}(\alpha^2)\,.
\label{eq:purity_nlo}
\end{align}
Since the linear entropy is $E_L = \frac{d}{d-1}(1 - \mathrm{Tr}((\hat\rho_A)^2))$, it follows that $E_L$ at NLO depends only on $\mathrm{Re}\!\big(\mathcal{M}^{\text{1-loop}}\big)$.

Equation~\eqref{eq:purity_nlo} also makes the normalization cancellation transparent: if the real part of the one-loop correction is proportional to the tree amplitude,
\begin{equation}
\mathrm{Re}\!\big(\mathcal{M}^{\text{1-loop}}_{kl}\big)
=
C\, \mathcal{M}^{\text{LO}}_{kl}
\end{equation}
for some constant $C$, then the NLO correction to the purity vanishes identically, and hence $E_L^{\text{NLO}} = E_L^{\text{LO}}$. A proportional correction is simply an overall rescaling of the amplitude, which is removed by the density-matrix normalization. Since the leading soft-photon virtual factor has precisely this universal multiplicative form~\cite{Weinberg:1965nx}, the same reasoning explains why the IR-sensitive multiplicative part of the fixed-Born virtual correction drops out of the normalized quantum observables. The residual unresolved-real contribution and hard collinear radiation require the separate treatment described in Sec.~\ref{sec:IR_structure}. The argument extends straightforwardly to the other smooth quantum observables used in this paper, including the stabilizer R\'enyi entropy $M_2$, which can be proved by showing a similar cancellation in  $\hat{\rho}$.

\subsection{Dependence of $M_2$ on ${\mathcal{M}}^{\text{1-loop}}$ }
For qubits and qutrits, we find that the stabilizer R\'enyi entropy at NLO is also independent of $\mathrm{Im} \big(\mathcal{M}^{\text{1-loop}}\big)$ away from unstable-particle poles, where width effects do not modify this real/imaginary separation.  To prove this, we have to treat qubits and qutrits separately.     In Eqns.~(\ref{eq:Malpha}, \ref{eq:Malpha_density}), for qubits, the Pauli operators are of the form $P=\{ {1\!\!1}, \sigma_x, \sigma_y, \sigma_z\}\otimes \{ {1\!\!1}, \sigma_x, \sigma_y, \sigma_z\}.$  From the properties of these operators, it is easy to see that $P^T=\pm P.$  If we think of the $kl$ final state as a single composite index, we can rewrite Eq.~\eqref{eq:rhomatrix} as 
\begin{align}
\rho = & \alpha^2 \mathcal{M}^\text{LO}\mathcal{M}^{\text{LO}\,T} + \nonumber\\ & \alpha^3\left[\mathrm{Re}\!\left(\mathcal{M}^\text{1-loop}\right)\mathcal{M}^{\text{LO}\,T}+\mathcal{M}^\text{LO}\mathrm{Re}\!\left(\mathcal{M}^{\text{1-loop}\,T}\right)\right]+\\& i \alpha^3\left[\mathrm{Im}\!\left(\mathcal{M}^\text{1-loop}\right)\mathcal{M}^{\text{LO}\,T}-\mathcal{M}^\text{LO}\mathrm{Im}\!\left(\mathcal{M}^{\text{1-loop}\,T}\right)\right] + \mathcal{O}(\alpha^4), \nonumber
\end{align}
where we've used the fact that the leading order amplitude is real and have split up the 1-loop amplitude into its real and imaginary components.  On the second line, the real part $\mathrm{Re}\!\left(\mathcal{M}^\text{1-loop}\right)$ is in a term that is symmetric under transpose, while the imaginary part on the third line is in an antisymmetric term.  Combined with the symmetry properties of $P$, we see that for qubits: 
\begin{align}
&  \mathrm{Tr}(\rho P) =\mathcal{O}(\alpha^4) + \\ & \left\{ \begin{array}{ll} \mathrm{Tr}\left[\left(\alpha^2 \mathcal{M}^\text{LO}\mathcal{M}^{\text{LO}\,T}+   \alpha^3 \mathrm{Re}\!\left(\mathcal{M}^\text{1-loop}\right)\mathcal{M}^{\text{LO}\,T}+   \alpha^3\mathcal{M}^\text{LO}\mathrm{Re}\!\left(\mathcal{M}^{\text{1-loop}\,T}\right)\right)P\right]  & \text{if } P^T = P \\
 i \alpha^3\mathrm{Tr}\left[\left(\mathrm{Im}\!\left(\mathcal{M}^\text{1-loop}\right)\mathcal{M}^{\text{LO}\,T}- \mathcal{M}^\text{LO}\mathrm{Im}\!\left(\mathcal{M}^{\text{1-loop}\,T}\right)\right)P\right] 
  & \text{if } P^T = -P\end{array} \right\} . \nonumber
\end{align}
Since there is no tree level piece surviving on the second line of the parentheses, the imaginary part of $\mathcal{M}^\text{1-loop}$ does not contribute to $|\mathrm{Tr}(\rho P)|^{2\alpha}$ (and thus to stabilizer entropy) at NLO order.  Turning to qutrits, the Pauli operators are parametrized as $P^{pqrs}=\left(X^p Z^q \right)\otimes\left(X^r Z^s \right),$ where $X, Z$ are the shift and clock operators, with powers $p,q,r,s=0,1,2.$  $X$ does not have well defined properties under transpose, but $Z^q$ is diagonal and thus symmetric.  Therefore, the above proof applies to the operators $P^{0q0s},$ but not to the remaining terms where $(p,r)\neq (0,0)$.  In fact, by numerical inspection we find that the imaginary amplitude contributes to individual $|\mathrm{Tr}(\rho P^{pqrs})|^{2\alpha}$ terms at NLO, but cancels in the sum over $p,q,r,s.$  To treat all of the terms, we write schematically $\mathrm{Tr}(\rho P^{pqrs}) = f(\mathrm{Re} (\mathcal{M}), i\, \mathrm{Im} (\mathcal{M}), \xi, i\chi)$ where $\xi$ and $i\chi$ are respectively the real and imaginary parameters in $P^{pqrs}$.  The only imaginary parameters needed are for $Z$ and $Z^2$ since they are defined with $\omega =e^{2\pi i/3}$, giving $i\chi = 2\pi i/3$.  Then $|\mathrm{Tr}(\rho P^{pqrs})|^{2\alpha} = |f(\mathrm{Re} (\mathcal{M}), i\, \mathrm{Im} (\mathcal{M}), \xi, i\chi)|^{2\alpha} = f(\mathrm{Re} (\mathcal{M}), i\, \mathrm{Im} (\mathcal{M}), \xi, i\chi)^{\alpha} f(\mathrm{Re} (\mathcal{M}), -i\, \mathrm{Im} (\mathcal{M}), \xi, -i\chi)^{\alpha}.$  For $(q,s)=(0,0)$, we only have $Z^0$, so no $\chi$ parameters are needed and $$|\mathrm{Tr}(\rho P^{p0r0})|^{2\alpha} =  f(\mathrm{Re} (\mathcal{M}), i\, \mathrm{Im} (\mathcal{M}), \xi, 0)^{\alpha} f(\mathrm{Re} (\mathcal{M}), -i\, \mathrm{Im} (\mathcal{M}), \xi, 0)^{\alpha}.$$ However, since this is even under $\mathrm{Im} (\mathcal{M})\to -\mathrm{Im} (\mathcal{M})$, there is no linear term in $\mathrm{Im} (\mathcal{M})$.  Now consider $(q,s)\neq(0,0).$  Define the tilde operation for $q,s$ as
\begin{align}
\tilde{q} \equiv \left\{\begin{matrix} 0 & \text{if } q=0 \\ 2 & \text{if } q=1 \\ 1 & \text{if } q=2\end{matrix} \right..    
\end{align}
This is useful because interchanging $Z=\mathrm{diag}\left(1,\omega,\omega^2\right)$ and $Z^2 = \mathrm{diag}\left(1,\omega^2,\omega^4\right)=\mathrm{diag}\left(1,\omega^{-1},\omega^{-2}\right)$, is equivalent to the exchange of $\omega =e^{2\pi i/3}\leftrightarrow e^{-2\pi i/3}=\omega^{-1}$ (and thus $\chi\to -\chi$).  Then combining terms related by the tilde operation
\begin{align}
|\mathrm{Tr}(\rho P^{pqrs})|^{2\alpha}+|\mathrm{Tr}(\rho P^{p\tilde{q}r\tilde{s}})|^{2\alpha} = & f(\mathrm{Re} (\mathcal{M}), i\, \mathrm{Im} (\mathcal{M}), \xi, i\chi)^{\alpha} f(\mathrm{Re} (\mathcal{M}), -i\, \mathrm{Im} (\mathcal{M}), \xi, -i\chi)^{\alpha}+ \nonumber \\ & f(\mathrm{Re} (\mathcal{M}), i\, \mathrm{Im} (\mathcal{M}), \xi, -i\chi)^{\alpha} f(\mathrm{Re} (\mathcal{M}), -i\, \mathrm{Im} (\mathcal{M}), \xi, i\chi)^{\alpha},
\end{align}
we find that these two terms together are also even under $\mathrm{Im} (\mathcal{M})\to -\mathrm{Im} (\mathcal{M})$, explaining the observed numerical cancellations. Hence, we have been able to prove that at NLO, $M_\alpha$ is independent of the imaginary part of the one-loop amplitude for qubits and qutrits. However, these proofs depended on the detailed properties of $P$ for these two cases and thus it is not clear if this result generalizes to higher dimensional qudits.

\clearpage
\section{One-Loop Computation Workflow}
\label{app:loop_Comp}
The one-loop electroweak amplitudes entering this work build on the classic calculations of radiative corrections to charged-lepton and diboson production~\cite{Passarino:1978jh,Denner:1988tv,Bohm:1987ck,Denner:2000bj} and are complementary to modern precision electroweak implementations and polarized calculations~\cite{Denner:2005nn,Denner:2026phn,Chen:2022dow,Dubovyk:2019szj,DelGratta:2025qyp,DelGratta:2025xjp,Dittmaier:2025ikh,Denner:2025xdz, Denner:1991kt}. For the quantum observables, however, we require the full helicity amplitudes for all initial-state configurations, including the helicity-suppressed $LL$ and $RR$ channels for which compact ready-to-use analytic expressions are not generally available. We therefore rederived the renormalized virtual amplitudes directly from the SM Lagrangian for $e^-e^+ \to \tau^-\tau^+$ and $ZZ$. The diagram count is of order a few hundred per process, so the computation was automated with \texttt{FeynArts}~\cite{Hahn:2000kx}, \texttt{FeynCalc}~\cite{Mertig:1990an,Shtabovenko:2023idz}, \texttt{FeynHelpers}~\cite{Shtabovenko:2016whf} and \texttt{LoopTools}~\cite{Hahn:1998yk}.

We renormalized the electroweak SM amplitudes in the on-shell renormalization scheme, using the conditions detailed in \cite{Bohm:1986rj} and applied in the one loop calculations \cite{Bohm:1987ck, Denner:1988tv}. To summarize the outline of the renormalization scheme: UV divergences are regulated via dimensional regularization, wherein the space time dimension $D$ is analytically continued to $D = 4  - 2 \epsilon$, with $\epsilon$ eventually taken to zero. Infrared divergences from virtual photon exchange are instead regulated by the fictitious photon mass $\lambda$ introduced in Sec.~\ref{sec:IR_structure}, set numerically to $\lambda=2m_e$. The IR divergent part of the one-loop amplitude, which is proportional to the Born amplitude, cancels in the normalized observables through the trace projection of Eq.~\eqref{eq:rhohat_at_NLO} for any value of $\lambda$; the residual regulator dependence enters only through power corrections not proportional to the Born amplitude, of relative size $\mathcal O(\lambda^2/s)$ and therefore numerically negligible. Infinite parts of the counterterms are then determined by the requirement that tadpole contributions and factors of $\Delta_M = \frac{1}{\epsilon} - \gamma_E - \log(\frac{M^2}{4\pi \mu^2})$ are canceled in the overall amplitude. Finite parts of the mass, charge, and wavefunction renormalization counterterms relevant to the above channels follow from the following prescriptions:\\

-- The poles of the renormalized propagators are located at the bare masses of the respective fields:
\begin{equation}\label{renCond1}
\text{Re}(\tilde\Sigma^W_{T/L}(M_W^2)) =\text{Re}(\tilde\Sigma^Z_{T/L}(M_Z^2)) = \text{Re}(\tilde\Sigma^H(M_H^2)) =\text{Re}( \tilde\Sigma^{\gamma Z}(0)) =\text{Re}(\tilde\Sigma^f(m_f^2)) = 0.
\end{equation}\\ \indent 
-- The residues of the renormalized Higgs, photon, and the left-chiral fermionic propagators are unity:
\begin{equation}\label{renCond2}
\frac{1}{k^2}\text{Re}(\tilde\Sigma^\gamma_{T}(k^2)) = \frac{\partial}{\partial \slashed p} \text{Re}((\tilde\Sigma^f_\textbf{L}(\slashed p))\Bigg|_{\slashed p = m_f} =\frac{\partial}{\partial p^2}\text{Re}(\tilde\Sigma^h(p^2))\Bigg|_{p^2= m_h^2} = 0
\end{equation}
where $\tilde\Sigma^A$ is the renormalized self energy of the field(s) indexed by $A$. The $T, L$ subscripts refer to the transverse and longitudinal components of the vector boson propagators. The \textbf{L} subscript refers to the left handed part of the fermionic self energy functions.  \\ \indent

--The charge renormalization counterterm is fixed such that the corrected photon vertex satisfies:\\
\begin{equation}
\Gamma^{\gamma e e}_\mu(k^2 = 0,\slashed{p} = m_e, \slashed{q} = m_e) = ie\gamma_\mu
\end{equation}
where $k, p,  \text{and } q$ are the respective momenta carried by the photon and two electron fields into the vertex. In our calculations, we use $\alpha$ (defined in the Thomson limit), $M_W$, $M_Z$, $M_H$, and the fermion masses as our input parameters, with numerical values \cite{ParticleDataGroup:2024cfk}: \\
\begin{equation*}
\alpha  = 0.00729735, M_W =  80.362 \text{ GeV}, M_Z = 91.188 \text{ GeV}, M_H = 125.13\text{ GeV,}
\end{equation*}
\begin{equation*}
m_e = .511 \text{ MeV}, m_\mu = 105.65 \text{ MeV}, m_\tau = 1.776 \text{ GeV,}
 \end{equation*}
 \begin{equation*}
m_u = 2.16 \text{ MeV}, m_d = 4.70 \text{ MeV}, m_s = 92.9 \text{ MeV},
 \end{equation*}
 \begin{equation*}
  m_c= 1.27 \text{ GeV}, m_b = 4.18 \text{ GeV}, m_t = 172.60\text{ GeV}.
 \end{equation*}

 Additionally, the complex-mass scheme (CMS) is invoked when performing computations near the resonances of the Higgs and $Z$-Boson \cite{Denner:2006ic, TranThanhVan:1990xi}. In this prescription, the bare masses of the unstable particles are substituted to include the imaginary NLO corrections to the poles of the corresponding propagators: 
 \begin{equation}
 M_Z^2 \rightarrow M_Z^2 - iM_Z \Gamma_z,\text{  } M_H^2 \rightarrow M_H^2 - iM_H\Gamma_H
 \end{equation}
 where $\Gamma_Z, \Gamma_H$ are the decay widths of the Higgs and Z-Boson, calculated to $\mathcal{O}(\alpha)$. To avoid double counting the NLO corrections, the counterterm lagrangian is modified to cancel this change in the bare lagrangian with the renormalization conditions above now applied to both the real and imaginary parts of the self energy corrections for the unstable particles. The complex mass substitutions are invoked in both the leading and next to leading results we report for the $e^-e^+ \rightarrow \tau^-\tau^+$ channel. Our leading order calculations are reported without any perturbative truncation while the NLO results are truncated according to the description in Sec.~\ref{sec:results_tautau}. 
 
The counterterms are solved for using the tools within \texttt{FeynCalc}, \texttt{FeynArts}, and \texttt{FeynHelpers}, the results of which are verified against \cite{Bohm:1986rj} and \cite{Denner:1991kt} where applicable. With these in hand, the computation of the overall loop results proceeds in two main steps: 

\paragraph{1. Diagram generation and tensor reduction.}
For each channel, \texttt{FeynArts}~\cite{Hahn:2000kx} is used to generate the tree and one-loop diagrams together with the corresponding counterterm insertions. The resulting amplitudes are imported into \texttt{FeynCalc}~\cite{Mertig:1990an,Shtabovenko:2023idz}, where the tensor loop integrals are reduced to the standard scalar basis $A_0$, $B_0$, $C_0$, and $D_0$ using the Passarino--Veltman framework~\cite{Passarino:1978jh,tHooft:1978jhc}. At this stage the amplitudes still carry the full spinorial and Lorentz structure, including external wavefunctions, gamma-matrix chains, metric contractions and Levi-Civita tensors. We then apply local routines that insert the external-state projectors appropriate to each helicity channel and simplify the resulting expressions to a form suitable for numerical evaluation.

\paragraph{2. Scalar-integral evaluation and renormalized amplitudes.}
After tensor reduction, the UV-divergent scalar functions $A_0$, $B_0$ and $C_0$ are combined with the counterterms and evaluated with the help of \texttt{FeynHelpers}~\cite{Shtabovenko:2016whf}. The cancellation of the $\Delta$ poles provides a stringent check of the implementation. The box integrals encoded in $D_0$ are UV finite, so we keep them in scalar form until numerical phase-space points are specified; they are then evaluated with \texttt{LoopTools}~\cite{Hahn:1998yk}, which proved more stable in the kinematic regimes relevant here. Largely inspired by the computational bottleneck of the box diagrams, the numerical evaluation was parallelized and run on a computing cluster.

The final output of this workflow is a set of renormalized one-loop helicity amplitudes as functions of $\sqrt{s}$ and $\cos\theta$. These amplitudes are then assembled into the normalized helicity density matrices discussed in Sec.~\ref{sec:scattering}, from which the entanglement and magic observables are extracted.

\clearpage
\bibliographystyle{JHEP}
{\footnotesize
\bibliography{citations}}
\end{document}